\documentclass[preprintnumbers, superscriptaddress, showpacs, nofootinbib, amsfonts, amsmath, amssymb, aps, prd, notitlepage, floatfix]{revtex4-2}
\usepackage[colorlinks=true, linkcolor=blue, citecolor=blue, urlcolor=blue]{hyperref}
\usepackage{xcolor}
\usepackage{graphicx}
\usepackage{subfigure}
\usepackage{multirow}
\usepackage{slashed}
\usepackage{physics}

\newcommand{\dhd}{{\textstyle d} \lower.03ex\hbox{\kern-0.38em$^{\scriptstyle-}$}\kern-0.05em{}}

\begin{document}

\title{A unified description of DGLAP, CSS, and BFKL: TMD factorization bridging large and small x}

\author{Swagato~Mukherjee}
\email{swagato@bnl.gov}
\affiliation{Physics Department, Brookhaven National Laboratory, Upton, New York 11973, USA}
\author{Vladimir~V.~Skokov}
\email{vskokov@ncsu.edu}
\affiliation{Department of Physics, North Carolina State University, Raleigh, NC 27695, USA}
\author{Andrey~Tarasov}
\email{ataraso@ncsu.edu}
\affiliation{Department of Physics, North Carolina State University, Raleigh, NC 27695, USA}
\affiliation{Joint BNL-SBU Center for Frontiers in Nuclear Science (CFNS) at Stony Brook University, Stony Brook, New York 11794, USA}
\author{Shaswat~Tiwari}
\email{sstiwari@ncsu.edu}
\affiliation{Department of Physics, North Carolina State University, Raleigh, NC 27695, USA}
\begin{abstract}
This paper introduces a transverse-momentum dependent (TMD) factorization scheme designed to unify both large and small Bjorken-x regimes. We compute the next-to-leading order (NLO) quantum chromodynamics (QCD) corrections to the gluon TMD operator for an unpolarized hadron within this proposed scheme. This leads to the emergence of a new TMD evolution, incorporating those in transverse momentum, rapidity, and Bjorken-x. When matched to the collinear factorization scheme, our factorization scheme faithfully reproduces the well-established Dokshitzer-Gribov-Lipatov-Altarelli-Parisi (DGLAP) and Collins-Soper-Sterman (CSS) evolutions. Conversely, matching with high-energy factorization not only yields the Balitsky-Fadin-Kuraev-Lipatov (BFKL) evolution but also reveals distinctive signatures of CSS logarithms. The development of this novel TMD factorization scheme, capable of seamlessly reconciling disparate Bjorken-x regimes and faithfully reproducing established QCD evolution equations, has the potential to significantly advance our comprehension of high-energy processes and three-dimensional parton structures of hadrons.
\end{abstract}
\date{\today}
\maketitle
\tableofcontents

\section{Introduction\label{sec:intro}}
Factorization properties of quantum chromodynamics (QCD)
, i.e., separation of QCD interactions into distinct dynamical modes,
are essential for understanding high-energy scattering \cite{Collins:2011zzd} phenomena. They provide a foundation for the quantitative description of a wide class of observables.
Factorization typically occurs in scattering problems with a hierarchy of scales. Depending on the details of this hierarchy, the structure of factorization can be quite different.

The factorization phenomenon can be formalized in terms of the so-called factorization theorems. Generally speaking, these theorems define an observable as a convolution of functions, each representing a particular dynamical mode. Each function depends on a set of factorization scales, which could be roughly understood as the boundaries between different modes. The physical observable, of course, does not depend on these scales.

In the recent decade, the transverse-momentum dependent (TMD) factorization \cite{Collins:1981uk,Collins:1984kg,Collins:1987pm,Collins:1989gx,Meng:1995yn,Ji:2004wu,Ji:2004xq,Boussarie:2023izj} has become one of the main avenues of research in this field. This factorization scheme can be used in analysis of high-energy scatterings with productions of a final state with transverse momentum, $q_\perp$,  much smaller than a hard scale of the interaction, $Q$, i.e., $q^2_\perp / Q^2 \ll 1$. The approach has been successfully implemented in analysis of a variety of scattering phenomena and observables, see e.g. Refs.~\cite{Echevarria:2014xaa,Kang:2015msa,Bacchetta:2017gcc,Bacchetta:2019sam,Scimemi:2019cmh,Bertone:2019nxa,Bacchetta:2019sam,Bacchetta:2020gko,Echevarria:2020hpy,Kang:2020xgk,Cammarota:2020qcw,Bury:2021sue}.

For example, in the TMD factorization approach, the production of a color-singlet state (Drell-Yan pair, Higgs and $W$ boson productions etc.) in an unpolarized hadron-hadron scattering reads
\begin{eqnarray}
&&\frac{d\sigma}{dQdyd^2q_\perp} = \sum_{ij} H_{ij}(Q, \mu) \int d^2b_\perp e^{iq_\perp b_\perp} f_i(x_a, b_\perp, \mu, \zeta_a)f_j(x_b, b_\perp, \mu, \zeta_b) + O\left(\frac{q^2_\perp}{Q^2}\right)
\,,
\label{int:TMDff}
\end{eqnarray}
where $Q$ is an invariant mass of the final state, $y$ is its rapidity, and $q_\perp$ is the measured transverse momentum\footnote{We use the following notation $q_\perp b_\perp \equiv q_m b_m$.}. The variables $x_{a(b)}$ and $\zeta_{a(b)}$ are defined as,
\begin{equation}
x_a = \frac{Qe^y}{E_{\rm cm}}, \ \ \ \ \ x_b = \frac{Qe^{-y}}{ E_{\rm cm}}
\end{equation}
\begin{equation}
\zeta_a = 2(x_aP^-_a)^2 e^{-2y_n},\ \ \ \ \zeta_b=2(x_bP^+_b)^2 e^{2y_n},\ \ \ \ \zeta_a\zeta_b=Q^4\,
\end{equation}
where $P_a$ and $P_b$ are the momenta of the two colliding hadrons such that the center of mass energy $E^2_{\rm cm} = 2 P^-_a P^+_b$. A parameter $y_n$ encodes scheme dependence and is usually chosen as $y_n=0$.

 For studies of the power corrections to the TMD factorization formula (\ref{int:TMDff}) see \cite{Balitsky:2017flc,Balitsky:2017gis,Ebert:2018gsn,Balitsky:2020jzt,Vladimirov:2023aot}. The sum goes over partons participating in the hard scattering defined by a hard function $H_{ij}$. The transverse-momentum dependent parton distribution functions (TMDPDFs) depend on an impact parameter variable $b_\perp$, the Fourier conjugate to the measured transverse momentum $q_\perp$. The $x_a$ and $x_b$ are the longitudinal-momentum fractions of the colliding hadrons carried by the partons involved in the hard scattering.

The functions in Eq. (\ref{int:TMDff}) depend on unphysical factorization scales, $\zeta$ and $\mu$, defining the separation of the dynamical modes in the TMD factorization scheme. Dependence on these parameters can be studied by perturbative methods and is governed by the anomalous dimensions $\gamma^i_\mu$ and $\gamma^i_\zeta$:
\begin{eqnarray}
&&\frac{d}{d\ln\mu}f_i(x, b_\perp, \mu, \zeta) = \gamma^i_\mu(\mu, \zeta)f_i(x, b_\perp, \mu, \zeta)
\,,
\label{ee:mu}
\end{eqnarray}
\begin{eqnarray}
&&\frac{d}{d\ln\zeta}f_i(x, b_\perp, \mu, \zeta) = \frac{1}{2}\gamma^i_\zeta(\mu, b_\perp)f_i(x, b_\perp, \mu, \zeta)
\,,
\label{ee:zeta}
\end{eqnarray}
where the second equation is
known as the Collins-Soper equation \cite{Collins:1981uk,Collins:1981va}. The all-order form of the anomalous dimensions is given by
\begin{eqnarray}
&&\gamma^i_\mu(\mu, \zeta) = \Gamma^i_{\rm cusp}[\alpha_s(\mu)]\ln\frac{\mu^2}{\zeta} + \gamma^i_s[\alpha_s(\mu)];\ \ \ \ \ \gamma^i_\zeta(\mu, b_\perp) = -2\int^\mu_{1/b_\perp}\frac{d\mu'}{\mu'}\Gamma^i_{\rm cusp}[\alpha_s(\mu')]+ \gamma^i_r[\alpha_s(1/b_\perp)] \,,
\end{eqnarray}
where $\Gamma^i_{\rm cusp}$ is the light-like cusp anomalous dimension \cite{Polyakov:1980ca,Korchemsky:1985xj}, which is known up to three loops in QCD \cite{Moch:2004pa}. The soft anomalous dimension $\gamma^i_s$ is known to three loops \cite{Baikov:2009bg,Li:2014afw,Moch:2004pa,Gehrmann:2010ue,Lee:2010cga,Blumlein:2021enk}, as well as the rapidity anomalous dimension $\gamma^i_r$ \cite{Li:2016axz,Li:2016ctv,Vladimirov:2016dll}.

The evolution equations, Eqs. (\ref{ee:mu}) and (\ref{ee:zeta}), 
can be used to give phenomenological predictions for TMDPDFs 
at different scales, see e.g. \cite{Scimemi:2018xaf}. The solution of these equations can be constructed, and the TMDPDF can be evolved from initial scales, $(\mu_0, \zeta_0)$, to final scales, $(\mu, \zeta)$, using
\begin{eqnarray}
&&f_i(x, b_\perp, \mu, \zeta) = \exp\Big(\int^\mu_{\mu_0}\frac{d\tilde{\mu}}{\tilde{\mu}}\gamma^i_\mu(\tilde{\mu}, \zeta_0)\Big)\exp\Big(\frac{1}{2}\gamma^i_\zeta(\mu, b_\perp)\ln\frac{\zeta}{\zeta_0}\Big)f_i(x, b_\perp, \mu_0, \zeta_0)
\,.
\label{tmd-evolved}
\end{eqnarray}
The above evolution resums large logarithms $\ln(\mu^2 b^2_\perp)$ and $\ln(\zeta b^2_\perp)$. As we will discuss later, these logarithms are of the ultraviolet (UV) origin, and the infrared (IR) structure of the TMD factorization is encoded in the boundary term $f_i(x, b_\perp, \mu_0, \zeta_0)$.

The TMDPDF, $f_i(x, b_\perp, \mu_0, \zeta_0)$, is intrinsically nonperturbative. However, for small values of $b_\perp \ll \Lambda^{-1}_{\rm QCD}$ the TMDPDF can be expanded around $b_\perp=0$ and matched
onto collinear parton distribution functions (PDFs) \cite{Collins:1984kg,Collins:1981uw,Kang:2012em,Sun:2013hua,Dai:2014ala,Braun:2009mi,Echevarria:2015byo,Scimemi:2019gge}. In this case $f_i(x, b_\perp, \mu_0, \zeta_0)$ is represented as an infinite series of PDFs with growing twists. These PDFs contain the non-perturbative structure of the TMDPDF. The coefficients of this expansion encode the perturbative IR component and give the evolution kernels for the corresponding collinear PDFs. For example, the leading twist-$2$ term of the expansion is known up to the NNLO \cite{Echevarria:2015byo,Catani:2011kr,Catani:2012qa,Gehrmann:2014yya,Lubbert:2016rku,Echevarria:2016scs} and contains the Dokshitzer–Gribov–Lipatov–Altarelli–Parisi  (DGLAP) evolution kernel~\cite{Dokshitzer:1977sg,Gribov:1972ri,Altarelli:1977zs}.

This approach is routinely used in phenomenological applications. Currently, only the leading term of the expansion is applied, see for details \cite{Echevarria:2014xaa,Scimemi:2017etj,Bacchetta:2017gcc}. However, there is a place for criticism. The basis of it is that the region of applicability of the TMD factorization approach is limited to small values of $q_\perp$, which means that a typical value of $b_\perp\sim 1/q_\perp$ is large. This means that the TMDPDFs genuinely contain contributions from all collinear twists. Hence, to obtain the correct IR structure of the TMDPDFs we would need to resum all terms of the collinear expansion, which is not feasible at the moment.
Thus, 
in practice, one simply extrapolates the leading twist term, which dominates in the small $b_\perp\ll \Lambda^{-1}_{\rm QCD}$ region to the desired region of applicability, by introducing some functions to model the nonperturbative features of TMDPDFs, see for instance Refs. \cite{Landry:2002ix,Konychev:2005iy,Becher:2011xn,DAlesio:2014mrz}.

While the collinear matching approach allows us to extract some information about the IR structure of TMDPDFs from the region of large $b_\perp \lesssim \Lambda^{-1}_{\rm QCD}$, there is an alternative approach that aims to look at the $x$ dependence of TMDPDFs, particularly in the region of small $x$.

It is well known that TMDPDFs are related to the dipole amplitudes which are the primary objects of consideration in the small $x$ regime \cite{Dominguez:2010xd,Dominguez:2011wm,Xiao:2017yya}. These amplitudes are analogous to PDFs in the collinear factorization approach, though their nature is absolutely different. First of all, the dipole amplitudes are functions of the impact parameter $b_\perp$ and, as a result, they contain contributions from all collinear twists. Since collinear PDFs don't have $b_\perp$ dependence, the dipole amplitudes are more similar to TMDPDFs.
However, the collinear PDFs depend on $x$ variable, while the dipole amplitudes have no such dependence. They should instead be understood as the small $x$ limit of TMDPDFs.

There are well-developed methods for studying the IR structure of the dipole amplitudes. One might thus consider instead of matching onto collinear PDFs matching onto dipole amplitudes. 
The formalism leading to the dipole amplitudes is based on the eikonal approximation with sub-eikonal corrections, which can be systematically extracted from expansion in eikonality~\cite{Altinoluk:2014oxa,Altinoluk:2015gia,Kovchegov:2015pbl,Altinoluk:2015xuy,Balitsky:2015qba,Kovchegov:2016zex,Balitsky:2016dgz,Balitsky:2017flc,Balitsky:2017gis,Agostini:2019avp,Agostini:2019hkj,Cougoulic:2019aja,Altinoluk:2020oyd,Altinoluk:2021lvu,Kovchegov:2021iyc,Chirilli:2021lif,Cougoulic:2022gbk,Agostini:2022oge}; computing sub-eikonal corrections is currently under active research in the small-$x$ community. Each term of this expansion contains the contribution of all collinear twists. The leading term of the expansion contains the Balitsky-Fadin-Kuraev-Lipatov (BFKL) evolution kernel ~\cite{Fadin:1975cb,Kuraev:1976ge,Kuraev:1977fs,Balitsky:1978ic}, which is the linear approximation of the Balitsky-Kovchegov (BK) ~\cite{Balitsky:1995ub,Balitsky:1997mk,Kovchegov:1999yj} and the Jalilian-Marian-Iancu-McLerran-Weigert-Leonidov-Kovner ~\cite{Jalilian-Marian:1997qno,Jalilian-Marian:1997ubg,Kovner:2000pt,Iancu:2000hn,Ferreiro:2001qy,Kovner:2013ona,Kovner:2014lca,Lublinsky:2016meo} evolution. The BFKL kernel is a counterpart of the DGLAP  kernel in the collinear factorization.

However, the matching onto dipole amplitudes is not a wholly satisfactory approach either. Its main flaw is similar to the collinear matching procedure. While each expansion term contains contributions from all twists, the dipole amplitudes do not depend on $x$. The eikonal expansion is effectively an expansion in $x$; hence, to reconstruct the full dependence of TMDPDFs on $x$, one has to resum all terms of the expansion, which is currently impossible. Similarly, each term in the collinear matching expansion contains a full dependence on $x$ but has no dependence on $b_\perp$. Therefore, to reconstruct the full $b_\perp$ dependence of TMDPDFs, one has to resum the whole series too.

As a result, we see that neither matching procedure allows us to reveal the full IR structure of the TMDPDFs. The main goal of this paper is to propose an alternative approach. To study the IR structure of the TMDPDFs we perform a next-to-leading order (NLO) computation of the TMDPDFs in the background field technique. Instead of expanding in either $b_\perp$ or $x$, we perform the calculation in full kinematics. In general, such calculation is challenging. To simplify the problem we perform the calculation in the dilute limit when the NLO correction is calculated in the background of only two partons. The gauge invariance provides us with a guiding principle to express our final result in terms of appropriate gauge links. 

An important element of our calculation is the separation between ``quantum'' and background field modes, which appear in the background field method~\cite{Abbott:1980hw,Abbott:1981ke}. While the initial TMD operator is constructed from both of these modes, the perturbative part of the IR structure is defined by the ``quantum" fields. To obtain this structure, we integrate over ``quantum'' fields perturbatively at NLO. At the same time, the non-perturbative part of the IR structure is encoded in the background fields which are fixed and give rise to TMD operators at an infrared scale.

\begin{figure}[htb]
 \begin{center}
\includegraphics[width=100mm]{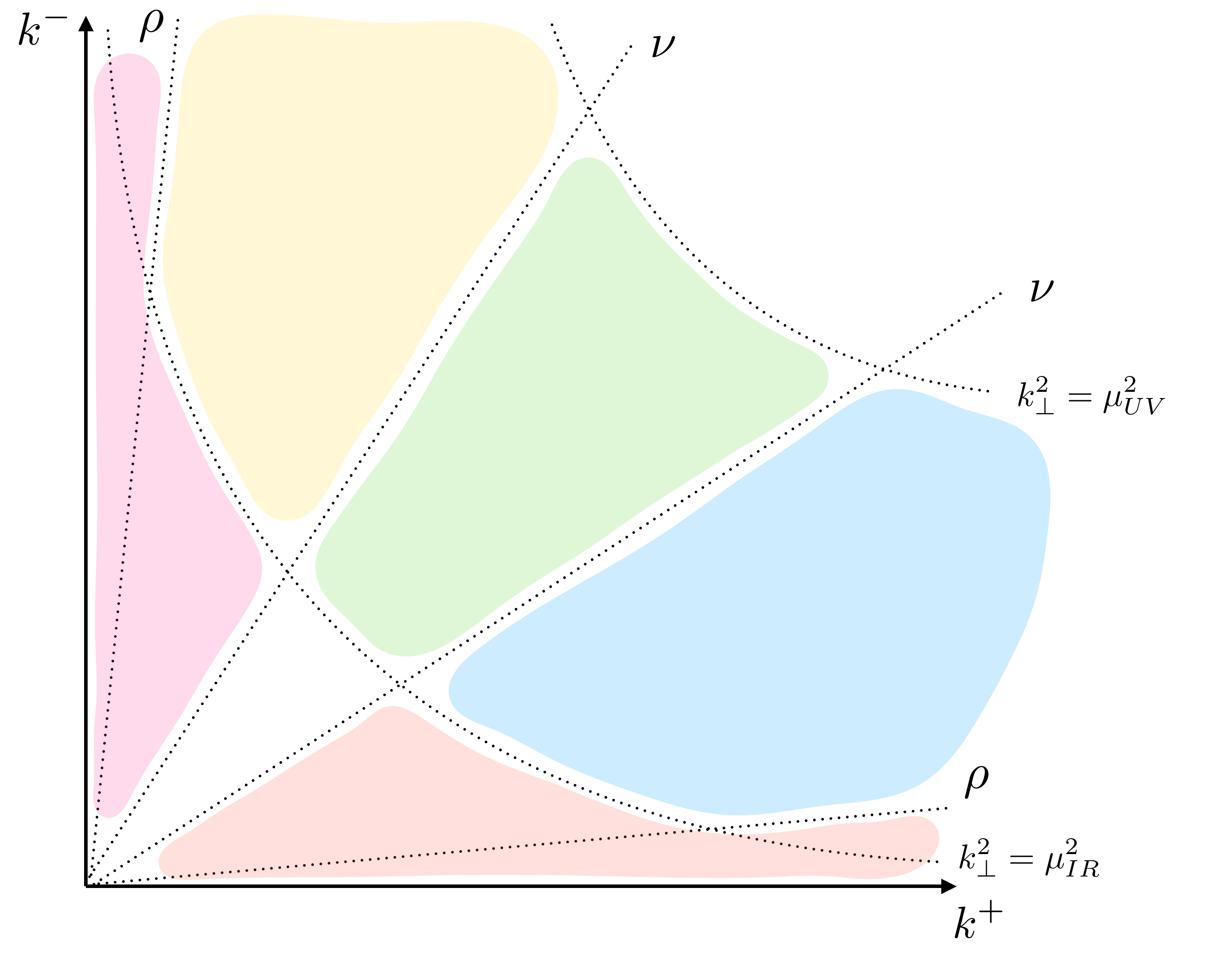}
 \end{center}
\caption{\label{fig:fnew}The new TMD factorization scheme. Green is soft, blue is collinear. $\mu_{IR}$ and $\rho$ are IR cut-offs. $\nu$ defines the region of intersection between two collinear modes. The double counting of this region is removed by inclusion of the the soft factor $\mathcal{S}(b_\perp)$ in the factorization formula. A mass-shell hyperbolas are defined by a scale $\mu^2$ via $2k^+ k^- = k^2_\perp = \mu^2$.}
 \end{figure}

It is very important to understand how the modes are separated. Schematically, the separation in our calculation is presented in Fig. \ref{fig:fnew}. The colorful regions in Fig. \ref{fig:fnew} correspond to a couple of TMD distributions in Eq. (\ref{int:TMDff}). To separate these modes from the modes of the hard function we introduce a cut-off $\mu^2_{\rm UV}$ scale,  which is of ultraviolet origin. This cut-off sets the $\mu$ scale of the TMDPDFs we start with. 

We will be mostly interested in the TMDPDFs of a hadron with a large $P^+$ component which are constructed from the blue and red regions in Fig. \ref{fig:fnew}. We will call these the collinear modes. Similarly, one could consider a second TMDPDFs which corresponds to a hadron with a large $P^-$ component. This TMDPDFs is constructed from the modes of yellow and purple regions to which we will refer to as   anti-collinear modes. There is an intersection region between the collinear and anti-collinear modes. This region is described by the soft modes which are denoted by the green color. The modes of this sector are parts of the definition of TMDPDFs. To separate the modes associated with different hadrons we introduce a cut-off scale $\nu$. This scale should be understood as an upper cut-off of rapidity for the collinear modes. In this sense it has UV nature since it separates the collinear modes from the large $k^-$ sector. The $\nu$ variable sets the $\zeta$ scale of the TMDPDF we start with. This corresponds to final scales $(\mu, \zeta)$ in the l.h.s of Eq. (\ref{tmd-evolved}).

To calculate the TMDPDFs at the NLO order we need to split the modes of the collinear sector into the ``quantum'' (blue) and ``background'' (red) fields. To do that we introduce a couple of factorization scales $\rho$ and $\mu^2_{\rm IR}$. As we mentioned above, in our calculation we aim to integrate over ``quantum'' (blue) fields, keeping the ``background'' (red) fields fixed. Note that the operators constructed from the background fields are at the initial scales $(\mu^2_{\rm IR}, \rho)$. This corresponds to initial scales $(\mu_0, \zeta_0)$ in the r.h.s. of Eq. (\ref{tmd-evolved}).

Let us clarify how we introduce the cut-offs for the regions. Instead of using explicit cut-offs, we apply a renormalization approach. The separation of modes leads to unphysical divergences in our perturbative integration over ``quantum" modes. We regulate these divergences by an appropriate regularization schemes. The associated scales play the role of the factorization scales in our calculation.

To regulate the divergences of integrals over transverse momenta we use the dimensional regularization with $d=2-2\epsilon$. To regulate the divergences of the integrals over longitudinal momenta fraction, the so-called rapidity divergences, we use a regularization scheme proposed in Refs.~\cite{Chiu:2011qc,Chiu:2012ir}. In particular, to regulate a divergent integral over $k^-$, we make the following replacement, 
\begin{eqnarray}
&&\int^\infty_0 \frac{d k^-}{k^-} \to \nu^\eta \int^\infty_0 \frac{d k^-}{k^-}  |k^+ |^{-\eta} \,,
\label{eta-reg}
\end{eqnarray}
where the renormalization scale $\nu$ plays the role of a cut-off in rapidity in our calculation, and $\eta$ is a regulator similar to $\epsilon$ in the dimensional regularization. For applications of this regulator in the small-$x$ calculations see Refs. \cite{Fleming:2014rea,Rothstein:2016bsq,Kang:2019ysm,Liu:2022ijp}. 

In our analysis, we pay close attention to the origin of divergences and explicitly distinguish the UV and the IR contributions. To achieve this we associate separate scales with the UV and IR divergences. Indeed, these scales should be understood as the upper and the lower cut-offs in our integrations. In particular, the divergence at $k_\perp\to\infty$ of the transverse integral corresponds to the $\mu^2_{\rm UV}$ scale, and the divergence at $k_\perp\to 0$  with $\mu^2_{\rm IR}$. Similarly, we relate the divergence of the integral over $k^-$ coming from the UV region of $k^-\to\infty$ with the $\nu$ scale and for the rapidity divergence at $k^-\to 0$ we introduce a new scale $\rho$, which is similar to $\nu$ in Eq.~(\ref{eta-reg}) (the corresponding regulator we denote as $\xi$).

In this paper, we consider the case of gluon TMDPDFs, but this approach can be easily extended to quarks as well. We find that the dependence of the TMDPDFs on the UV scales is standard, given by the RGEs (\ref{ee:mu}) and (\ref{ee:zeta}). But, we observe that the IR structure is more involved compared to the leading order results obtained in the collinear matching or with the eikonal expansion. In particular, we find a double logarithmic contribution, in addition to single logarithmic terms. These terms have not been observed before. The single logarithmic terms have kernels that are, in general, different from either the DGLAP or BFKL evolution kernels.

We also reproduce the finite terms. The role of these terms should not be underestimated as they become large in some kinematic limits. For example, to compare our results with the collinear matching approach, we perform an expansion onto collinear PDFs and consider the leading twist contribution. In this case we find that the finite terms start to diverge and develop a large single logarithmic contribution. In combination with other terms, this yields a single logarithmic contribution with the DGLAP evolution kernel which is in full agreement with the leading twist term of the collinear matching expansion.

Similarly, to compare our equations to the eikonal expansion approach, we expand our result in the powers of $x$. We consider the leading term of expansion and find that the finite terms have a large logarithm of rapidity, which in combination with other terms leads to a single logarithmic contribution with the BFKL evolution kernel. This is in agreement with the eikonal expansion approach.

As a result, we find that our approach yields a general description of the IR structure of the TMDPDFs valid in a broad kinematic range. At the same time in particular limits, it is in agreement with the previously obtained results.

The paper is organized as follows. In Sec. \ref{sec:bfm}, we review the background field method in the context of the QCD factorization. After this we focus on the TMD factorization in Sec. \ref{sec:TMD} where we give the key definitions and concepts. Sec. \ref{sec:NLO} is the main part of the part. In Sec. \ref{sec:NLO-real} we discuss contribution of the real emission diagrams at the NLO order. This is followed by Sec. \ref{sec:NLO-virt} where we consider virtual diagrams. In Sec. \ref{sec:NLO-soft} we provide details on the calculation of the soft factor. Then in Sec. \ref{sec:NLO-fin} we combine all parts of the NLO calculation to get our final result. We compare our final result with the collinear matching approach in Sec. \ref{sec:cmatch} and with the eikonal expansion in Sec. \ref{sec:eikexp}. We provide the details of calculation of the real emission diagrams in App. \ref{Ap:evL} and virtual emission diagrams in App. \ref{Ap:vd}. 

\section{TMD factorization and the background field method\label{sec:two}}
The fundamental approach that we use in our calculation is the background field method~\cite{Abbott:1980hw,Abbott:1981ke}. Within this approach, the QCD factorization can be introduced and interpreted as the separation of the QCD fields into distinct dynamical modes interacting with one another. In this section, we give a short review of the background field method and introduce the TMD factorization from this point of view.

\subsection{Background field method\label{sec:bfm}}
The concept of factorization can be elegantly introduced within the background field method. Starting with a matrix element of an arbitrary operator $\mathcal{O}(\hat{C})$ which depends on quark and gluon fields,\footnote{Here we understand $C$ as a general notation for the quark and gluon fields.} one can write it as a functional integral over those fields
\begin{eqnarray}
&&\langle P_1 | \mathcal{O}(\hat{C}) |P_2\rangle = \int \mathcal{D}C~ \Psi^\ast_{P_1}(\vec{C}(t_f)) \mathcal{O}(C) \Psi_{P_2}(\vec{C}(t_i)) e^{iS_{\rm QCD}(C)} \,,
\label{int:mel-form1}
\end{eqnarray}
where $\Psi_{P_2}$ and $\Psi_{P_1}$ are initial and final state wave functions at $t_i\to-\infty$ and $t_f\to \infty$.

\begin{figure}[htb]
 \begin{center}
\includegraphics[width=170mm]{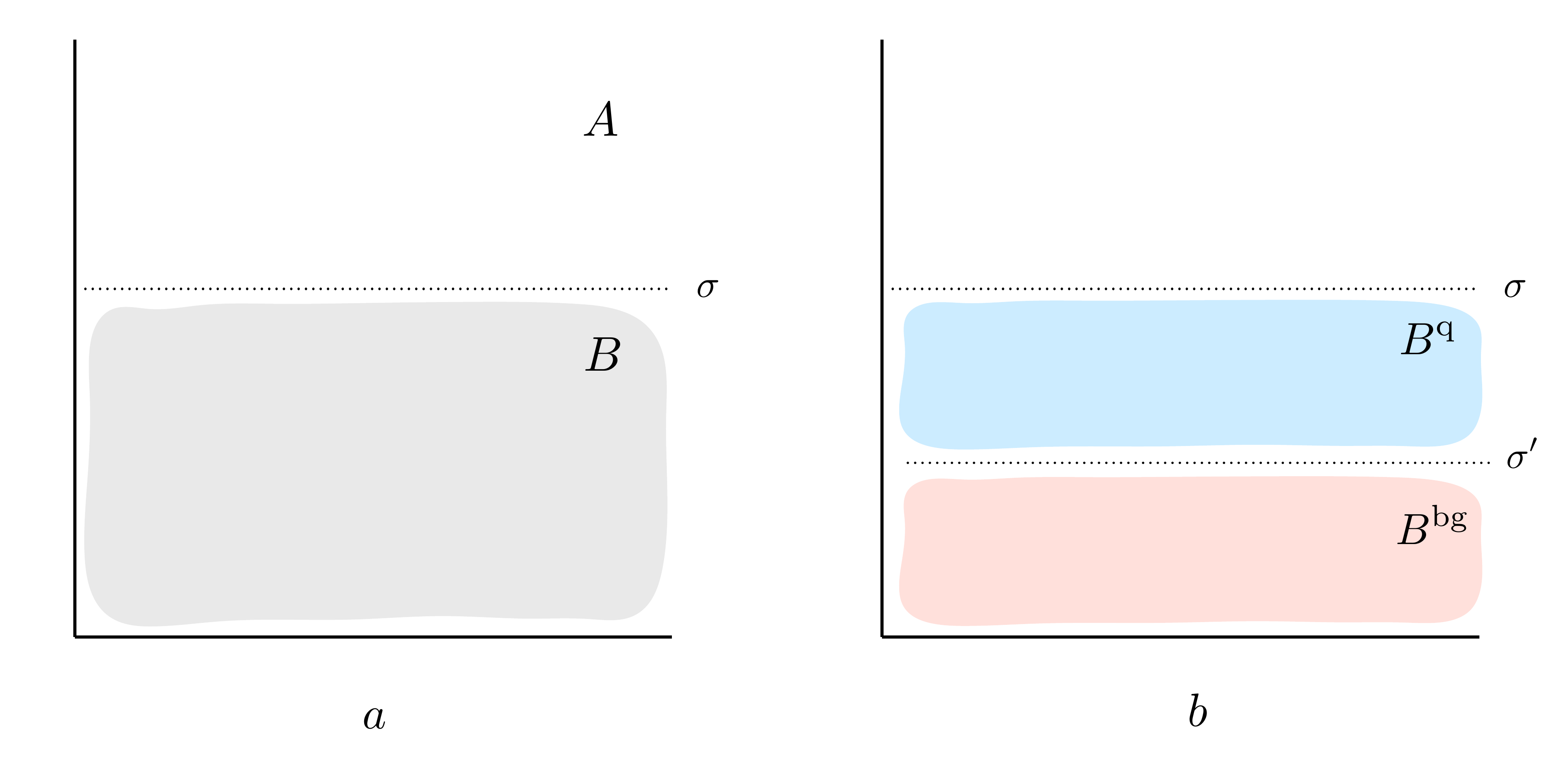}
 \end{center}
\caption{\label{fig:frap}Schematic representation of the separation of modes in the background field method. a) The ``quantum" fields $A$ are above the $\sigma$ cut-off. The background fields $B$ (grey) are below the cut-off. b) To study the dependence on the factorization scale, we split the background fields into two components $B \to B^{\rm q} + B^{\rm bg}$. We integrate over $B^{\rm q}$ (blue) fields in a fixed background of $B^{\rm bg}$ (red) fields.}
 \end{figure}

Evaluation of this functional integral is hard in general. Hence, we split the field into different components, and rewrite the integral as independent functional integrals over those components. This is the logic of the background field method. This separation is arbitrary in general, but in the context of the QCD factorization, we shall assume that there is a set of factorization scales $\sigma$ that separate the components from each other, see Fig. \ref{fig:frap}a.

For example, in the case of deep inelastic scattering (DIS) in the Bjorken limit, it is sufficient to split the field into two field modes $A$ and $B$,
\begin{eqnarray}
&&C_\mu \to A_\mu + B_\mu \,.
\label{int:Bj-spt}
\end{eqnarray}
The fields are separated by the value of the transverse momenta, i.e. one can introduce a cut-off for the transverse momenta $\sigma=\mu$, such that the fields $A$ are defined as having $k^2_\perp > \mu^2$, and fields $B$ are characterized with $k^2_\perp < \mu^2$. This is called the collinear factorization scheme which is defined by a strong ordering of emission in transverse momenta.
Another example where the factorized QCD medium can be described by splitting the field into two modes is the high energy rapidity factorization in DIS. In this case, the field mode $A$ has longitudinal momentum fraction $k^- > \sigma$ and the mode $B$ has $k^- < \sigma$. The factorization scale $\sigma$ in this case is the rigid cut-off in $k^-$.

For brevity, let's assume that the factorization can be described with only two modes as in (\ref{int:Bj-spt}). And a scale $\sigma$ is a factorization scale separating components $A$ and $B$. In general, the structure of dynamical field modes can be more complicated and depend on the nuances of the factorization in the process. For example, as we will see in the next subsection, to describe the TMD factorization one has to introduce four modes: one hard mode (``quantum''), two collinear modes (background), and a soft mode for an intersection between the collinear modes.

We will call the $A$ fields ``quantum'' to indicate that we aim to integrate over this mode. The fields $B$ are usually called the background fields. The functional integral over this mode contains the non-perturbative part of the scattering process. One can think about the background fields as fields associated with the target, and the ``quantum" fields as fields of the hard scattering.

After splitting (\ref{int:Bj-spt}) we can rewrite the functional integral in Eq. (\ref{int:mel-form1}) as
\begin{eqnarray}
&&\langle P_1 | \mathcal{O}(\hat{C}) |P_2\rangle = \int \mathcal{D}B~ \Psi^\ast_{P_1}(\vec{B}(t_f)) \tilde{\mathcal{O}}(B, \sigma) \Psi_{P_2}(\vec{B}(t_i)) e^{iS_{\rm QCD}(B)} \,,
\label{int:bBqA}
\end{eqnarray}
where the integral over ``quantum'' mode $A$ is,
\begin{eqnarray}
&&\tilde{\mathcal{O}}(B, \sigma) = \int \mathcal{D}A ~\mathcal{O}(A + B) e^{iS_{\rm bQCD}(A, B)} \,.
\label{int:AbInt}
\end{eqnarray}
The integration over $A$ should be performed with the QCD action in the background field,
\begin{eqnarray}
&&S_{\rm bQCD}(A, B) = S_{\rm QCD}(A+B) - S_{\rm QCD}(B) \,.
\end{eqnarray}

In Eq. (\ref{int:bBqA}) we assume that due to the factorization regime in the high-energy scattering, the wave functions depend only on the $B$ fields, which, as we mentioned above, should be associated with the fields of the target.

The integral over ``quantum" fields $A$ in Eq. (\ref{int:AbInt}), i.e. fields of the projectile, can be evaluated perturbatively. However, it usually requires some sort of expansion in factorization scales, e.g. twist expansion in the collinear factorization, etc. The result of integration has a general form
\begin{eqnarray}
&&\tilde{\mathcal{O}}(B, \sigma) = \sum_i H_i(\sigma)\otimes\mathcal{V}_i(B, \sigma)\,.
\label{int:aAi}
\end{eqnarray}
where $\mathcal{V}_i$ are operators constructed from background mode $B$. Since the background fields $B$ depend on the factorization scale $\sigma$, so do the operators, as we explicitly indicate. $H_i(\sigma)$ is a hard function that we obtain after integration over the ``quantum" fields $A$.

After this, Eq. (\ref{int:bBqA}) reads
\begin{eqnarray}
&&\langle P_1 | \mathcal{O}(\hat{C}) |P_2\rangle = \sum_i H_i(\sigma)\otimes \langle P_1|\mathcal{V}_i(\sigma)|P_2\rangle \,,
\label{int:fact}
\end{eqnarray}
where the matrix elements of operators $\mathcal{V}_i$ are generated by the functional integral over background fields $B$,
\begin{eqnarray}
&&\langle P_1|\mathcal{V}_i(\sigma)|P_2\rangle \equiv \int \mathcal{D}B~ \Psi^\ast_{P_1}(\vec{B}(t_f)) \mathcal{V}_i(B, \sigma) \Psi_{P_2}(\vec{B}(t_i)).
\label{int:meB}
\end{eqnarray}
Here, the index $i$ enumerates different types of operators. These matrix elements can be further parameterized in terms of distribution functions. For example, in the collinear factorization, those are the well-known collinear PDFs. 

Eq.(\ref{int:fact}) is a factorization formula that we obtain by assuming the QCD medium develops different dynamical modes, so the separation (\ref{int:Bj-spt}) can be done and the integration over ``quantum" modes $A$ can be considered independently in a fixed configuration of background fields $B$.

Note that the structure of factorization depends on the scattering process and kinematics. However, due to a wide separation of scales, each type of factorization is characterized by a set of large logarithms. These logarithms should be resummed. This is usually done by solving the RGEs which are differential equations with respect to the factorization scales, see for example Eqs. (\ref{ee:mu}) and (\ref{ee:zeta}) for the TMD factorization.

One can use the following approach to extract the dependence on the factorization scales in the background field method. Starting with a matrix element (\ref{int:meB}), we introduce a new set of factorization scales $\sigma'$ and split the background field $B$ as $B \to B^{\rm q} + B^{\rm bg}$. Here the fields $B^{\rm q}$ correspond to dynamical modes between scales $\sigma$ and $\sigma'$, and $B^{\rm bg}$ is a new background field corresponding to fields below the scale $\sigma'$, see Fig. \ref{fig:frap}b.

Since the cut-offs $\sigma$ act as upper cut-offs for the field modes, we will refer to them as the UV scales. Similarly, since the cut-offs $\sigma'$  act as lower cut-offs for the $B^{\rm q}$ modes, we will refer to them as the IR scales.

To study the dependence on the set of UV factorization scales $\sigma$, we need to integrate over $B^{\rm q}$ fields. This integration is done perturbatively and leads to the answer
\begin{eqnarray}
&&\langle P_1|\mathcal{V}_i( \sigma)|P_2\rangle = \sum_j C_{ij}(\sigma, \sigma')\otimes \langle P_1|\mathcal{V}_j(\sigma')|P_2\rangle \,,
\label{mel:ptb}
\end{eqnarray}
where coefficients $C_{ij}(\sigma, \sigma')$ are obtained by integrating over the $B^{\rm q}$ modes and describe dynamics between the UV scales $\sigma$ and the IR scales $\sigma'$. One can now differentiate with respect to the UV scales $\sigma$ and obtain a system of evolution equations that define the dependence of the matrix element (\ref{int:meB}) and the corresponding distribution functions on the set of factorization scales $\sigma$, e.g. Eqs. (\ref{ee:mu}) and (\ref{ee:zeta}) in the TMD factorization.

At the same time, the coefficients $C_{ij}(\sigma, \sigma')$ also describe the perturbative component of the IR structure of the matrix element through the dependence on the IR cut-off $\sigma'$. The non-perturbative part of the IR structure is encoded in the matrix element of the operators $\mathcal{V}_j(\sigma')$.

This paper aims to calculate the $C_{ij}(\sigma, \sigma')$ coefficients for the gluon TMDPDFs. The role of the UV scales $\sigma$ is played by the cut-offs $(\mu^2_{\rm UV}, \nu)$, while the IR scales $\sigma'$ correspond to the cut-offs $(\mu^2_{\rm IR}, \rho)$, see Fig. \ref{fig:fnew}. More details can be found in sections \ref{sec:intro} and \ref{sec:NLO}.

Another crucial but technical point is how we can actually introduce different field modes in practice. We can use rigid cut-offs, for example, as it is done in the high-energy rapidity factorization approach. These cut-offs subsequently appear in the perturbative calculation of functional integrals, e.g. in Eq. (\ref{int:AbInt}), as cut-offs in loop integrals over momenta. These cut-offs regulate all divergences of the loop integrals. However, these rigid cut-offs are rarely used due to technical difficulties.

We use a different method in this paper, known as the renormalization approach.
In this scheme, each divergent integral is regulated with a regulator accompanied by a renormalization scale. This scale plays the role of a cut-off separating different modes. As mentioned before, we shall use dimensional regularization to separate the modes in the transverse momentum. To separate the modes in rapidity, we will use the regularization of the longitudinal momentum fraction integrals as in Eq. (\ref{eta-reg}).\footnote{There is a number of alternative methods: deviation from the light-cone direction \cite{Collins:2011zzd,Ji:2004wu}, $\delta$-regulator \cite{Echevarria:2011epo,Chiu:2009yx}, exponential regulator \cite{Li:2016axz}, and analytic regulator \cite{Beneke:2003pa,Chiu:2007yn,Becher:2011dz}.} Note that in the renormalization approach, we will find some unphysical poles which is a consequence of the separation of different dynamical field modes. These poles need to be removed by a proper renormalization of operators.

\subsection{TMD factorization\label{sec:TMD}}
In the previous section, we showed how the background field method can be used to construct a factorization scheme. While our discussion was general, different factorization schemes can be substantially distinct. This can be most easily seen in the structure of the dynamical modes, e.g., compare the structure of the high-energy rapidity factorization in DIS, see Fig. \ref{fig:frap}, and the TMD factorization presented in Fig. \ref{fig:fnew}.

In this section, we will give an overview of the transverse momentum dependent (TMD) factorization. In particular, we will discuss the structure of modes in Fig. \ref{fig:fnew} using the framework of the background field approach and give a formal definition of the corresponding operators.

The TMD factorization is applied to high-energy scattering reactions when a final state with a small transverse momentum $q_\perp$ is measured. The scattering reaction includes two energetic hadrons moving along $n^\mu = (1, 0, 0, -1)$ and $\bar{n}^\mu = (1, 0, 0, 1)$ directions, e.g. two colliding hadrons in the Drell-Yan scattering with momenta $\bar{P}^\mu = P^-n^\mu + M^2 \bar{n}^\mu /P^-$ and $P^\mu = P^+ \bar{n}^\mu + M^2 n^\mu / P^+$ in the center of mass frame, where $M^2$ is hadron's mass.

Let's now introduce the TMD factorization scheme in the background field approach. Following the logic of the previous section, we start with the background field $B$. Since the scattering process involves two hadrons, we need to introduce two types of background field modes. We will refer to the modes associated with the $P^\mu$ hadron as collinear modes and modes of the $\bar{P}^\mu$ hadron as anti-collinear. Both of these modes describe the dynamics of the corresponding hadron.

The hard scattering process is described by the ``quantum" mode $A$. This mode is separated from the background modes by a factorization scale $\mu^2_{\rm UV}$, see Fig. \ref{fig:fnew}. The integration over fields of this mode yields a hard function in the factorization formula, c.f. Eqs. (\ref{int:TMDff}) and (\ref{int:fact}). In integration over ``quantum" modes collinear and anti-collinear background modes yield two TMD operators, i.e. operators $\mathcal{V}_i$ in the notation of Eq. (\ref{int:fact}). The matrix elements of these operators can be parameterized in terms of the TMDPDFs, see Eq. (\ref{int:TMDff}). For brevity, we refer to these matrix elements as TMDPDFs.

The resulting matrix element of the TMD operator constructed from the gluon collinear modes has the form
\begin{eqnarray}
&&\mathcal{B}_{ij}(x_B, b_\perp) = \int^\infty_{-\infty} dz^- e^{-i x_BP^+ z^-} \langle P,S|\bar{T}\{F^m_{-i}(z^-, b_\perp) [z^-, \infty]^{ma}_b\} T\{[\infty, 0^-]^{an}_0 F^n_{-j}(0^-, 0_\perp)\}|P,S\rangle\,.
\label{gop-def}
\end{eqnarray}
Although there is a similar quark TMD operator and our approach can be applied to it as well in this paper we focus only on the gluon case. Moreover, one can also write a similar set of operators for the anti-collinear background modes. Since the analysis is identical, we do not explicitly present it in this paper. Finally, we will only consider the unpolarized matrix elements to simplify the discussion.

In Eq. (\ref{gop-def}), $i$ and $j$ are Lorentz  indices. The adjoint Wilson lines are along the light-cone direction
\begin{eqnarray}
&&[x^-, y^-]_{z_\perp} = \mathcal{P}\exp\Big[ig \int^{x^-}_{y^-} dz^- A_-(z^-, z_\perp)\Big]\,.
\end{eqnarray}
We assume that the Wilson lines are connected at infinity with a transverse gauge link, which makes the operator gauge invariant. We choose future-point Wilson lines which correspond to SIDIS. The $\tilde{T}$ notation is to indicate that the corresponding operators should be inverse-time ordered. The Fourier transformation of the matrix element is defined as
\begin{eqnarray}
&&\mathcal{B}_{ij}(x_B, p_\perp) = \int d^2b_\perp e^{ip_\perp b_\perp} \mathcal{B}_{ij}(x_B, b_\perp)\,.   
\end{eqnarray}

In the TMD factorization scheme, the collinear and anti-collinear modes are only separated by their rapidity $y=1/2\ln(k^-/k^+)$. There might be an intersection region depending on how these modes are separated. In the TMD factorization framework, this is called the soft region. From the point of view of the background field approach, the fields of this mode should still be considered as the background fields.

To resolve the potential double counting in the intersection region, one must introduce a rapidity cut-off $\nu$ and assume that the collinear fields in Eq. (\ref{gop-def}) are constructed only from the modes below this cut-off. In Fig. \ref{fig:fnew}, these fields correspond to the blue and red regions. Similarly, the anti-collinear modes correspond to the yellow and purple sectors.

To take into account the contribution of the soft region, we need to multiply (\ref{gop-def}) with a soft function constructed from the soft modes. Note that the structure of the soft factor depends on the details of separation in rapidity between the fields of the collinear and anti-colliner modes.\footnote{In principle, one can construct the collinear and anti-collinear modes to avoid the intersection and thus render the soft factor trivial, see Refs. \cite{Becher:2010tm,Becher:2012yn}.} We use a renormalization approach and define the collinear fields using the $\eta$ regulator, see Eq. (\ref{eta-reg}). In this case the soft factor $\sqrt{\mathcal{S}(b_\perp)}$ is given by a vacuum matrix element
\begin{eqnarray}
&&\mathcal{S}(b_\perp) = \frac{1}{N^2_c - 1}\langle 0|{\rm Tr}[S^\dag_{\bar{n}}(b_\perp)S_{n}(b_\perp) S^\dag_{n}(0_\perp) S_{\bar{n}}(0_\perp)]|0\rangle \,,
\label{soft_func}
\end{eqnarray}
where the Wilson lines,
\begin{eqnarray}
&&S_{n}(b_\perp) = P\exp\Big[ig\int^\infty_0 d(x\cdot \bar{n}) n\cdot A(x\cdot \bar{n}, b_\perp)\Big]
\end{eqnarray}
are constructed from soft modes defined by the regularization
\begin{eqnarray}
\int \frac{dk^+dk^-}{k^+k^-} = \nu^\eta 2^{-\eta/2} \int \frac{dk^+dk^-}{k^+k^-} |k^+ - k^-|^{-\eta} \,.
\end{eqnarray}
The fields of the soft mode correspond to the green sector in Fig. \ref{fig:fnew}.

The full matrix element of the background modes is defined as
\begin{eqnarray}
&&f_{ij}(x_B, b_\perp) = \sqrt{\mathcal{S}(b_\perp)} \mathcal{B}_{ij}(x_B, b_\perp)\,.
\label{fTMD}
\end{eqnarray} 

To study the dependence of this matrix element on the factorization scales $\sigma=(\mu^2_{\rm UV}, \nu)$, we implement the approach discussed in the previous section. We introduce a new set of factorization scales $\sigma'=(\mu^2_{\rm IR}, \rho)$ to split the background collinear mode into the $B^{\rm q}$ mode which lies between the scales $\sigma$ and $\sigma'$, see the blue region in Fig. \ref{fig:fnew}, and the $B^{\rm bg}$ mode which lies below the scales $\sigma'$, i.e. the red sector in Fig. \ref{fig:fnew}. 

We aim to integrate over $B^{\rm q}$ and soft modes to obtain a hard function containing the perturbative structure of the matrix element (\ref{fTMD}) (see Eq. (\ref{mel:ptb})). The non-perturbative structure is encoded in the matrix element constructed from the fields of the $B^{\rm bg}$ mode. The details and results of this calculation are presented in the next section.

\begin{figure}[htb]
 \begin{center}
\includegraphics[width=100mm]{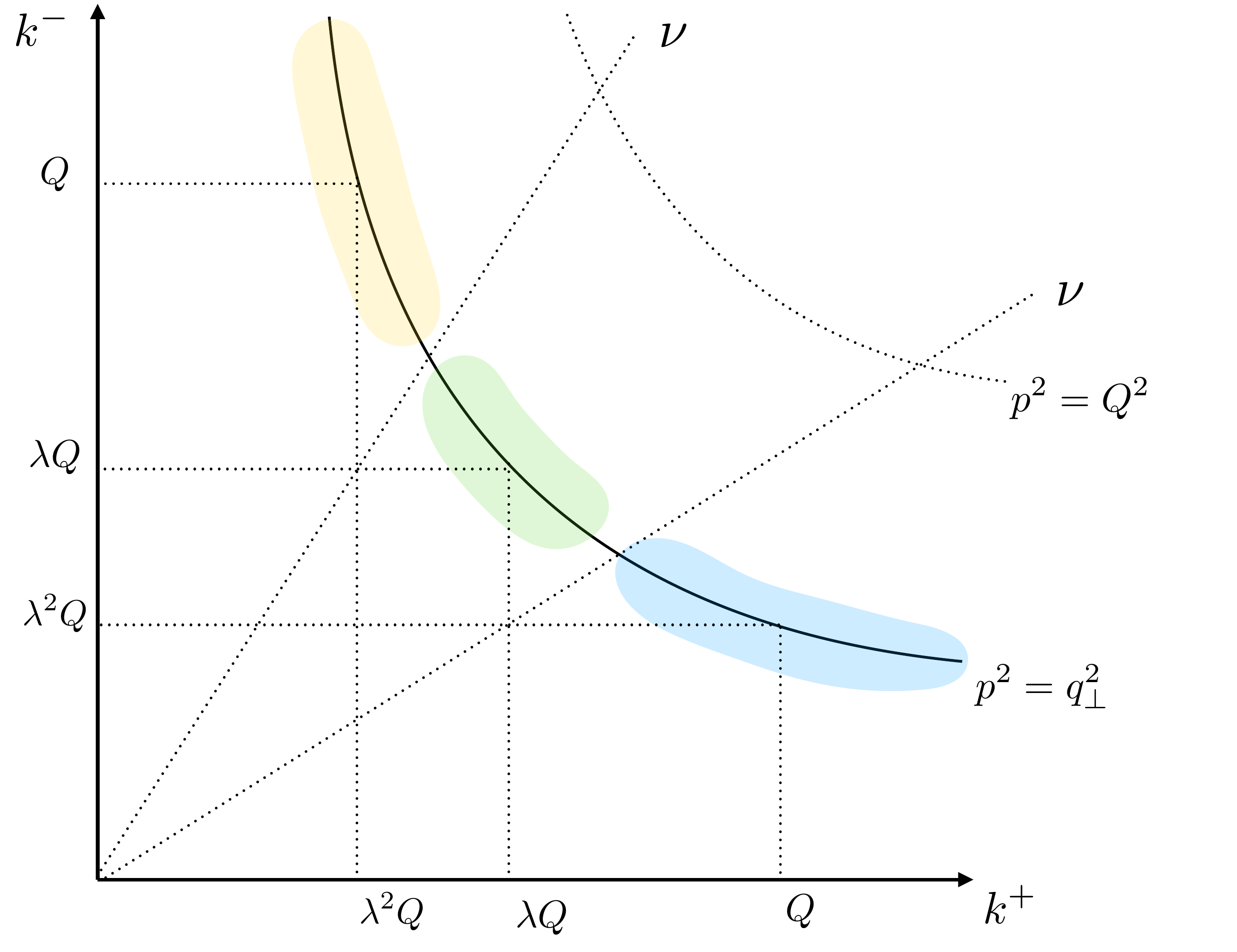}
 \end{center}
\caption{\label{fig:fSCET}SCET factorization scheme. The collinear (blue) and soft (green) modes are around the mass-shell parabola $p^2 = q^2_\perp \sim 1/b^2_\perp$.}
 \end{figure}

Before moving to the main part of the paper, we would like to point out that our definition of these modes differs from the Soft Collinear Effective Theory (SCET). The structure of modes in SCET is summarized in Fig. \ref{fig:fSCET}.

The collinear modes are defined to have the following scaling of momenta: $p_n = (p_n^+,p_n^-,\vec{p}_{n\perp}) \sim Q(\lambda^2, 1, \lambda)$ and $p_{\bar{n}}\sim Q(1, \lambda^2, \lambda)$, where $\lambda\sim q_\perp/Q^2\ll 1$. Both collinear fields are approximately on the mass-shell $p^2_n \sim p^2_{\bar{n}} \sim q^2_\perp$, so they lie on the same hyperbola. The hard fields are defined as having $p^2 \geq Q^2$. The soft emission is defined as having $p_s \sim Q(\lambda, \lambda, \lambda)$. The main difference with the background field approach is that in the latter the field modes are not necessarily on the mass-shell; thus in principle, we consider more general kinematics. Also, to define the modes we do not use any scaling of momenta but rather use a set of cut-offs to {delineate regions in momentum space.}

\section{Calculation of the TMDPDF at the NLO order\label{sec:NLO}}
The calculation of the gluon TMDPDFs, defined by Eq. (\ref{fTMD}) at the NLO order contains three parts: calculation of the real emission diagrams, virtual emission diagrams, and the soft factor. The details of the calculation of real and virtual emission diagrams are presented in Apps. \ref{Ap:evL} and \ref{Ap:vd}. In the next two sections, we examine the structure of singularities of the diagrams involved. While the soft factor at the NLO order is well known, we discuss it for completeness in Sec. \ref{sec:NLO-soft}. Finally, we combine all parts of the calculation in Sec. \ref{sec:NLO-fin}, where we discuss the renormalization of the solution and present our final result.

\subsection{Real emission diagrams\label{sec:NLO-real}}
\begin{figure}[htb]
 \begin{center}
\includegraphics[width=170mm]{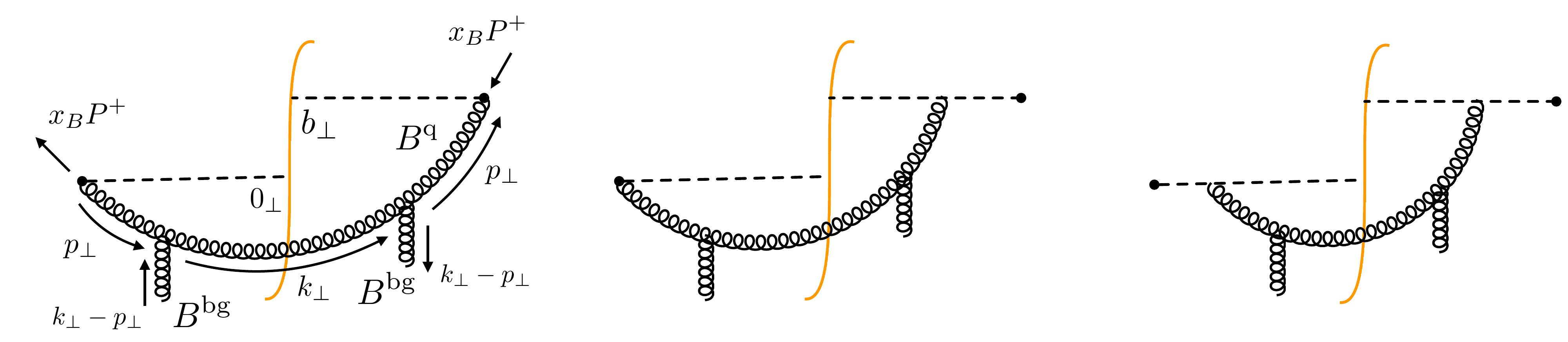}
 \end{center} \caption{\label{fig:red}Real emission diagrams.}
 \end{figure}

In this section, we present an analysis of the real emission diagrams. Some typical diagrams are given in Fig. \ref{fig:red}. As per the background field approach, the loop is constructed from the ``quantum" fields $B^{\rm q}$, and the calculation is done in the background field $B^{\rm bg}$. 

\begin{figure}[htb]
 \begin{center} \includegraphics[width=160mm]{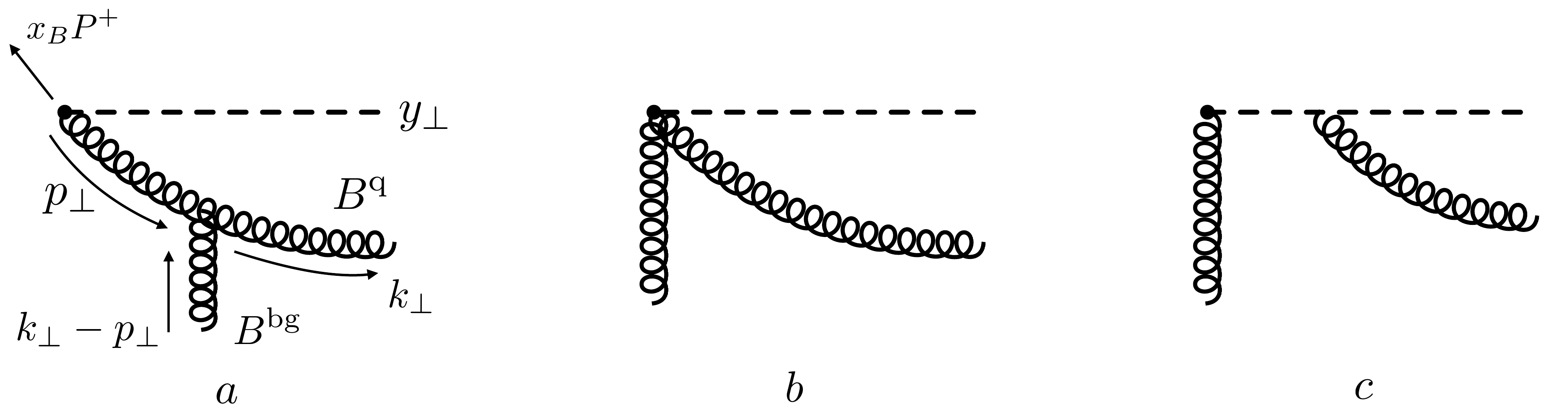}
 \end{center}
\caption{\label{fig:evL}Diagrams contributing to the emission vertex $L_{\mu j}(k,y_\perp,x_B)$.}
 \end{figure}

It is convenient to start the derivation with a calculation of the emission vertex
\begin{eqnarray}
&&L^{ab}_{\mu j}(k,y_\perp,x_B)\equiv i\lim_{k^2\to 0}k^2\langle B^{{\rm q}a}_\mu(k)\int^{\infty}_{-\infty} dy^- e^{ix_BP^+y^-} [\infty,y^-]^{bd}_y F^{{\rm q+bg};d}_{-j}(y^-,y_\perp)\rangle_{B^{\rm bg}}\,,
\label{def:evL}
\end{eqnarray}
which corresponds to a sum of diagrams in Fig. \ref{fig:evL}. Once this vertex is calculated, the sum of real emission diagrams can be easily obtained by taking a product of two vertices as
\begin{eqnarray}
&&\langle\tilde{\mathcal{F}}^a_i(x_B, x_\perp)\mathcal{F}^{a}_j(x_B, y_\perp )\rangle^{\rm real} = -\int\frac{\dhd k^-}{2k^-} \int \dhd^2k_\perp \tilde{L}^{\ \mu ba}_i(k,x_\perp,x_B) L^{ab}_{\mu j}(k,y_\perp,x_B) \,.
\label{re:twoLprod}
\end{eqnarray}

The details of the calculation of the emission vertex can be found in Appendix \ref{Ap:evL}. Taking the product, Eq.(\ref{re:twoLprod}), of two emission vertices Eq.(\ref{evL:final}) and Eq.(\ref{evL:final-conj}), it is easy to obtain an expression for the matrix element in Eq.(\ref{gop-def}),
\begin{eqnarray}
&&\mathcal{B}^{\rm q(1)+bg;real}_{ij}(x_B, b_\perp)
\label{rem:bres}
\\
&&= - 16\pi \alpha_s N_c \int \dhd^2p_\perp e^{ip_\perp b_\perp}\int\frac{\dhd k^-}{2k^-} \int \dhd^2k_\perp \Big[ \Big( \frac{(p + k)_l}{2 x_B P^+ k^-+k^2_\perp} \frac{ x_B P^+ k^-\delta^k_i - k^k k_i}{2 x_B P^+ k^-+p^2_\perp} - \frac{\delta^k_l p_i + g_{li} k^k }{2 x_B P^+ k^- + p^2_\perp} \Big)
\nonumber\\
&&\times \Big( \frac{(p + k)_m}{2x_BP^+k^- + k^2_\perp} \frac{ x_BP^+ k^- g_{kj} - k_k k_j}{ 2x_BP^+k^- + p^2_\perp} - \frac{ g_{mk} p_j + g_{mj}k_k}{2x_BP^+k^- + p^2_\perp} \Big) + \frac{g_{il} }{k^2_\perp} \Big( \frac{(p + k)_m}{2x_BP^+k^- + k^2_\perp} \frac{ x_BP^+ k^- k_j + k^2_\perp k_j}{ 2x_BP^+k^- + p^2_\perp}
\nonumber\\
&& - \frac{ k_m p_j - g_{mj}k^2_\perp }{2x_BP^+k^- + p^2_\perp} \Big) + \Big( \frac{(p + k)_l}{2 x_B P^+ k^-+k^2_\perp} \frac{ x_B P^+ k^- k_i+ k^2_\perp k_i}{2 x_B P^+ k^-+p^2_\perp} - \frac{k_l p_i - g_{li} k^2_\perp }{2 x_B P^+ k^- + p^2_\perp} \Big) \frac{g_{mj} }{k^2_\perp} - \frac{g_{il}g_{mj} }{k^2_\perp} \Big]
\nonumber\\
&&\times \int d^2z_\perp e^{i( k - p )_\perp z_\perp} \mathcal{B}^{\rm bg}_{lm}(x_B + \frac{k^2_\perp}{2 P^+ k^-}, z_\perp) \,,
\nonumber
\end{eqnarray}
where we indicate that the matrix element on the right has operators constructed from  $B^{\rm bg}$ fields, but the one on the left has contributions from both $B^{\rm bg}$ and $B^{\rm q}$. Superscript $q(1)$ indicates that ``quantum" fields are evaluated at order $\alpha_s$. 

To separate the rapidity divergences appearing in the integral over $k^-$ from the divergences in the transverse integral over $k_\perp$, it is  convenient to introduce the variable
\begin{eqnarray}
&&z \equiv \frac{x_B}{x_B + \frac{k^2_\perp}{2P^+k^- }}\,.
\label{var:z}
\end{eqnarray}
In terms of this variable, the equation can be re-written in a compact form,
\begin{eqnarray}
&&\mathcal{B}^{\rm q(1)+bg; real}_{ij}(x_B, b_\perp) = - 4\alpha_s N_c \int \dhd^2p_\perp e^{ip_\perp b_\perp} \int^1_0 \frac{dz }{z(1-z)}
\label{rem:comp}\\
&&\times \int \dhd^2k_\perp \Big[ \mathcal{R}^{a}_{ij;lm}(z, p_\perp, k_\perp) + \mathcal{R}^{b}_{ij;lm}(z, p_\perp, k_\perp) + \mathcal{R}^{c}_{ij;lm}(k_\perp) \Big] \int d^2z_\perp e^{i( k - p )_\perp z_\perp} \mathcal{B}^{\rm bg}_{lm}(\frac{x_B}{z}, z_\perp)\,,
\nonumber
\end{eqnarray}
where the real emission kernels are coming from Eq. (\ref{rem:bres}),
\begin{eqnarray}
&&\mathcal{R}^{a}_{ij;lm}(z, p_\perp, k_\perp)\equiv (1-z)^2 \Big( \frac{1}{2}\frac{(p + k)_l}{k^2_\perp} \frac{ zk^2_\perp\delta^k_i - 2(1-z)k^k k_i}{z k^2_\perp + (1-z)p^2_\perp} - \frac{\delta^k_l p_i + g_{li} k^k }{zk^2_\perp + (1-z)p^2_\perp} \Big) 
\nonumber\\
&&\times\Big( \frac{1}{2} \frac{(p + k)_m}{ k^2_\perp } \frac{ zk^2_\perp g_{kj} - 2(1-z)k_k k_j}{ zk^2_\perp + (1-z)p^2_\perp} - \frac{ g_{mk} p_j + g_{mj}k_k}{zk^2_\perp + (1-z)p^2_\perp} \Big),
\end{eqnarray}
\begin{eqnarray}
&&\mathcal{R}^{b}_{ij;lm}(z, p_\perp, k_\perp)\equiv (1-z)\frac{g_{il} }{k^2_\perp} \Big( \frac{(p + k)_m}{ 2} \frac{ z k_j + 2 (1-z) k_j}{ zk^2_\perp + (1-z)p^2_\perp} - \frac{ k_m p_j - g_{mj}k^2_\perp }{z k^2_\perp + (1-z) p^2_\perp} \Big)
\nonumber\\
&&  + (1-z)\Big( \frac{(p + k)_l}{2 } \frac{ z k_i+ 2(1-z) k_i}{zk^2_\perp + (1-z)p^2_\perp} - \frac{k_l p_i - g_{li} k^2_\perp }{zk^2_\perp + (1-z)p^2_\perp} \Big) \frac{g_{mj} }{k^2_\perp},
\end{eqnarray}
\begin{eqnarray}
&&\mathcal{R}^{c}_{ij;lm}(k_\perp) \equiv - \frac{g_{il}g_{mj} }{k^2_\perp}.
\end{eqnarray}
The above kernels are generated by different real emission diagrams. Specifically, the functions $\mathcal{R}^a$ and $\mathcal{R}^b$ represent diagrams with the same topology as the first two diagrams in Fig. \ref{fig:red}.

It is straightforward to see that these terms do not contain any singularities. In particular, a potential rapidity divergence at $z\to 0$ is regulated by a non-zero value of $x_B$. Therefore, these terms do not provide any logarithmic contribution and yield finite terms in our final result.

Yet, in Sec. \ref{mat:cl}, we will find that although these terms are  finite for a generic kinematics, in the collinear limit (when $k_\perp - p_\perp \to 0$), they become large and diverge logarithmically. These terms ultimately generate a part of the DGLAP splitting function.

At the same time, we find that the last term in Eq. (\ref{rem:comp}), i.e. $\mathcal{R}^c$, contains divergences. This term is given by the last diagram in Fig. \ref{fig:red}.\footnote{Note that we perform the calculation in the axial gauge. In other gauges, the distribution of terms between diagrams can differ, but the sum of diagrams should not change. In particular, we also calculated the background Feynman gauge with the same result, see Appendix~\ref{App:FG}.} The singularities are regulated by the factorization cut-offs. As discussed in the section \ref{sec:intro}, we introduce these cut-offs using the renormalization approach. Now, let's understand how this procedure works.

The contribution of the last term in Eq. (\ref{rem:comp}) has a simple form,
\begin{eqnarray}
&&\mathcal{B}^{\rm q(1)+bg;real, c}_{ij}(x_B, b_\perp) = 4\alpha_s N_c \int \dhd^2p_\perp e^{ip_\perp b_\perp} \int^1_0 \frac{dz }{z(1-z)} \int \frac{ \dhd^2k_\perp }{k^2_\perp} \int d^2z_\perp e^{i( k - p )_\perp z_\perp} \mathcal{B}^{\rm bg}_{ij}(\frac{x_B}{z}, z_\perp)\,.
\label{Rc:form1}
\end{eqnarray}

Consider with the rapidity divergence first. While the contribution does not contain a  divergence as $z\to 0$, which is regulated by a non-zero value of $x_B$, there is a divergence as $z \to 1$. In terms of the $k^-$ variable, this divergence corresponds to $k^-\to\infty$. To separate this divergence, we rewrite the contribution as
\begin{eqnarray}
&&\mathcal{B}^{\rm q(1)+bg;real, c}_{ij}\left(x_B, b_\perp\right) 
\label{rc:tki}\\
&&= 4\alpha_s N_c \int \frac{ \dhd^2k_\perp }{k^2_\perp} e^{ik_\perp b_\perp}\Big( \int^1_0\frac{dz}{(1-z)_+}\mathcal{B}^{\rm bg}_{ij}(\frac{x_B}{z}, b_\perp) + \int^1_0\frac{dz}{z}\mathcal{B}^{\rm bg}_{ij}(\frac{x_B}{z}, b_\perp ) + \int^1_0\frac{dz}{1-z}\mathcal{B}^{\rm bg}_{ij}(x_B, b_\perp)\Big) \,,
\nonumber
\end{eqnarray}
where the plus distribution is defined in the usual way,
\begin{eqnarray}
&&(f(z))_+ \equiv f(z) - \delta(1-z)\int^1_0 dz' f(z').
\end{eqnarray}

Now, the rapidity divergence is in the last term of Eq. (\ref{rc:tki}), which we regulate by making the replacement (\ref{eta-reg}). In terms of the $z$ variable, this corresponds to
\begin{eqnarray}
&&\int^1_0 \frac{dz }{1-z} \to \Big(\frac{\nu}{x_B P^+}\Big)^\eta \int^1_0 \frac{dz}{1 - z}  \Big( \frac{1 - z}{z} \Big)^{-\eta} \,,
\end{eqnarray}
where the factorization scale $\nu$ should be understood as the upper cut-off in rapidity. Performing integration over $z$ we obtain
\begin{eqnarray}
&&\mathcal{B}^{\rm q(1)+bg;real, c}_{ij}(x_B, b_\perp) = 4\alpha_s N_c \int \frac{ \dhd^2k_\perp }{k^2_\perp} e^{ik_\perp b_\perp}
\\
&&\times \Big( \int^1_0\frac{dz}{(1-z)_+}\mathcal{B}^{\rm bg}_{ij}(\frac{x_B}{z}, b_\perp ) + \int^1_0\frac{dz}{z}\mathcal{B}^{\rm bg}_{ij}(\frac{x_B}{z}, b_\perp ) - \Big(\frac{\nu}{x_B P^+}\Big)^\eta \Gamma(1-\eta)\Gamma(\eta)   \mathcal{B}^{\rm bg}_{ij}(x_B, b_\perp)\Big) \,.
\nonumber
\end{eqnarray}

Now we notice that apart from the rapidity divergence, there is a divergence in the integral over transverse momenta. We regulate it using the dimensional regularization,
\begin{eqnarray}
&&\mathcal{B}^{\rm q(1)+bg;real, c}_{ij}(x_B, b_\perp) = \frac{\alpha_s N_c}{\pi } \Big( \frac{\mu^2 b^2_\perp }{ 4 e^{-2\gamma_E}} \Big)^{\epsilon} e^{-\epsilon \gamma_E} \Gamma(-\epsilon )
\\
&&\times \Big( \int^1_0\frac{dz}{(1-z)_+}\mathcal{B}^{\rm bg}_{ij}(\frac{x_B}{z}, b_\perp ) + \int^1_0\frac{dz}{z}\mathcal{B}^{\rm bg}_{ij}(\frac{x_B}{z}, b_\perp ) - \Big(\frac{\nu}{x_B P^+}\Big)^\eta \Gamma(1-\eta)\Gamma(\eta)   \mathcal{B}^{\rm bg}_{ij}(x_B, b_\perp)\Big)\,,
\nonumber
\end{eqnarray}
where $\mu$ is the $\overline{\rm{MS}}$ scale,
\begin{eqnarray}
&&\mu^2 \equiv \frac{4\pi}{e^{\gamma_E}}\mu^2_0\,.
\end{eqnarray}

Expanding in powers of $\epsilon$ and introducing the notation,
\begin{eqnarray}
&& L^\mu_b \equiv \ln\Big(\frac{b^2_\perp \mu^2}{4e^{-2\gamma_E}}\Big)\,,
\end{eqnarray}
one can easily obtain
\begin{eqnarray}
&&\mathcal{B}^{\rm q(1)+bg;real, c}_{ij}(x_B, b_\perp) = - \frac{\alpha_s N_c}{\pi } \Big(\frac{1}{\epsilon} + L^\mu_b\Big)
\\
&&\times \Big( \int^1_0\frac{dz}{(1-z)_+}\mathcal{B}^{\rm bg}_{ij}(\frac{x_B}{z}, b_\perp ) + \int^1_0\frac{dz}{z}\mathcal{B}^{\rm bg}_{ij}(\frac{x_B}{z}, b_\perp ) - \Big(\frac{\nu}{x_B P^+}\Big)^\eta \Gamma(1-\eta)\Gamma(\eta)   \mathcal{B}^{\rm bg}_{ij}(x_B, b_\perp)\Big)\,.
\nonumber
\end{eqnarray}

Also expanding in $\eta$, the equation reads
\begin{eqnarray}
&&\mathcal{B}^{\rm q(1)+bg;real, c}_{ij}(x_B, b_\perp) = - \frac{\alpha_s N_c}{\pi } \Big(\frac{1}{\epsilon} + L^\mu_b\Big)  
\label{rc:tc}\\
&&\times \Big( \int^1_0 \frac{dz }{(1-z)_+} \mathcal{B}^{\rm bg}_{ij}(\frac{x_B}{z}, b_\perp) + \int^1_0 \frac{dz }{z} \mathcal{B}^{\rm bg}_{ij}(\frac{x_B}{z}, b_\perp) - \left(\frac{1}{\eta} + \ln(\frac{\nu}{x_B P^+})\right)   \mathcal{B}^{\rm bg}_{ij}(x_B, b_\perp) \Big)\,.
\nonumber
\end{eqnarray}

While the origin of the pole $1/\eta$ is obvious, i.e. it corresponds to the contribution of the UV modes $k^-\to\infty$ with $\nu$ to be interpreted as a corresponding UV cut-off, the $1/\epsilon$ pole requires some explanation.  
Looking at the integral over transverse momentum $k_\perp$, in Eq.(\ref{rem:comp}), one initially concludes that it is of the IR origin since the UV part of the integral is regulated by a non-zero value of $b_\perp$. This is indeed the case for the first two terms in Eq. (\ref{rc:tc}). However, this is not correct for the last term.

In the next section, we will see that the virtual correction (check Eq. (\ref{virt:form1})) contains a similar contribution,
\begin{eqnarray}
&&- 4\alpha_s N_c \int^1_0 \frac{dz}{1 - z} \int \frac{ \dhd^2k_\perp }{k^2_\perp }  \mathcal{B}^{\rm bg}_{ij}(x_B, b_\perp)\,,
\label{rc:vt}
\end{eqnarray}
which should be set to zero in the dimensional regularization, because it contains a scaleless integral. In other words, the UV and IR regions in this integral cancel each other $1/\epsilon_{\rm UV} - 1/\epsilon_{\rm IR} = 0$.

However, while being zero, this term of the virtual emission plays an important role. Taking the sum of this term with the last term of Eq. (\ref{rc:tki}), which contains the rapidity divergence, we get the contribution
\begin{eqnarray}
&&4\alpha_s N_c \int^1_0\frac{dz}{1-z} \int \frac{ \dhd^2k_\perp }{k^2_\perp} (  e^{ik_\perp b_\perp} - 1 ) \mathcal{B}^{\rm bg}_{ij}(x_B, b_\perp )\,.
\end{eqnarray}
We see that two terms of this equation cancel each other in the IR region and the integral is defined by the UV part of the virtual term (\ref{rc:vt}).

Having this in mind, we relate the $1/\epsilon$ singularity in the last term of Eq. (\ref{rc:tc}) with the UV contribution and set the corresponding factorization scale to $\mu_{\rm UV}$, which should be interpreted as a UV cut-off of the transverse momentum integral. At the same time, the $1/\epsilon$ singularity in the first two terms is of IR origin, so we set the corresponding scales to the factorization scale  $\mu_{\rm IR}$, which should be interpreted as the lower cut-off of the transverse momentum integral.

As a result, the equation reads
\begin{eqnarray}
&&\mathcal{B}^{\rm q(1)+bg;real, c}_{ij}(x_B, b_\perp) = - \frac{\alpha_s N_c}{\pi } \Big(\frac{1}{\epsilon_{\rm IR}} + L^{\mu_{\rm IR}}_b\Big) \int^1_0 dz \Big[ \frac{1 }{(1-z)_+} + \frac{1 }{z} \Big] \mathcal{B}^{\rm bg}_{ij}(\frac{x_B}{z}, b_\perp)
\nonumber\\
&&+ \frac{\alpha_s N_c}{\pi } \Big(\frac{1}{\epsilon_{\rm UV}} + L^{\mu_{\rm UV}}_b\Big) \left(\frac{1}{\eta} + \ln(\frac{\nu}{x_B P^+})\right)   \mathcal{B}^{\rm bg}_{ij}(x_B, b_\perp) \,.
\end{eqnarray}

Finally, summing all real emission diagrams, we get
\begin{eqnarray}
&&\mathcal{B}^{\rm q(1)+bg;real}_{ij}(x_B, b_\perp) = - 4\alpha_s N_c \int \dhd^2p_\perp e^{ip_\perp b_\perp} \int^1_0 \frac{dz }{z(1-z)} \int \dhd^2k_\perp \Big[ \mathcal{R}^{a}_{ij;lm}(z, p_\perp, k_\perp) + \mathcal{R}^{b}_{ij;lm}(z, p_\perp, k_\perp) \Big]
\nonumber\\
&&\times \int d^2z_\perp e^{i(k - p)_\perp z_\perp}\mathcal{B}^{\rm bg}_{lm}(\frac{x_B}{z}, z_\perp)- \frac{\alpha_s N_c}{\pi } \Big(\frac{1}{\epsilon_{\rm IR}} + L^{\mu_{\rm IR}}_b\Big) \int^1_0 dz \Big[ \frac{1 }{(1-z)_+} + \frac{1 }{z} \Big] \mathcal{B}^{\rm bg}_{ij}(\frac{x_B}{z}, b_\perp)
\nonumber\\
&&+ \frac{\alpha_s N_c}{\pi } \Big(\frac{1}{\epsilon_{\rm UV}} + L^{\mu_{\rm UV}}_b\Big) \left(\frac{1}{\eta} + \ln(\frac{\nu}{x_B P^+})\right)   \mathcal{B}^{\rm bg}_{ij}(x_B, b_\perp)\,.
\label{real:cs-full}
\end{eqnarray}

\subsection{Virtual emission diagrams\label{sec:NLO-virt}}
The virtual emission contribution consists of diagrams presented in Fig. \ref{fig:ved} and their complex conjugates. The details of the calculation of these diagrams are presented in Appendix \ref{Ap:vd}. Let's consider the final form of the one-loop correction to the matrix element, see Eq. (\ref{virt:z}),
\begin{eqnarray}
&&\mathcal{B}^{\rm q(1)+bg;virt}_{ij}(x_B, b_\perp) = - 2\alpha_sN_c \int^1_0 \frac{dz}{z} \int \dhd^2 p_\perp e^{ip_\perp b_\perp}\int \dhd^2k_\perp e^{-i k_\perp b_\perp}\mathcal{V}_{ij;lm}(z, p_\perp - k_\perp, k_\perp)
\nonumber\\
&&\times\int d^2z_\perp e^{i(k-p)_\perp z_\perp} \mathcal{B}^{\rm bg}_{lm}(x_B, z_\perp) - 4\alpha_sN_c \int^1_0 \frac{dz}{1 - z} \int \frac{ \dhd^2k_\perp }{k^2_\perp }  \mathcal{B}^{\rm bg}_{ij}(x_B, b_\perp)\,,
\label{virt:form1}
\end{eqnarray}
 where
\begin{eqnarray}
&&\mathcal{V}_{ij;lm}(z, l_\perp, k_\perp) \equiv \frac{g_{il} (2 l_j k_m - l_m k_j ) + (2 l_i k_l - l_l k_i ) g_{mj} }{k^2_\perp(z k^2_\perp + (1-z)(l + k)^2_\perp)}\,.
\end{eqnarray}

Note that the last term of Eq. (\ref{virt:form1}) contains a scaleless integral which in the dimensional regularization should be replaced with zero. However, as we discussed in the previous section, this term, while being zero, plays an important role and generates a UV singularity in a sum with the real emission diagrams. 
Considering this, we drop this term and consider only the first one. It is convenient to make the shift, $p_\perp \to p_\perp + k_\perp$, which yields
\begin{eqnarray}
&&\mathcal{B}^{\rm q(1)+bg;virt}_{ij}(x_B, b_\perp) = - 2\alpha_sN_c \int^1_0 \frac{dz}{z} \int \dhd^2 p_\perp e^{ip_\perp b_\perp}\int \dhd^2k_\perp \mathcal{V}_{ij;lm}(z, p_\perp, k_\perp) \int d^2z_\perp e^{-ip_\perp z_\perp} \mathcal{B}^{\rm bg}_{lm}(x_B, z_\perp) \,.
\label{virt:form11}
\end{eqnarray}

\begin{figure}[htb]
 \begin{center}
\includegraphics[width=170mm]{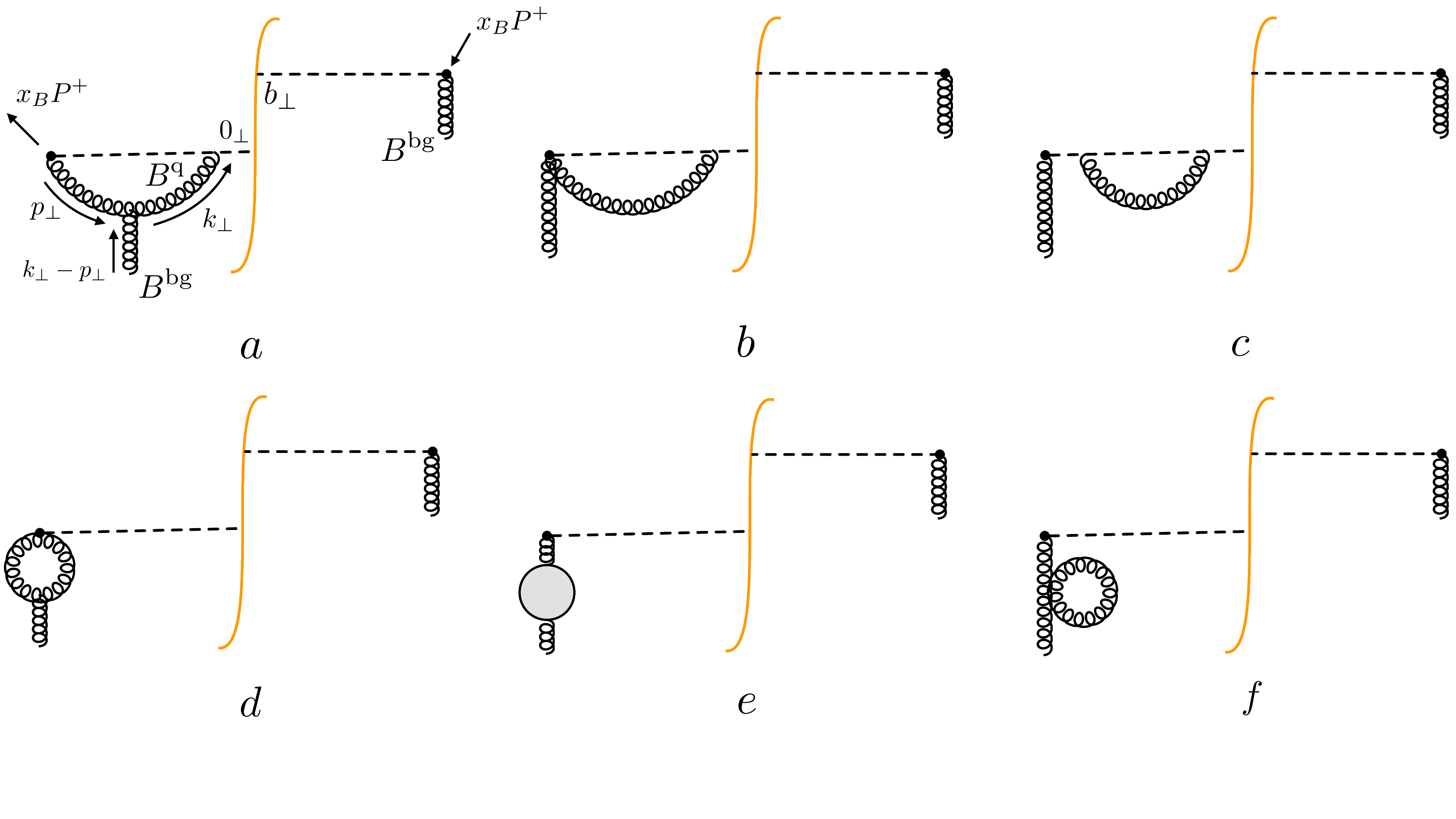}
 \end{center}
\caption{\label{fig:ved}Virtual emission diagrams.}
 \end{figure}

It is easy to see that the equation contains a rapidity divergence at $z\to 0$. This divergence corresponds to the emission of gluons with longitudinal momentum fraction $k^-\to 0$. According to our factorization scheme, we regulate this divergence by modifying the integral over $k^-$ as,
\begin{eqnarray}
&&\int^\infty_0 \frac{\dhd k^-}{k^-} \to \rho^\xi \int^\infty_0 \frac{\dhd k^-}{k^-}  |k^+ |^{-\xi} \,,
\end{eqnarray}
where $\xi$ is the regulator of this divergence, and $\rho$ is the corresponding scale, which should be understood as the lower cut-off in rapidity. We interpret $\rho$ as an IR cut-off. In terms of the $z$ variable, this corresponds to the replacement
\begin{eqnarray}
&&\int^1_0 \frac{dz }{z} \to \Big(\frac{\rho}{x_B P^+}\Big)^\xi \int^1_0 \frac{dz}{z}  \Big( \frac{1 - z}{z} \Big)^{-\xi}\,.
\label{reg:z0}
\end{eqnarray}

Making this replacement in Eq. (\ref{virt:form1}) and integrating over $z$, we write
\begin{eqnarray}
&&\mathcal{B}^{\rm q(1)+bg;virt}_{ij}(x_B, b_\perp) = - 2\alpha_sN_c \Gamma(1-\xi) \Gamma(\xi)\Big(\frac{\rho}{x_B P^+}\Big)^\xi \int \dhd^2 p_\perp e^{ip_\perp b_\perp}
\nonumber\\
&&\times \int \dhd^2k_\perp \frac{g_{il} (2 p_j k_m - p_m k_j ) + (2 p_i k_l - p_l k_i ) g_{mj} }{k^2_\perp (p + k)^2_\perp} \Big(\frac{k^2_\perp}{(p + k)^2_\perp}\Big)^{-\xi} \int d^2z_\perp e^{-ip_\perp z_\perp} \mathcal{B}^{\rm bg}_{lm}(x_B, z_\perp)\,.
\label{virt:form2}
\end{eqnarray}

The integral over transverse momentum $k_\perp$ contains a divergence of the IR origin corresponding to $p_\perp + k_\perp \to 0$.\footnote{Note the shift of variables that we did after Eq. (\ref{virt:form1}).} The integration can be performed using the dimensional regularization,
\begin{eqnarray}
&&\int \dhd^{2-2\epsilon}k_\perp \frac{k_m }{k^2_\perp (p + k)^2_\perp} \Big(\frac{k^2_\perp}{(p + k)^2_\perp}\Big)^{-\xi} 
= - \frac{1}{(4\pi)^{1-\epsilon} } \frac{\Gamma(\xi - \epsilon)\Gamma(1-\xi - \epsilon)\Gamma(1 +\epsilon)}{\Gamma(1 + \xi)\Gamma(1 - \xi ) \Gamma( 1 - 2 \epsilon)} \frac{p_m}{  (p^2_\perp)^{1 +\epsilon} }\,,
\label{virt:kint}
\end{eqnarray}
which yields the following form of the matrix element,
\begin{eqnarray}
&&\mathcal{B}^{\rm q(1)+bg;virt}_{ij}(x_B, b_\perp) = \frac{2\alpha_sN_c}{(4\pi)^{1-\epsilon} } \Big(\frac{\rho}{x_B P^+}\Big)^\xi \Big(\frac{\mu^2 e^{\gamma_E}}{4\pi}\Big)^{\epsilon} \frac{\Gamma(\xi) \Gamma(\xi - \epsilon)\Gamma(1-\xi - \epsilon)\Gamma(1 +\epsilon)}{\Gamma(1 + \xi) \Gamma( 1 - 2 \epsilon)}
\nonumber\\
&&\times \int \dhd^2 p_\perp e^{ip_\perp b_\perp} \frac{g_{il} p_j p_m + p_i p_l g_{mj}}{  (p^2_\perp)^{1 +\epsilon} } \int d^2z_\perp e^{-ip_\perp z_\perp} \mathcal{B}^{\rm bg}_{lm}(x_B, z_\perp)\,.
\label{virt:form3}
\end{eqnarray}

Expanding first in $\xi$,
\begin{eqnarray}
&&\mathcal{B}^{\rm q(1)+bg;virt}_{ij}(x_B, b_\perp) = \frac{2\alpha_sN_c}{(4\pi)^{1-\epsilon} }  \Big(\frac{\mu^2 e^{\gamma_E}}{4\pi}\Big)^{\epsilon} \frac{\Gamma(1 +\epsilon) \Gamma (1-\epsilon) \Gamma (-\epsilon)}{\Gamma( 1 - 2 \epsilon)} \left( \frac{1 }{\xi} + \ln (\frac{\rho}{x_B P^+}) +  \psi ^{(0)}(-\epsilon) - \psi ^{(0)}(1-\epsilon) \right)
\nonumber\\
&&\times \int \dhd^2 p_\perp e^{ip_\perp b_\perp} \frac{g_{il} p_j p_m + p_i p_l g_{mj}}{  (p^2_\perp)^{1 +\epsilon} } \int d^2z_\perp e^{-ip_\perp z_\perp} \mathcal{B}^{\rm bg}_{lm}(x_B, z_\perp)\,,
\label{virt:form4}
\end{eqnarray}
and then in $\epsilon$, we find
\begin{eqnarray}
&&\mathcal{B}^{\rm q(1)+bg;virt}_{ij}(x_B, b_\perp) = - \frac{\alpha_sN_c}{2\pi } \left( \frac{1}{\epsilon^2_{\rm IR}} + \frac{1}{\epsilon_{\rm IR}} \left(  \frac{1 }{\xi} + \ln (\frac{\rho}{x_B P^+}) \right) - \frac{\pi^2 }{12} \right)
\nonumber\\
&&\times \int d^{2}z_\perp \int \dhd^{2} p_\perp e^{ip_\perp (b-z)_\perp} \Big(\frac{\mu^2_{\rm IR}}{p^2_\perp } \Big)^{\epsilon_{\rm IR}}\frac{g_{il} p_j p_m + p_i p_l g_{mj}}{ p^2_\perp } \mathcal{B}^{\rm bg}_{lm}(x_B, z_\perp)
\label{virt:cs-full-norun}\,,
\end{eqnarray}
where we indicate that the divergence in $\epsilon$ is of the IR origin. As we mentioned above, we also interpret the $1/\xi$ pole as originating from the IR region of $k^-\to 0$.

Finally, adding the contribution of diagrams in Figs. \ref{fig:ved}e,f\footnote{See also Refs. \cite{Leibbrandt:1983pj,Altinoluk:2023dww}.} we obtain
\begin{eqnarray}
&&\mathcal{B}^{\rm q(1)+bg;virt}_{ij}(x_B, b_\perp) = - \frac{\alpha_sN_c}{2\pi } \left( \frac{1}{\epsilon^2_{\rm IR}} + \frac{1}{\epsilon_{\rm IR}} \left(  \frac{1 }{\xi} + \ln (\frac{\rho}{x_B P^+}) \right) - \frac{\pi^2 }{12} \right)
\nonumber\\
&&\times \int d^{2}z_\perp \int \dhd^{2} p_\perp e^{ip_\perp (b-z)_\perp} \Big(\frac{\mu^2_{\rm IR}}{p^2_\perp } \Big)^{\epsilon_{\rm IR}}\frac{g_{il} p_j p_m + p_i p_l g_{mj}}{ p^2_\perp } \mathcal{B}^{\rm bg}_{lm}(x_B, z_\perp)
\nonumber\\
&&+\frac{\alpha_s N_c}{2\pi} \Big(  \frac{1}{\epsilon_{\rm UV}}\frac{\beta_0}{2N_c} + \frac{67}{18} - \frac{5N_f}{9N_c} \Big) \int d^2z_\perp \int \dhd^2p_\perp e^{ip_\perp (b - z )_\perp} \Big( \frac{\mu^2_{\rm UV}}{p^2_\perp}\Big)^{\epsilon_{\rm UV}} \mathcal{B}^{\rm bg}_{ij}(x_B, z_\perp)\,,
\label{virt:cs-full}
\end{eqnarray}
where
\begin{eqnarray}
&&\beta_0 = \frac{11N_c}{3} - \frac{2N_f}{3}\,.
\end{eqnarray}

Note that in the collinear approximation often used in NLO calculations of TMDPDFs, if the transverse momentum of background gluons is much smaller than the transverse momentum inside the loop, the virtual correction in dimensional regularization should be set to zero. Indeed, replacing $\mathcal{B}^{\rm bg}_{lm}(x_B, z_\perp) \to \mathcal{B}^{\rm bg}_{lm}(x_B, 0_\perp)$ in the first term of Eq. (\ref{virt:form1}) we obtain a scaleless integral over $k_\perp$, which is zero in the dimensional regularization. We will use this observation later when we compare our result with the collinear matching procedure. However, as found in our calculation, the virtual correction generally has a non-trivial form.

\subsection{Calculation of the soft factor\label{sec:NLO-soft}}
To calculate the TMDPDF (\ref{fTMD}), we also need to find the soft function. Expanding the soft function (\ref{soft_func}) in the coupling constant, it is straightforward to see that $\mathcal{S}^{(0)}(b_\perp) = 1$.

\begin{figure}[htb]
 \begin{center}
\includegraphics[width=160mm]{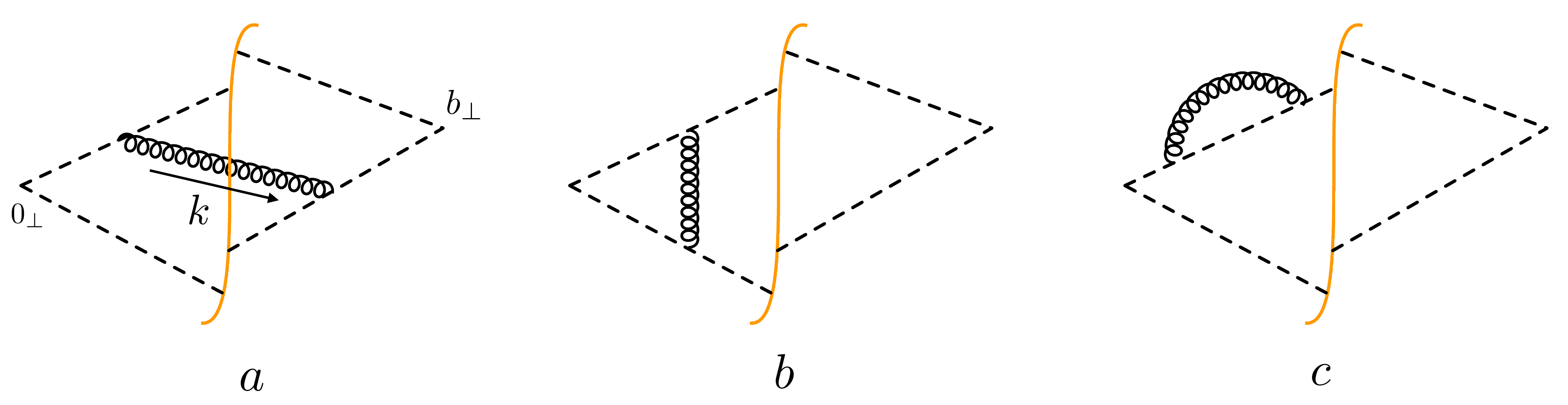}
 \end{center}
\caption{\label{fig:soft}The soft function at the NLO order. It is understood that there are also conjugated diagrams that we don't draw explicitly.}
 \end{figure}

At the order $\alpha_s$, the contributing diagrams are given in Fig. \ref{fig:soft}. It is easy to calculate these diagrams in the Feynman gauge. In this gauge, the diagram in Fig. \ref{fig:soft}c and its complex conjugate do not contribute.

Let's start with a diagram in Fig. \ref{fig:soft}a. Expanding the soft Wilson lines in Eq. (\ref{soft_func}) to the $g^2$ order we get
\begin{eqnarray}
&&\mathcal{S}^{(1)}_a(b_\perp) = g^2 N_c \int \dhd^4k 2\pi\delta(k^2) \Big[ \frac{1}{k^+-i\epsilon} \frac{1}{k^-+i\epsilon} + \frac{1}{k^--i\epsilon} \frac{1}{k^+ + i\epsilon} \Big] e^{ik_\perp b_\perp}\,,
\end{eqnarray}
where the second term corresponds to the conjugated diagram.

The equation contains a rapidity divergence, which we regulate by inserting $ \nu^\eta 2^{-\eta/2} |k^+ - k^-|^{-\eta}$. Note that this regularization differs from the rapidity regularization of the matrix element. Using the delta function to evaluate the integral over $k^+$ and the dimensional regularization for the transverse momentum integral we rewrite the equation as
\begin{eqnarray}
&&\mathcal{S}^{(1)}_a(b_\perp) = 4 \alpha_s N_c \nu^\eta 2^{-\eta/2} \Big( \frac{\mu^2 e^{\gamma_E}}{4\pi} \Big)^\epsilon \int^\infty_0 \frac{dk^-}{k^-}\int \dhd^{2-2\epsilon}k_\perp \Big|k^- - \frac{k^2_\perp}{2k^-}\Big|^{-\eta} \frac{1}{k^2_\perp} e^{ik_\perp b_\perp} \,.
\label{ap:sf-form1}
\end{eqnarray}

Integrating over the longitudinal momentum $k^-$,
\begin{eqnarray}
&&\mathcal{S}^{(1)}_a(b_\perp) = 4 \alpha_s N_c \nu^\eta \Big( \frac{\mu^2 }{4\pi e^{-\gamma_E}} \Big)^\epsilon  2^{-\eta}\frac{\Gamma(\frac{1}{2}-\frac{\eta}{2})\Gamma(\frac{\eta}{2})}{\sqrt{\pi}} \int \dhd^{2-2\epsilon}k_\perp \frac{1}{(k^2_\perp)^{1+\eta/2}} e^{ik_\perp b_\perp}\,,
\end{eqnarray}
and the transverse momentum $k_\perp$, we obtain
\begin{eqnarray}
&&\mathcal{S}^{(1)}_a(b_\perp) = \frac{ \alpha_s N_c \nu^\eta}{\pi^{3/2}} \Big( \frac{\mu^2 b^2_\perp}{4 e^{-\gamma_E}} \Big)^\epsilon  2^{- 2\eta } \frac{ \Gamma(\frac{1}{2}-\frac{\eta}{2})\Gamma(\frac{\eta}{2}) \Gamma (-\epsilon-\frac{\eta}{2}) }{\Gamma(1 + \frac{\eta}{2})} (b^2_\perp)^{ \frac{\eta}{2} }\,.
\end{eqnarray}

Expanding first in $\eta$
\begin{eqnarray}
&&\mathcal{S}^{(1)}_a(b_\perp) 
= \frac{ \alpha_s N_c }{\pi} \Big( \frac{\mu^2 b^2_\perp}{4 e^{-2\gamma_E}} \Big)^\epsilon e^{-\epsilon\gamma_E}\Gamma (-\epsilon) \left( \frac{2 }{\eta} + \ln (\frac{\mu^2 b^2_\perp}{4 e^{-2\gamma_E}}) +  2\ln (\frac{\nu}{\mu}) - \gamma_E - \psi^{(0)}(-\epsilon) \right) \,,
\end{eqnarray}
and second in $\epsilon$, we write
\begin{eqnarray}
&&\mathcal{S}^{(1)}_a(b_\perp) 
= \frac{ \alpha_s N_c }{2\pi} \left( \frac{2}{\epsilon^2} + 4\left(\frac{1}{\epsilon} + L_b\right)\left( -\frac{1 }{\eta}  +  \ln \frac{\mu}{\nu}  \right) -   L^2_b - \frac{\pi ^2 }{6}\right) \,.
\label{ap:sfa}
\end{eqnarray}

Now let's discuss the origin of the poles in Eq. (\ref{ap:sfa}). From Eq. (\ref{ap:sf-form1}) one may initially conclude that the UV region of integration is regulated by the $b_\perp$ variable, and the integral over transverse momentum is divergent in the IR region, hence one may associate $1/\epsilon$ divergence with the IR contribution.

However, there is another diagram and its complex conjugate, presented in Fig. \ref{fig:soft}b. The sum of these diagrams has the following expression,
\begin{eqnarray}
&&\mathcal{S}^{(1)}_b(b_\perp) = -i g^2 N_c \int^\infty_0 dx^+ \int^\infty_0 dx^- \Big( (x^-, b_\perp|\frac{1}{p^2 + i\epsilon}|x^+, b_\perp) + (x^-, 0_\perp|\frac{1}{p^2 + i\epsilon}|x^+, 0_\perp)\Big) \,,
\end{eqnarray}
which can be rewritten as
\begin{eqnarray}
&&\mathcal{S}^{(1)}_b(b_\perp) = - 2i g^2 N_c  \int^\infty_{-\infty} \frac{dk^+}{2\pi} \int^\infty_{-\infty} \frac{dk^-}{2\pi} \int \dhd^2k_\perp \frac{1}{k^- + i\epsilon} \frac{1}{k^+ - i\epsilon}\frac{1}{2k^+ k^- - k^2_\perp + i\epsilon} \,.
\end{eqnarray}

Taking a pole in the $k^+$ variable, we get the result
\begin{eqnarray}
&&\mathcal{S}^{(1)}_b(b_\perp) = - 4\alpha_s N_c   \int^\infty_0 \frac{dk^-}{k^-} \int  \frac{\dhd^2k_\perp}{ k^2_\perp  } \,,
\end{eqnarray}
which contains a scaleless integral over transverse momentum. In the dimensional regularization, this integral should be set to zero, which corresponds to a cancellation between the UV and IR regions: $1/\epsilon_{\rm UV} - 1/\epsilon_{\rm IR} = 0$.

However, the role of this contribution is not trivial. Taking a sum of diagrams in Fig. \ref{fig:soft}a and \ref{fig:soft}b we obtain
\begin{eqnarray}
&&\mathcal{S}^{(1)}_{a+b}(b_\perp) = 4 \alpha_s N_c   \int^\infty_0 \frac{dk^-}{k^-}\int  \frac{\dhd^{2}k_\perp}{k^2_\perp} (e^{ik_\perp b_\perp} - 1)\,.
\end{eqnarray}
We see that in the IR region, two contributions cancel each other. The only surviving piece is the UV part of the diagram in Fig. \ref{fig:soft}b.

Hence, we find that the singularity in Eq. (\ref{ap:sfa}) should be associated with the UV region. As a result, we get the following final result for the soft function at the NLO order,
\begin{eqnarray}
&&\mathcal{S}^{(1)}(b_\perp) 
= \frac{ \alpha_s N_c }{2\pi} \left( \frac{2}{\epsilon^2_{\rm UV}} + 4(\frac{1}{\epsilon_{\rm UV}} + L_b)\left( -\frac{1 }{\eta}  +  \ln \frac{\mu_{UV}}{\nu}  \right) -   L^2_b - \frac{\pi ^2 }{6}\right)\,.
\label{ap:sf-fin}
\end{eqnarray}

\subsection{Gluon TMDPDFs at the NLO order\label{sec:NLO-fin}}
Taking a sum of the real (\ref{real:cs-full}) and virtual (\ref{virt:cs-full}) contributions we obtain the following form of the matrix element (\ref{gop-def}) at NLO 
\begin{eqnarray}
&&\mathcal{B}^{\rm q(1)+bg}_{ij}(x_B, b_\perp) = - 4\alpha_s N_c \int \dhd^2p_\perp e^{ip_\perp b_\perp} \int^1_0 \frac{dz }{z(1-z)} \int \dhd^2k_\perp \Big[ \mathcal{R}^{a}_{ij;lm}(z, p_\perp, k_\perp) + \mathcal{R}^{b}_{ij;lm}(z, p_\perp, k_\perp) \Big]
\nonumber\\
&&\times \int d^2z_\perp e^{i(k - p)_\perp z_\perp}\mathcal{B}^{\rm bg}_{lm}(\frac{x_B}{z}, z_\perp) + \frac{\alpha_s N_c}{\pi } \Big(\frac{1}{\epsilon_{\rm UV}} + L_b\Big) \Big(\frac{1}{\eta} + \ln(\frac{\nu}{x_B P^+})\Big)   \mathcal{B}^{\rm bg}_{ij}(x_B, b_\perp)
\nonumber\\
&&- \frac{\alpha_s N_c}{\pi } \Big(\frac{1}{\epsilon_{\rm IR}} + L_b\Big) \int^1_0 dz \Big[  \frac{1 }{(1-z)_+} + \frac{1 }{z} \Big] \mathcal{B}^{\rm bg}_{ij}(\frac{x_B}{z}, b_\perp)
 - \frac{\alpha_sN_c}{2\pi } \Big( \frac{1}{\epsilon^2_{\rm IR}} + \frac{1}{\epsilon_{\rm IR}} \Big(  \frac{1 }{\xi} + \ln (\frac{\rho}{x_B P^+}) \Big) - \frac{\pi^2 }{12} \Big)
\nonumber\\
&&\times \int d^{2}z_\perp \int \dhd^{2} p_\perp e^{ip_\perp (b-z)_\perp} \Big(\frac{\mu^2}{p^2_\perp } \Big)^{\epsilon_{\rm IR}}\frac{g_{il} p_j p_m + p_i p_l g_{mj}}{ p^2_\perp } \mathcal{B}^{\rm bg}_{lm}(x_B, z_\perp) +\frac{\alpha_s N_c}{2\pi} \Big(  \frac{1}{\epsilon_{\rm UV}}\frac{\beta_0}{2N_c} + \frac{67}{18} - \frac{5N_f}{9N_c} \Big)
\nonumber\\
&&\times \int d^2z_\perp \int \dhd^2p_\perp e^{ip_\perp (b - z )_\perp} \Big( \frac{\mu^2_{\rm UV}}{p^2_\perp}\Big)^{\epsilon_{\rm UV}} \mathcal{B}^{\rm bg}_{ij}(x_B, z_\perp)\,.
\label{bareNLO}
\end{eqnarray}

The matrix element contains different types of divergences. Let's first consider the $1/\eta$ divergence, which is generated with the emission at $k^-\to \infty$. This divergence appears due to factorization between the soft function and the matrix element. To remove this divergence, we consider the TMDPDF (\ref{fTMD}), which at NLO takes the form
\begin{eqnarray}
&&f^{(1)}_{ij}(x_B, b_\perp) = \mathcal{B}^{\rm q(1)+bg}_{ij}(x_B, b_\perp) + \frac{1}{2}\mathcal{S}^{(1)}(b_\perp)\mathcal{B}^{\rm q(0)+bg}_{ij}(x_B, b_\perp)\,,
\end{eqnarray}
where we have taken into account that the soft factor is trivial in the leading order in the coupling constant.

Combining Eqs. (\ref{bareNLO}) and (\ref{ap:sf-fin}) we find that the $1/\eta$ poles cancel between the soft function and the matrix element,
\begin{eqnarray}
&&f^{(1)}_{ij}(x_B, b_\perp) = - 4\alpha_s N_c \int \dhd^2p_\perp e^{ip_\perp b_\perp} \int^1_0 \frac{dz }{z(1-z)} \int \dhd^2k_\perp \Big[ \mathcal{R}^{a}_{ij;lm}(z, p_\perp, k_\perp)+ \mathcal{R}^{b}_{ij;lm}(z, p_\perp, k_\perp) \Big]
\nonumber\\
&& \times \int d^2z_\perp e^{i(k_\perp-p_\perp)z_\perp}f^{(0)}_{lm}(\frac{x_B}{z}, z_\perp) + \frac{ \alpha_s N_c }{4\pi} \Big( \frac{2}{\epsilon^2_{\rm UV}} + 4(\frac{1}{\epsilon_{\rm UV}} + L^{\mu_{\rm UV}}_b) \ln \frac{\mu_{\rm UV}}{x_B P^+}  - (L^{\mu_{\rm UV}}_b)^2 - \frac{\pi ^2 }{6}\Big)f^{(0)}_{ij}(x_B, b_\perp)
\nonumber\\
&&- \frac{\alpha_s N_c}{\pi } \Big(\frac{1}{\epsilon_{\rm IR}} + L^{\mu_{\rm IR}}_b\Big) \int^1_0 dz \Big[  \frac{1 }{(1-z)_+} + \frac{1 }{z} \Big] f^{(0)}_{ij}(\frac{x_B}{z}, b_\perp)  - \frac{\alpha_sN_c}{2\pi } \Big( \frac{1}{\epsilon^2_{\rm IR}} + \frac{1}{\epsilon_{\rm IR}} \Big(  \frac{1 }{\xi} + \ln (\frac{\rho}{x_B P^+}) \Big) - \frac{\pi^2 }{12} \Big)
\nonumber\\
&&\times \int d^{2}z_\perp \int \dhd^{2} p_\perp e^{ip_\perp (b-z)_\perp} \Big(\frac{\mu^2_{\rm IR}}{p^2_\perp } \Big)^{\epsilon_{\rm IR}}\frac{g_{il} p_j p_m + p_i p_l g_{mj}}{ p^2_\perp } f^{(0)}_{lm}(x_B, z_\perp) +\frac{\alpha_s N_c}{2\pi} \Big(  \frac{1}{\epsilon_{\rm UV}}\frac{\beta_0}{2N_c} + \frac{67}{18} - \frac{5N_f}{9N_c} \Big)
\nonumber\\
&&\times \int d^2z_\perp \int \dhd^2p_\perp e^{ip_\perp (b - z )_\perp} \Big( \frac{\mu^2_{\rm UV}}{p^2_\perp}\Big)^{\epsilon_{\rm UV}} f^{(0)}_{ij}(x_B, z_\perp)\,,
\end{eqnarray}
where we have taken into account that both the soft function and contribution of the ``quantum" fields $B^q$ in the matrix element (\ref{gop-def}) are trivial at the leading order in the coupling constant, so the TMDPDF (\ref{fTMD}) at this order is $f^{(0)}_{ij}(x_B, b_\perp) = \mathcal{B}^{\rm bg}_{ij}(x_B, b_\perp)$.

The UV divergences can be removed by performing the UV renormalization with the universal UV renormalization factor for the TMDPDFs
\begin{eqnarray}
&&Z_{\rm UV} = 1 - \frac{\alpha_s N_c}{2\pi}\Big[\frac{1}{\epsilon^2_{\rm UV}} + \frac{1}{\epsilon_{\rm UV}}\ln\Big(\frac{\mu^2_{\rm UV}}{(x_B P^+)^2}\Big) + \frac{1}{\epsilon_{\rm UV}}\frac{\beta_0}{2N_c}\Big]\,.
\end{eqnarray}

The equation reads,
\begin{eqnarray}
&&f^{(1)}_{ij}(x_B, b_\perp) = - 4\alpha_s N_c \int \dhd^2p_\perp e^{ip_\perp b_\perp} \int^1_0 \frac{dz }{z(1-z)} \int \dhd^2k_\perp \Big[ \mathcal{R}^{a}_{ij;lm}(z, p_\perp, k_\perp) + \mathcal{R}^{b}_{ij;lm}(z, p_\perp, k_\perp) \Big]
\nonumber\\
&&\times \int d^2z_\perp e^{i(k_\perp-p_\perp)z_\perp}f^{(0)}_{lm}(\frac{x_B}{z}, z_\perp) + \frac{ \alpha_s N_c }{2\pi} \Big( - \frac{1}{2}(L^{\mu_{\rm UV}}_b)^2 + L^{\mu_{\rm UV}}_b \ln \frac{\mu^2_{\rm UV}}{\zeta^2}   - \frac{\pi ^2 }{12}\Big)f^{(0)}_{ij}(x_B, b_\perp)
\nonumber\\
&&- \frac{\alpha_s N_c}{\pi } \Big(\frac{1}{\epsilon_{\rm IR}} + L^{\mu_{\rm IR}}_b\Big) \int^1_0 dz \Big[  \frac{1 }{(1-z)_+} + \frac{1 }{z} \Big] f^{(0)}_{ij}(\frac{x_B}{z}, b_\perp)
 - \frac{\alpha_sN_c}{2\pi } \Big( \frac{1}{\epsilon^2_{\rm IR}} + \frac{1}{\epsilon_{\rm IR}} \Big(  \frac{1 }{\xi} + \ln (\frac{\rho}{x_B P^+}) \Big) - \frac{\pi^2 }{12} \Big)
\nonumber\\
&&\times \int d^{2}z_\perp \int \dhd^{2} p_\perp e^{ip_\perp (b-z)_\perp} \Big(\frac{\mu^2_{\rm IR}}{p^2_\perp } \Big)^{\epsilon_{\rm IR}}\frac{g_{il} p_j p_m + p_i p_l g_{mj}}{ p^2_\perp } f^{(0)}_{lm}(x_B, z_\perp)+\frac{\alpha_s N_c}{2\pi} \int d^2z_\perp \int \dhd^2p_\perp e^{ip_\perp (b - z )_\perp}
\nonumber\\
&&\times \Big( \frac{\beta_0}{2N_c} \ln \frac{\mu^2_{\rm UV}}{p^2_\perp} + \frac{67}{18} - \frac{5N_f}{9N_c} \Big) f^{(0)}_{ij}(x_B, z_\perp)\,,
\label{res:beforeren}
\end{eqnarray}
where we introduced a scale $\zeta = x_B P^+$, following the prescription of the TMD factorization.

The remaining IR divergences and the rapidity pole $1/\xi$ are of no concern since they should be associated with the initial operator constructed from background fields. Absorbing these poles into the initial distribution, we obtain
\begin{eqnarray}
&&f_{ij}(x_B, b_\perp, \mu^2_{\rm UV}, \zeta) = f_{ij}(x_B, b_\perp, \mu^2_{\rm IR}, \rho) - 4\alpha_s N_c \int \dhd^2p_\perp e^{ip_\perp b_\perp} \int^1_0 \frac{dz }{z(1-z)} \int \dhd^2k_\perp \Big[ \mathcal{R}^{a}_{ij;lm}(z, p_\perp, k_\perp)
\nonumber\\
&&+ \mathcal{R}^{b}_{ij;lm}(z, p_\perp, k_\perp) \Big] \int d^2z_\perp e^{-i(p_\perp - k_\perp)z_\perp}f_{lm}(\frac{x_B}{z}, z_\perp, \mu^2_{\rm IR}, \rho) + \frac{ \alpha_s N_c }{2\pi} \Big( - \frac{1}{2}(L^{\mu_{\rm UV}}_b)^2 + L^{\mu_{\rm UV}}_b \ln \frac{\mu^2_{\rm UV}}{\zeta^2}   - \frac{\pi ^2 }{12}\Big)
\nonumber\\
&&\times f_{ij}(x_B, b_\perp, \mu^2_{\rm IR}, \rho) - \frac{\alpha_s N_c}{\pi } L^{\mu_{\rm IR}}_b \int^1_0 dz \Big[ \frac{1 }{(1-z)_+} + \frac{1 }{z} \Big] f_{ij}(\frac{x_B}{z}, b_\perp, \mu^2_{\rm IR}, \rho)  - \frac{\alpha_sN_c}{2\pi } \int d^2z_\perp \int \dhd^{2} p_\perp e^{ip_\perp (b-z)_\perp}
\nonumber\\
&&\times \Big( \frac{1}{2} \ln^2 \frac{\mu^2_{\rm IR}}{p^2_\perp } + \ln \frac{\mu^2_{\rm IR}}{p^2_\perp } \ln \frac{\rho}{x_B P^+} - \frac{\pi^2 }{12} \Big) \frac{g_{il} p_j p_m + p_i p_l g_{mj}}{ p^2_\perp } f_{lm}(x_B, z_\perp, \mu^2_{\rm IR}, \rho)
\nonumber\\
&&+\frac{\alpha_s N_c}{2\pi} \int d^2z_\perp \int \dhd^2p_\perp e^{ip_\perp (b - z )_\perp} \Big( \frac{\beta_0}{2N_c} \ln \frac{\mu^2_{\rm UV}}{p^2_\perp} + \frac{67}{18} - \frac{5N_f}{9N_c} \Big) f_{ij}(x_B, z_\perp, \mu^2_{\rm IR}, \rho) + O(\alpha^2_s)\,.
\label{fr:form1}
\end{eqnarray}
Equation (\ref{fr:form1}) is our final result for the NLO correction to the TMD distribution obtained in our factorization scheme. In the next section, we will discuss how our factorization scheme reduces to other factorization schemes in certain kinematic regions.

\section{Matching to other factorization schemes\label{mat:cl}}
In the previous section, we derived the gluon TMDPDFs at NLO in the TMD factorization scheme defined by a set of factorization scales introduced in Fig. \ref{fig:fnew}. These scales define the factorization in both the rapidity and transverse momentum. However, there are kinematical regions where the factorization is dominated by either a wide separation in the transverse momenta or longitudinal momenta (rapidity). In these regions, our factorization effectively matches other schemes. We will explore this matching in this section.

In particular, we will consider two scenarios. Firstly, we will look at the limit of large $x_B$ and small $b_\perp \ll \Lambda^{-1}_{\rm QCD}$, and relate our TMD factorization scheme to the collinear factorization through the so-called collinear matching procedure. In this procedure, we will expand the TMDPDFs in terms of the collinear PDFs. As we will see, this implies that the TMD factorization is dominated by a wide separation in transverse momentum and thus reduces to collinear factorization.

Secondly, we will consider an opposite kinematic limit of small $x_B$ and large $b_\perp \lesssim \Lambda^{-1}_{\rm QCD} $ in which the TMDPDFs can be expanded in terms of the unintegrated gluon distributions. We will show that in this limit, the TMD factorization scheme matches the small-x rapidity factorization, which is characterized by a wide separation in rapidity.

\subsection{Matching onto the collinear factorization scheme\label{sec:cmatch}}
Our result (\ref{fr:form1}) for the gluon TMDPDFs is applicable in a wide range of kinematic variables $x_B$ and $b_\perp$. In particular, it describes the perturbative structure of the TMDPDFs for large values of $b_\perp \lesssim \Lambda^{-1}_{\rm QCD}$. In this limit, the transverse momenta inside the ``quantum" loops of $B^{\rm q}$ fields are of the order of the transverse momenta of the background fields $B^{\rm bg}$. For example, in the real emission diagrams, see Fig. \ref{fig:red}, in this kinematic limit we have $p_\perp\sim b^{-1}_\perp \gtrsim l_\perp = k_\perp - p_\perp$, where $l_\perp\sim \mu_{\rm IR}$ is a transverse momentum of the background fields. In this case, the transverse logarithms $L^{\mu_{\rm IR}}_b$ are not large. Similarly, the transverse integrals in the finite terms do not generate a large contribution.

In this section, we consider an opposite case, when the kinematic variable $b_\perp \ll \Lambda^{-1}_{\rm QCD}$ is very small. In this case, the transverse momenta in the ``quantum" loops are much larger than the typical transverse momenta of the background fields: $p_\perp \sim b^{-1}_\perp \gg l_\perp \sim \mu_{\rm IR}$. So, the corresponding transverse logarithms are large, and the transverse integrals in the finite terms provide a large contribution. This is a consequence of a wide separation in the transverse momentum between the $B^{\rm q}$ and $B^{\rm bg}$ fields. This separation dominates our TMD factorization scheme in this particular kinematic limit. As a result, we conclude that our factorization matches the collinear factorization scheme which is defined by the same wide separation in the transverse momentum.

To formalize this statement, let's perform the collinear matching procedure in Eq. (\ref{res:beforeren}) by expanding the TMDPDFs in terms of the collinear PDFs. This expansion enforces the transverse momenta of the background gluons to be exactly zero, which reflects a strict ordering in the transverse momenta. One consequence of this procedure, see discussion after Eq. (\ref{virt:cs-full}), is that the virtual emission diagrams don't contribute. In this case, the TMDPDFs at the NLO order take the form
\begin{eqnarray}
&&f^{(1)}_{ij}(x_B, b_\perp) = - 4\alpha_s N_c \int^1_0 \frac{dz }{z(1-z)} \int \dhd^2k_\perp e^{ik_\perp b_\perp} \Big[ \mathcal{R}^{a}_{ij;lm}(z, k_\perp, k_\perp) + \mathcal{R}^{b}_{ij;lm}(z, k_\perp, k_\perp) \Big] f^{(0)}_{lm}(\frac{x_B}{z}, 0_\perp)
\nonumber\\
&& - \frac{\alpha_s N_c}{\pi } \Big(\frac{1}{\epsilon_{\rm IR}} + L^{\mu_{\rm IR}}_b\Big) \int^1_0 dz \Big[  \frac{1 }{(1-z)_+} + \frac{1 }{z} \Big] f^{(0)}_{ij}(\frac{x_B}{z}, 0_\perp) - \frac{1}{\epsilon_{\rm IR}} \frac{\alpha_s \beta_0}{4\pi}   f^{(0)}_{ij}(x_B, 0_\perp)
\nonumber\\
&&+ \frac{ \alpha_s N_c }{2\pi} \Big( - \frac{1}{2}(L^{\mu_{\rm UV}}_b)^2 + L^{\mu_{\rm UV}}_b \ln \frac{\mu^2_{\rm UV}}{\zeta^2}   - \frac{\pi ^2 }{12}\Big)f^{(0)}_{ij}(x_B, 0_\perp)
 +  \dots\,,
\end{eqnarray}
where the ellipsis stands for the higher twist contributions, and
\begin{eqnarray}
&&\mathcal{R}^{a}_{ij;lm}(z, k_\perp, k_\perp)
\\
&&= (1-z)^2 \Big( \frac{k_l}{k^2_\perp} \frac{ zk^2_\perp\delta^k_i - 2(1-z)k^k k_i}{k^2_\perp} - \frac{\delta^k_l k_i + g_{li} k^k }{k^2_\perp} \Big) 
\Big( \frac{k_m}{ k^2_\perp } \frac{ zk^2_\perp g_{kj} - 2(1-z)k_k k_j}{ k^2_\perp} - \frac{ g_{mk} k_j + g_{mj}k_k}{k^2_\perp} \Big)\,,
\nonumber
\end{eqnarray}
and
\begin{eqnarray}
&&\mathcal{R}^{b}_{ij;lm}(z, k_\perp, k_\perp)= (1-z)\frac{g_{il} }{k^2_\perp} \Big( (1 - z)\frac{k_m k_j }{ k^2_\perp} + g_{mj} \Big)
 + (1-z)\Big( (1 - z)\frac{ k_i k_l }{k^2_\perp} + g_{li} \Big) \frac{g_{mj} }{k^2_\perp} \,.
\end{eqnarray}

As we mentioned above, at small values of $b_\perp$ the transverse integrals in the finite terms, i.e. terms with kernels $\mathcal{R}^{a}$ and $\mathcal{R}^{b}$, provide a large contribution. In the limit $b_\perp\to0$, these terms become logarithmically divergent. This is a consequence of an additional factorization condition, i.e. the strict ordering of the transverse momenta, which is imposed by the collinear matching procedure.

At this point, it is  convenient to introduce a parametrization of the TMDPDFs (matrix elements) in terms of the distribution functions
\begin{eqnarray}
&&f_{ij}(x_B, b_\perp)  = x_BP^+\Big[ -\frac{g_{ij}}{2}f_1(x_B, b_\perp) + \Big(\frac{g_{ij}}{2} + \frac{b_ib_j}{b^2_\perp}\Big)h_1(x_B, b_\perp)\Big] \,,
\label{dc:cr}
\end{eqnarray}
which in the momentum space takes the form,
\begin{eqnarray}
&&f_{ij}(x_B, p_\perp) = x_B P^+ \Big[-\frac{g_{ij}}{2} \tilde{f}_1(x_B, p_\perp) + \Big(\frac{g_{ij}}{2} + \frac{p_ip_j}{p^2_\perp} \Big)  \tilde{h}_1(x_B, p_\perp)\Big]\,.
\label{prm:mt}
\end{eqnarray}
Let's consider a projection onto the unpolarized TMDPDF $f_1(x_B, b_\perp)$ and neglect mixing with the $h_1(x_B, b_\perp)$ distribution. In this case, the equation reads
\begin{eqnarray}
&&f^{(1)}_1(x_B, b_\perp) 
= 4\alpha_s N_c \int^1_0 \frac{dz }{z} ( - 2 + z - z^2 ) \int \frac{\dhd^2k_\perp}{k^2_\perp} e^{ik_\perp b_\perp}  f^{(0)}_1(\frac{x_B}{z}, 0_\perp) - \frac{\alpha_s N_c}{\pi } \Big(\frac{1}{\epsilon_{\rm IR}} + L^{\mu_{\rm IR}}_b\Big)
\nonumber\\
&&\times \int^1_0 \frac{dz}{z} \Big[  \frac{1 }{(1-z)_+} + \frac{1 }{z} \Big] f^{(0)}_1(\frac{x_B}{z}, 0_\perp) - \frac{1}{\epsilon_{\rm IR}} \frac{\alpha_s \beta_0}{4\pi}   f^{(0)}_1(x_B, 0_\perp)
\nonumber\\
&&+ \frac{ \alpha_s N_c }{2\pi} \Big( - \frac{1}{2}(L^{\mu_{\rm UV}}_b)^2 + L^{\mu_{\rm UV}}_b \ln \frac{\mu^2_{\rm UV}}{\zeta^2}   - \frac{\pi ^2 }{12}\Big)f^{(0)}_1(x_B, 0_\perp)
 + \dots\,.
\label{cm:form2}
\end{eqnarray}

Performing the integration over transverse momentum $k_\perp$ in the first term and expanding in $\epsilon$ we obtain
\begin{eqnarray}
&&f^{(1)}_1(x_B, b_\perp) 
= - \frac{\alpha_s N_c}{\pi} \Big(\frac{1}{\epsilon_{\rm IR}} + L^{\mu_{\rm IR}}_b\Big) \int^1_0 \frac{dz }{z} ( - 2 + z - z^2 ) f^{(0)}_1(\frac{x_B}{z}, 0_\perp) - \frac{\alpha_s N_c}{\pi } \Big(\frac{1}{\epsilon_{\rm IR}} + L^{\mu_{\rm IR}}_b\Big)
\nonumber\\
&&\times \int^1_0 \frac{dz}{z} \Big[  \frac{1 }{(1-z)_+} + \frac{1 }{z} \Big] f^{(0)}_1(\frac{x_B}{z}, 0_\perp) - \frac{1}{\epsilon_{\rm IR}} \frac{\alpha_s \beta_0}{4\pi}   f^{(0)}_1(x_B, 0_\perp)
\nonumber\\
&&+ \frac{ \alpha_s N_c }{2\pi} \Big( - \frac{1}{2}(L^{\mu_{\rm UV}}_b)^2 + L^{\mu_{\rm UV}}_b \ln \frac{\mu^2_{\rm UV}}{\zeta^2}   - \frac{\pi ^2 }{12}\Big)f^{(0)}_1(x_B, 0_\perp)
 + \dots\,.
\label{cm:form3}
\end{eqnarray}

Note that the first term in Eq. (\ref{cm:form3}) corresponds to the finite terms in our result (\ref{res:beforeren}). In Eq. (\ref{res:beforeren}) these terms are finite because the corresponding transverse integrals are regularized by a non-zero value $l_\perp = k_\perp - p_\perp$ of the transverse momentum of the background fields. In the small $b_\perp$ limit the transverse momentum in the ``quantum" loop becomes large. In this case, the dominant contribution to the integral over $k_\perp$ in the finite terms corresponds to the ordering $k_\perp \sim p_\perp \gg l_\perp$. Hence, in this regime, our factorization matches with the collinear factorization, which is defined by the strict ordering $k_\perp = p_\perp \gg l_\perp = 0$. With this condition the integral over $k_\perp$ becomes divergent and thus gives a pole $1/\epsilon_{\rm IR}$ in the first term of the Eq. (\ref{cm:form3}). We emphasize again that this pole is a consequence of the matching procedure with the collinear factorization, which dominates the solution at small $b_\perp$.

Absorbing the infrared poles into the distribution constructed from the background fields, we finally obtain
\begin{eqnarray}
&&f_1(x_B, b_\perp, \mu^2_{\rm UV}, \zeta) = f_1(x_B, 0_\perp, \mu^2_{\rm IR})
\label{cm:form4}\\
&& - \frac{\alpha_s N_c}{\pi}  L^{\mu_{\rm IR}}_b \int^1_0 \frac{dz }{z} P_{gg}(z) f_1(\frac{x_B}{z}, 0_\perp, \mu^2_{\rm IR}) + \frac{\alpha_s N_c}{2\pi } \Big( - \frac{1}{2}(L^{\mu_{\rm UV}}_b)^2 + L^{\mu_{\rm UV}}_b \ln \frac{\mu^2_{\rm UV}}{\zeta^2}   - \frac{\pi ^2 }{12}\Big) f_1(x_B, 0_\perp, \mu^2_{\rm IR})+\dots\,,
\nonumber
\end{eqnarray}
where,
\begin{eqnarray}
&&P_{gg}(z) = \frac{1}{(1 - z)_+} + \frac{1 }{z}  -2 + z  - z^2
\end{eqnarray}
is the DGLAP splitting function. Note that the dependence on the IR scale $\rho$ disappears. The matching equation (\ref{cm:form4}) is a well-known result, see e.g. \cite{Chiu:2012ir}. We conclude that our calculation provides the correct IR structure of the TMDPDFs in the large $x_B$ and small $b_\perp$ approximation.

\subsection{Matching onto the high-energy rapidity factorization scheme\label{sec:eikexp}}
In this section, we will study a kinematic limit that is opposite to the one considered in the previous section. We'll consider a limit of a small value of the $x_B$ variable and a large value of the $b_\perp \lesssim \Lambda^{-1}_{\rm QCD}$. We will demonstrate that in this kinematic limit, our TMD factorization scheme matches the high-energy rapidity factorization, which is characterized by a strong ordering of the longitudinal momenta $p^- \gg l^-$, where $p^-$ is a typical longitudinal momentum of the ``quantum" fields $B^{\rm q}$ and $l^-$ is a corresponding momentum of the background fields $B^{\rm bg}$.

Firstly, since the value of the $b_\perp$ variable is large, it is easy to see that the transverse integrals and logarithms in Eq. (\ref{res:beforeren}) do not acquire large values, which is opposite to the collinear limit explored in the previous section. At the same time, the integrals over $z$ variable contain a potential divergence at $z\to 0$. Though this divergence is regulated by a non-zero value of $x_B$, in the limit of small $x_B$, it leads to a large value of the corresponding integrals. The $x_B$ variable effectively plays the role of a cut-off in rapidity between the ``quantum" and collinear modes. When $x_B\to 0$, the typical value of the background momenta $l^- \to 0$ as well, leading to a wide separation in the longitudianl momenta between the modes, i.e. $p^- \gg l^- \to 0$, where $l^-\neq 0$. As a result, with these kinematical approximations, the TMD factorization scheme can be matched onto the high-energy rapidity factorization which is characterized by a strict ordering $p^- \gg l^- = 0$.

The matching procedure can be easily constructed by the eikonal expansion of the matrix element in Eq. (\ref{fr:form1}), which is effectively an expansion in powers of $x_B$. In this section, we will consider only the leading term of the expansion; it can be obtained by neglecting the $x_B$ dependence in the matrix element (\ref{fr:form1}). Note that by introducing a strict ordering $p^- \gg l^- = 0$, we effectively impose an additional factorization condition between the kinematical modes, which leads to the appearance of some new unphysical poles in the finite terms, which are supposed to be absorbed into the matrix element of the background fields.

To understand the structure of Eq. (\ref{fr:form1}) in the limit of small $x_B$ let's first consider the matrix element of the TMD operator. Taking a limit $x_B \to 0$, and assuming that at small-$x_B$ the background fields are dominated by a classical configuration with the only non-zero component $A_-$, we can rewrite the matrix element of the TMD operator as
\begin{eqnarray}
&&\lim_{x_B\to 0}\mathcal{B}_{ij}(x_B, b_\perp) \propto \langle p | {\rm Tr} (U_b \partial_i U^\dag_b)(U_0\partial_jU^\dag_0) |p\rangle \,,
\label{opatsmx}
\end{eqnarray}
where the infinite Wilson line $U_x\equiv [\infty, -\infty]_x$.

Comparing (\ref{opatsmx}) with the decomposition (\ref{prm:mt}) we find that in the limit $x_B\to0$ two unpolarized TMD distributions coincide,
\begin{eqnarray}
&&\lim_{x_B\to0}\tilde{f}_1(x_B, p_\perp) = \lim_{x_B\to0}\tilde{h}_1(x_B, p_\perp)\,.
\label{background:smallx}
\end{eqnarray}
As a result, in this limit, the matrix element can be parametrized as
\begin{eqnarray}
&&\lim_{x_B\to 0}f_{ij}(x_B, p_\perp) = \frac{p_i p_j}{p^2_\perp} \mathcal{H}_1(p_\perp)\,,
\end{eqnarray}
where $\mathcal{H}_1$ is a dipole amplitude defined by a matrix element of a product of two Wilson lines.

As a result, after the eikonal expansion, Eq. (\ref{res:beforeren}) in the momentum space reads
\begin{eqnarray}
&& f^{(1)}_{ij}(x_B, p_\perp) 
\nonumber\\
&&= - 4\alpha_s N_c \int^1_0 \frac{dz }{z(1-z)} \int \dhd^2k_\perp \Big[ \mathcal{R}^{a}_{ij;lm}(z, p_\perp, k_\perp) + \mathcal{R}^{b}_{ij;lm}(z, p_\perp, k_\perp) \Big] \frac{(p-k)_l(p-k)_m}{(p-k)^2_\perp}\mathcal{H}^{(0)}_1(p_\perp - k_\perp) 
\nonumber\\
&&+ \frac{ \alpha_s N_c }{2\pi} \int d^2b_\perp \Big( - \frac{1}{2}(L^{\mu_{\rm UV}}_b)^2 + L^{\mu_{\rm UV}}_b \ln \frac{\mu^2_{\rm UV}}{\zeta^2}   - \frac{\pi ^2 }{12}\Big) \int \dhd^2 k_\perp e^{ib_\perp k_\perp}\frac{(p-k)_i(p-k)_j}{(p-k)^2_\perp}\mathcal{H}^{(0)}_1(p_\perp - k_\perp)
\nonumber\\
&&- \frac{\alpha_s N_c}{\pi } \int d^2b_\perp \Big(\frac{1}{\epsilon_{\rm IR}} + L^{\mu_{\rm IR}}_b\Big) \int^1_0  \frac{dz}{z} \int \dhd^2 k_\perp e^{ib_\perp k_\perp} \frac{(p-k)_i(p-k)_j}{(p-k)^2_\perp}\mathcal{H}^{(0)}_1(p_\perp - k_\perp)
 \nonumber\\
 &&+ \frac{\alpha_sN_c}{\pi }  \Big( \frac{1}{\epsilon^2_{\rm IR}} + \frac{1}{\epsilon_{\rm IR}} \Big(  \frac{1 }{\xi} + \ln (\frac{\rho}{x_B P^+}) \Big) - \frac{\pi^2 }{12} \Big)
    \Big(\frac{\mu^2_{\rm IR}}{p^2_\perp } \Big)^{\epsilon_{\rm IR}}\frac{ p_i p_j }{ p^2_\perp } \mathcal{H}^{(0)}_1(p_\perp)
\nonumber\\
&&+ \frac{\alpha_s N_c}{2\pi} \Big( \frac{\beta_0}{2N_c} \ln \frac{\mu^2_{\rm UV}}{p^2_\perp} + \frac{67}{18} - \frac{5N_f}{9N_c} \Big) \frac{p_i p_j}{p^2_\perp} \mathcal{H}^{(0)}_1(p_\perp) + \dots\, ,
\label{eik-exp}
\end{eqnarray}
where ellipsis stands for the higher order terms of expansion in eikonality.

From Eq. (\ref{eik-exp}), one can see that the leading term of expansion contains a rapidity divergent term which has an integral $\int^1_0  \frac{dz}{z}$ with a divergence at $z\to 0$. We need to regulate this divergence using a replacement (\ref{reg:z0}). For brevity, let's focus on a projection onto the unpolarized TMD distribution
$\tilde{f}_1(x_B, p_\perp)$,\footnote{ Similarly, one can consider a projection onto the $\tilde{h}_1(x_B, p_\perp)$ distribution.}\footnote{ Here we use a notation $f_1(x_B, p_\perp)\equiv x_BP^+\tilde{f}_1(x_B, p_\perp)$.}
\begin{eqnarray}
&& f^{(1)}_1(x_B, p_\perp) 
\nonumber\\
&&= - 4\alpha_s N_c \int^1_0 \frac{dz }{z(1-z)} \int \dhd^2k_\perp \Big[ \mathcal{R}^{a}_{ii;lm}(z, p_\perp, k_\perp) + \mathcal{R}^{b}_{ii;lm}(z, p_\perp, k_\perp) \Big] \frac{(p-k)_l(p-k)_m}{(p-k)^2_\perp}\mathcal{H}^{(0)}_1(p_\perp - k_\perp) 
\nonumber\\
&&+ \frac{ \alpha_s N_c }{2\pi} \int d^2b_\perp \Big( - \frac{1}{2}(L^{\mu_{\rm UV}}_b)^2 + L^{\mu_{\rm UV}}_b \ln \frac{\mu^2_{\rm UV}}{\zeta^2}   - \frac{\pi ^2 }{12}\Big) \int \dhd^2 k_\perp e^{ib_\perp k_\perp} \mathcal{H}^{(0)}_1(p_\perp - k_\perp) - \frac{\alpha_s N_c}{\pi} \int d^2b_\perp \Big(\frac{1}{\epsilon_{\rm IR}} + L^{\mu_{\rm IR}}_b\Big)
\nonumber\\
&&\times \Big(\frac{1}{\xi} + \ln \frac{\rho}{x_B P^+}\Big)  \int \dhd^2 k_\perp e^{ib_\perp k_\perp} \mathcal{H}^{(0)}_1(p_\perp - k_\perp)
+ \frac{\alpha_sN_c}{\pi}  \Big( \frac{1}{\epsilon^2_{\rm IR}} + \frac{1}{\epsilon_{\rm IR}} \Big(  \frac{1 }{\xi} + \ln (\frac{\rho}{x_B P^+}) \Big) - \frac{\pi^2 }{12} \Big)
    \Big(\frac{\mu^2_{\rm IR}}{p^2_\perp } \Big)^{\epsilon_{\rm IR}} \mathcal{H}^{(0)}_1(p_\perp)
\nonumber\\
&&+ \frac{\alpha_s N_c}{2\pi} \Big( \frac{\beta_0}{2N_c} \ln \frac{\mu^2_{\rm UV}}{p^2_\perp} + \frac{67}{18} - \frac{5N_f}{9N_c} \Big) \mathcal{H}^{(0)}_1(p_\perp) + \dots\:\,.
\label{sx:form2}
\end{eqnarray}
Note that since
\begin{eqnarray}
&&\lim_{z\to0}\Big[ \mathcal{R}^{a}_{ii;lm}(z, p_\perp, k_\perp) + \mathcal{R}^{b}_{ii;lm}(z, p_\perp, k_\perp) \Big] \frac{(p-k)_l(p-k)_m}{(p-k)^2_\perp} = 0 \,,
\end{eqnarray}
the integral over $z$ in the first term is finite and doesn't require any regularization.

Eq. (\ref{sx:form2}) contains IR poles which should be absorbed into the distribution constructed from the background fields. As a result, we finally obtain,
\begin{eqnarray}
&& f_1(x_B, p_\perp, \mu^2_{\rm UV}, \zeta) = \mathcal{H}_1(p_\perp, \mu^2_{\rm IR}, \rho) + \ln \frac{\rho}{x_B P^+} \int \dhd^2 k_\perp K_{\rm BFKL}(p_\perp, k_\perp) \mathcal{H}_1(p_\perp - k_\perp, \mu^2_{\rm IR}, \rho)
\nonumber\\
&& + \frac{ \alpha_s N_c }{2\pi} \int d^2b_\perp \Big( - \frac{1}{2}(L^{\mu_{\rm UV}}_b)^2 + L^{\mu_{\rm UV}}_b \ln \frac{\mu^2_{\rm UV}}{\zeta^2}   - \frac{\pi ^2 }{12}\Big) \int \dhd^2 k_\perp e^{i k_\perp b_\perp } \mathcal{H}_1(p_\perp - k_\perp, \mu^2_{\rm IR}, \rho)
\nonumber\\
&&- 4\alpha_s N_c \int^1_0 \frac{dz }{z(1-z)} \int \dhd^2k_\perp \Big[ \mathcal{R}^{a}_{ii;lm}(z, p_\perp, k_\perp) + \mathcal{R}^{b}_{ii;lm}(z, p_\perp, k_\perp) \Big] \frac{(p-k)_l(p-k)_m}{(p-k)^2_\perp}\mathcal{H}^{(0)}_1(p_\perp - k_\perp, \mu^2_{\rm IR}, \rho) 
\nonumber\\
&&+ \frac{\alpha_sN_c}{\pi}  \Big(  \frac{1}{2}\ln^2\frac{\mu^2_{\rm IR}}{p^2_\perp } - \frac{\pi^2 }{12} \Big) \mathcal{H}_1(p_\perp, \mu^2_{\rm IR}, \rho) + \frac{\alpha_s N_c}{2\pi} \Big( \frac{\beta_0}{2N_c} \ln \frac{\mu^2_{\rm UV}}{p^2_\perp} + \frac{67}{18} - \frac{5N_f}{9N_c} \Big) \mathcal{H}_1(p_\perp, \mu^2_{\rm IR}, \rho) + \dots\,,
\label{sx:form3}
\end{eqnarray}
where
\begin{eqnarray}
&&K_{\rm BFKL}(p_\perp, k_\perp) = - \frac{\alpha_s N_c}{\pi} \int d^2b_\perp  e^{i k_\perp b_\perp } L^{\mu_{\rm IR}}_b + \frac{\alpha_sN_c}{\pi} (2\pi)^2\delta^2(k_\perp)\ln\frac{\mu^2_{\rm IR}}{p^2_\perp }
\end{eqnarray}
is the BFKL evolution kernel. Note that the last term of the kernel originates from the virtual emission.

Equation (\ref{sx:form3}) contains a double logarithmic term $\ln^2(\mu^2_{\rm IR}/p^2_\perp)$. Since we assume that $b_\perp \sim 1/p_\perp \lesssim \Lambda^{-1}_{\rm QCD}$, we expect that the IR scale $\mu^2_{\rm IR}$ is chosen in a way that $\mu^2_{\rm IR}\sim p^2_\perp$, and the logarithm doesn't take large values. This is different from the first term in Eq. (\ref{sx:form3}), which, for $\rho\gg x_B P^+ \to 0$, has a large rapidity logarithm $\ln (\rho / x_B P^+)$. As a result, we can, in a practical manner, neglect the IR double logarithmic term, as well as the $\mu_{\rm IR}$ dependence of the dipole amplitude. Removing the double logarithmic term, along with some finite terms, we write,
\begin{eqnarray}
&& f_1(x_B, p_\perp, \mu^2_{\rm UV}, \zeta) \simeq \mathcal{H}_1(p_\perp, \rho) + \ln \frac{\rho}{x_B P^+} \int \dhd^2 k_\perp K_{\rm BFKL}(p_\perp, k_\perp) \mathcal{H}_1(p_\perp - k_\perp, \rho)
\nonumber\\
&& + \frac{ \alpha_s N_c }{2\pi} \int d^2b_\perp \Big( - \frac{1}{2}(L^{\mu_{\rm UV}}_b)^2 + L^{\mu_{\rm UV}}_b \ln \frac{\mu^2_{\rm UV}}{\zeta^2}   - \frac{\pi ^2 }{12}\Big) \int \dhd^2 k_\perp e^{i k_\perp b_\perp } \mathcal{H}_1(p_\perp - k_\perp, \rho)
\nonumber\\
&&+ \frac{\alpha_s N_c}{2\pi} \Big( \frac{\beta_0}{2N_c} \ln \frac{\mu^2_{\rm UV}}{p^2_\perp} + \frac{67}{18} - \frac{5N_f}{9N_c} \Big) \mathcal{H}_1(p_\perp, \rho) \,.
\label{sx:form4}
\end{eqnarray}

Equation (\ref{sx:form4}) has a form similar to Eq. (\ref{cm:form4}), where the first term describes the IR structure of the TMD distribution, and the rest of the terms describe its UV content. From Eq. (\ref{sx:form4}), we find that in the limit of large $b_\perp \lesssim \Lambda^{-1}_{\rm QCD}$ and small $x_B$ the IR structure of the distribution is dominated by the rapidity logarithm and is described by the BFKL kernel. At the same time, the UV part is governed by the CSS evolution. This structure is consistent with a form of the cross-section of the back-to-back inclusive di-jets production in DIS calculated in the small x framework, see \cite{Mueller:2012uf,Mueller:2013wwa,Xiao:2017yya,Caucal:2021ent,Caucal:2022ulg,Caucal:2023nci,Caucal:2023fsf}.

\section{Conclusions}
In this paper, we study the IR structure of the gluon TMDPDFs. We go beyond the standard collinear matching procedure, which is performed in the $b_\perp \ll \Lambda^{-1}_{\rm QCD}$ approximation, and develop a new approach that is valid in a wide region of $x_B$ and $b_\perp \lesssim \Lambda^{-1}_{\rm QCD}$. To do that, we employ the background field method and perform the calculation of the TMDPDFs in the dilute limit at the NLO order.

We find that at this order the perturbative structure of the TMDPDFs is described by Eq. (\ref{fr:form1}). The equation describes the dependence of the TMDPDFs on various factorization scales. In our approach, we explicitly distinguish the scales corresponding to the UV and IR physics. In particular, our approach yields the standard dependence on the UV scales, described by the UV logarithms that can be resummed into the CSS evolution.

At the same time, the dependence on the IR scales is a new result. In particular, apart from the transverse logarithm $\ln (b^2_\perp \mu^2_{\rm IR})$, our result contains a rapidity logarithm $\ln (\rho/x_B P^+)$, which dependence on a factorization scale $\rho$ of the IR origin. This dependence comes from the virtual emission diagrams, which are not trivial in our approach.

In general, the IR part is not dominated by a single logarithm. As a result, the corresponding kernels are neither the DGLAP nor the BFKL ones. We also keep finite non-logarithmic terms. Again, in our general kinematics the contribution of these terms can be comparable with the logarithmic contribution. As is evident from our result, the IR dynamics of the TMDPDFs is governed by an interplay between the aforementioned terms.

We also demonstrate that our result is consistent with the collinear matching procedure. We argue that in the kinematic limit of $b_\perp \ll \Lambda^{-1}_{\rm QCD}$ our result is dominated by the transverse logarithm $\ln (b^2_\perp \mu^2_{\rm IR})$. We also observe that non-logarithmic terms play an important role in this limit. These terms are enhanced, and in the collinear matching, they are described by the same transverse logarithm. At the same time, the virtual correction is trivial. Combining all terms we find that the IR structure is governed by the DGLAP kernel; we then recover the standard collinear matching formula for the gluon TMDPDFs.

Meanwhile, our result is derived in general kinematics, so it describes the IR physics of the TMDPDFs in the region of large $b_\perp \lesssim \Lambda^{-1}_{\rm QCD}$ as well. To better understand this structure, we particularly consider a limit of small values of $x_B$. We argue that in this limit, the IR physics is dominated by a rapidity logarithm $\ln (\rho/x_B P^+)$, while the transverse integral is suppressed. At the same time, some finite terms are enhanced in this limit and are described by the same rapidity logarithm. In analogy to the collinear matching, one can perform a matching procedure expanding the TMDPDFs in terms of dipole amplitudes. The expansion converges at small values of the $x_B$ variable. Combining all terms  we find that the dominant contribution in the IR sector is described by the BFKL kernel. This is evident from the matching equation (\ref{sx:form4}), which is also a new result. We want to emphasize the critical role played by the virtual emission, which, as we argue, is not trivial at any finite value of $b_\perp$,  and is essential for reproducing the BFKL kernel at small values of $x_B$. 

As a result, we conclude that the TMD factorization provides a unified description of the CSS, DGLAP, and BFKL evolutions. 

The possibility to construct such formalism has been explored before in the Color Glass Condensate (CGC) Effective Field Theory \cite{Mueller:2012uf,Mueller:2013wwa,Xiao:2017yya,Stasto:2018rci,Caucal:2021ent,Caucal:2022ulg,Caucal:2023nci,Caucal:2023fsf}, high-energy rapidity factorization \cite{Balitsky:2015qba,Balitsky:2016dgz,Balitsky:2022vnb,Balitsky:2023hmh}, high-energy factorization \cite{Zhou:2016tfe,Zhou:2018lfq,Hentschinski:2020tbi,Hentschinski:2021lsh,Celiberto:2022fgx}, and SCET framework \cite{Fleming:2014rea,Rothstein:2016bsq,Neill:2023jcd,Stewart:2023lwz}. The main difference between all methods is essentially how the QCD medium is factorized into different dynamic modes and what types of modes are taken into account.

Our calculation shares some similarities with the aforementioned works. For example, in Refs. \cite{Mueller:2012uf,Zhou:2016tfe,Xiao:2017yya,Hentschinski:2020tbi,Hentschinski:2021lsh} the authors use the operator definition of the gluon TMD distribution, and in Refs. \cite{Zhou:2016tfe,Hentschinski:2020tbi,Hentschinski:2021lsh} the calculation is done in the dilute limit. However, in Refs. \cite{Mueller:2012uf,Mueller:2013wwa,Xiao:2017yya,Stasto:2018rci,Caucal:2021ent,Caucal:2022ulg,Caucal:2023nci,Caucal:2023fsf,Zhou:2016tfe,Hentschinski:2020tbi,Hentschinski:2021lsh} the form of the background field is circumscribed to the small-$x$ partons. As a result, the authors recovered the CSS evolution scheme in combination with the BFKL evolution. It was also observed that CSS and BFKL evolutions correspond to distinct regions of phase space. This is consistent with our results described in Sec. \ref{sec:eikexp}, where we consider the limit of small values of the $x_B$ variable. 

However, our form of the background field is more general. It is defined by a system of UV and IR factorization scales, which we introduced in Sec. \ref{sec:two}. The proposed scheme allows us to link not only the CSS and BFKL evolution but also include the DGLAP evolution; in our calculations, the latter dominates the IR structure of the TMDPDFs in the $b_\perp \ll \Lambda^{-1}_{\rm QCD}$ limit. This is in full agreement with the well-known results obtained in the collinear matching procedure, see discussion in Sec. \ref{sec:cmatch}. Hence, our scheme is more general than previously proposed methods.

A calculation with a general form of the background field was previously done in the high-energy rapidity factorization approach \cite{Balitsky:2015qba,Balitsky:2016dgz,Balitsky:2022vnb,Balitsky:2023hmh}. However, this approach is different from the TMD factorization we use in this paper. Essentially, the TMDPDFs that we study, which are the standard TMDPDFs \cite{Collins:2011zzd,Collins:1981uk,Collins:1984kg,Collins:1987pm,Collins:1989gx,Meng:1995yn,Ji:2004xq,Ji:2004wu,Boussarie:2023izj}, are completely different from the distributions considered in Refs. \cite{Balitsky:2015qba,Balitsky:2016dgz,Balitsky:2022vnb,Balitsky:2023hmh}. These two types of the TMD distributions cannot be directly compared. Indeed, while the TMDPDFs in the rapidity factorization approach depend only on a single rapidity factorization scale, the TMDPDFs in the standard TMD factorization scheme contain dependence on two scales, which we study in detail in this paper.

A somewhat alternative approach to constructing a general formalism was also proposed in SCET \cite{Fleming:2014rea,Rothstein:2016bsq,Neill:2023jcd,Stewart:2023lwz}. Apart from many technical differences, the main distinction with our approach comes from the fact that in the SCET framework, different dynamical modes are, by construction, well separated from each other, which is done using momentum scaling. In particular, the small-x effects are associated with the contribution of the so-called Glauber mode, see for instance, Ref. \cite{Rothstein:2016bsq}. This significantly differs from our approach based on the background field method. The background field method allows us to consider scattering in a general background field, which we define using a set of factorization scales. This general field does not necessarily correspond to a particular kinematic mode in SCET. However, as one can conclude from Sec. \ref{mat:cl}, the field effectively reduces to the SCET modes when a particular kinematic limit is imposed. As a result, our calculation in a general background field allows us to describe a smooth transition between different momenta regions, which seems rather challenging in the SCET framework.

In this paper, we relate the BFKL and DGLAP evolution using the TMD factorization approach. An alternative method, which is based on the analysis of the parton emission in the $k_T$-factorization \cite{Catani:1990eg,Catani:1990xk,Catani:1994sq} was presented in Refs. \cite{Salam:1998tj,Altarelli:1999vw,Ciafaloni:1999au,Ciafaloni:1999yw,Altarelli:2001ji,Altarelli:2003hk,Ciafaloni:2003rd,Ciafaloni:2003ek,Ciafaloni:2003kd,Marzani:2015oyb,Colferai:2023dcf}. In this method, a collinear resummation of the DGLAP terms is performed using analysis of the behavior of the anomalous dimensions in Mellin space. While at first sight this is quite different from our approach, it would be interesting to investigate a potential link to our method. We leave this analysis for the future.

Though in this paper, we mainly study the gluon TMDPDFs, our formalism can be easily applied to the case of quark distributions. We plan to do that in the future. This will allow us to implement the formalism in phenomenological applications. Since our formalism provides a general description of the IR structure of the TMDPDFs, including the region of large $b_\perp \lesssim \Lambda^{-1}_{\rm QCD}$, which dominates the TMD factorization, it would be necessary and interesting to see how this effects the global analysis of data compared with the solution based on the collinear matching technique.

Another obvious extension of our formalism is the study of the IR structure of the quasi-TMDPDFs \cite{Ji:2014hxa,Ji:2018hvs,Ebert:2018gzl,Ebert:2019okf,Ebert:2019tvc,Ji:2019sxk,Ji:2019ewn,Ebert:2020gxr,Ji:2020jeb,Ji:2021znw,Ebert:2022fmh,Deng:2022gzi}. We plan to perform this analysis and implement it in lattice computations of the TMDPDFs.

The latter would be especially beneficial for the small-$x$ physics. Indeed, our result demonstrates a deep relation between the TMD and small-$x$ physics. In particular, it suggests that our solution for the TMDPDFs can be used to construct an initial condition (defined at large $x$) for the small-$x$ evolution. Currently, the initial condition used in phenomenological applications is model-dependent, so the possibility of constraining it using lattice computations is of great interest.

Finally, since our approach bridges different types of kinematic effects, it can be applied in a variety of problems where the interplay between small- and large-x physics is important, e.g. spin effects at small-$x$ \cite{Tarasov:2021yll,Tarasov:2020cwl,Kovchegov:2015pbl,Kovchegov:2016zex,Cougoulic:2022gbk,Adamiak:2023yhz}.

\section{Acknowledgnents}
We thank Ian Balitsky, Paul Caucal, Florian Cougoulic, Yuri V. Kovchegov, Farid Salazar, Bj\"orn Schenke, Tomasz Stebel, Yossathorn Tawabutr, and Raju Venugopalan for the discussions. 

This material is based upon work supported by The U.S. Department of Energy, Office of Science, Office of Nuclear Physics through Contract Nos.~DE-SC0012704 and DE-SC0020081, and within the frameworks of Saturated Glue (SURGE) Topical Collaboration in Nuclear Theory. V.S. thanks the Binational Science Foundation grant \#2021789 for support.

\appendix
\section{Calculation of the real emission vertex\label{Ap:evL}}
In this Appendix we present details of the calculation of the emission vertex (\ref{def:evL}) which corresponds to resummation of diagrams presented in Fig. \ref{fig:evL}. We will perform our derivation in the background field method. As we mentioned before, one of the big advantages of the background field method is the ability to independently choose gauges of the background $B^{\rm bg}_\mu$ and ``quantum" field $B^{\rm q}_\mu$. In particular, in our calculation, we will use an axial gauge for the ``quantum" field $B^{\rm q}_\mu$, and choose the background field to be of a form $B^{\rm bg}_\mu(x^-, x_\perp)$ with a gauge fixed condition
\begin{eqnarray}
&&B^{\rm bg}_+ = 0
\label{gauge1}
\end{eqnarray}
supplemented by a boundary condition for the transverse component of the field,
\begin{eqnarray}
&&\lim_{x^-\to\infty}B^{\rm bg}_i = 0 \,.
\end{eqnarray}

To determine the operator structure of the background fields, we will use the following approach. We will work in the dilute regime assuming that there are only two gluons in the background (i.e. $g^2$ order in the coupling constant). Some typical diagrams contributing to the NLO correction to the TMD operator are presented in Fig. \ref{fig:red}. We will find that a background field representing each gluon combines into an operator structure $\partial_\mu B^{\rm bg}_\nu - \partial_\nu B^{\rm bg}_\mu$. This structure is nothing else but the abelian part of the strength tensor $F_{\mu\nu}\equiv F^{\rm bg}_{\mu\nu}$, constructed from the background field $B^{\rm bg}_\mu$, which is a proper gauge covariant representation of a single gluon insertion. As a result, in our dilute approximation, it's sufficient, without any loss of generality, to replace this operator structure with a full non-abelian strength tensor $F_{\mu\nu}$. To restore gauge invariance we will also insert gauge links connecting the two strength tensors. It is important to emphasize that the background field method allows us to restore the non-abelian part of the strength tensor and Wilson lines as well. However, to do that one has to consider diagrams beyond the $g^2$ order in the coupling constant. To simplify our discussion, we do not do it explicitly. Indeed, this is sufficient in the dilute limit since the abelian part of the strength tensors is unambiguously defined by the leading order diagrams in the coupling constant, and the higher order diagrams merely restore the non-abelian parts of the strength tensors or gauge links.

However, we should mention that this is not the case in general when multiple gluon insertions become important (i.e. in the dense limit). In this case, corresponding operators of the background fields might contain, in contrast to the dilute limit, more than two strength tensors. The main difficulty, in this case, is that a given gluon insertion might correspond either to a strength tensor or a gauge link, so the operator structure in general cannot be found. The standard approach to overcome this difficulty is to expand all background fields onto a given direction defined by the kinematics of the problem (e.g. the light-cone direction). After this expansion, the structure of the strength tensors and gauge links can be unambiguously restored. However, this choice of direction is not universal. In particular, it is different in the Bjorken and Regge limits. At this point, it is not clear whether a universal expansion can be constructed. We leave this problem for future research.

First, we rewrite the emission vertex (\ref{def:evL}) using the following equations for derivatives of the Wilson lines constructed from the field $B^{\rm q+bg}_\mu = B^{\rm q}_\mu + B^{\rm bg}_\mu$,
\begin{eqnarray}
&&\partial_-[\infty,y^-]_y = - i g[\infty,y^-]_y B^{\rm q+bg}_-(y^-, y_\perp) \,,
\end{eqnarray}
and
\begin{eqnarray}
&&\partial_i [\infty, y^-]_y = - ig[\infty, y^-]_y B^{\rm q+bg}_i(y^-, y_\perp) - ig\int^\infty_{y^-} dz^- [\infty, z^-]_y F^{\rm q+bg}_{- i}(z^-, y_\perp)[z^-, y^-]_y \,.
\end{eqnarray}

Taking into account these equations, the emission vertex reads,
\begin{eqnarray}
&&L^{ab}_{\mu j}(k,y_\perp,x_B)=i\lim_{k^2\to 0}k^2 \int^{\infty}_{-\infty} dy^- e^{ix_BP^+y^-} \Big\{ -ix_BP^+ [\infty,y^-]^{bd}_y \langle B^{{\rm q}a}_\mu(k) B^{{\rm q}d}_j(y^-,y_\perp)\rangle_{B^{\rm bg}}
\nonumber\\
&&- \partial_j \Big( [\infty,y^-]^{bd}_y \langle B^{{\rm q}a}_\mu(k) B^{{\rm q}d}_-(y^-,y_\perp)\rangle_{B^{\rm bg}} \Big)
 - ig\int^\infty_{y^-} dz^- ([\infty, z^-]_y F_{- j}(z^-, y_\perp)[z^-, y^-]_y)^{bd} \langle B^{{\rm q}a}_\mu(k) B^{{\rm q}d}_-(y^-,y_\perp)\rangle_{B^{\rm bg}}
\nonumber\\
&&+ig\int_{y^-}^\infty dz^- ([\infty,z^-] T^e [z^-,y^-])^{bm}_{y_\perp}F^m_{- j}(y^-, y_\perp) \langle B^{{\rm q}a}_\mu(k) B^{{\rm q}e}_-(z^-, y_\perp)\rangle_{B^{\rm bg}} \Big\}\,,
\label{ev:initial}
\end{eqnarray}
where the last term corresponds to an emission from the Wilson line, see Fig. \ref{fig:evL}c, and the strength tensor $F_{\mu\nu}\equiv F^{\rm bg}_{\mu\nu}$ is constructed from the background field $B^{\rm bg}_\mu$. Now we need to substitute here a contraction of two ``quantum fields" $B^{\rm q}_\mu$ in a background $B^{\rm bg}_\mu$. This contraction in the axial gauge of the field $B^{\rm q}_\mu$ has the form
\begin{eqnarray}
&&\langle B^{{\rm q}a}_\mu(x) B^{{\rm q}b}_\nu(y) \rangle_{B^{\rm bg}} = (x|\frac{-id_{\mu\nu}(\hat{p})\delta^{ab}}{\hat{p}^2} |y)
\nonumber\\
&&+ (- i g) (x| \frac{-id_{\mu\rho}(\hat{p})}{\hat{p}^2} \Big[ g^{\rho\sigma} \{ \hat{p}_\alpha,  B^{{\rm bg}\alpha} \} + 2 i (\partial^\rho B^{{\rm bg}\sigma} - \partial^\sigma B^{{\rm bg}\rho}) - \hat{p}^\rho B^{{\rm bg}\sigma} - B^{{\rm bg}\rho} \hat{p}^\sigma \Big] \frac{-id_{\sigma\nu}(\hat{p})}{\hat{p}^2} |y)^{ab} + O(g^2) \,,\label{prax}
\end{eqnarray}
where the expression in squared brackets is the standard three-gluon vertex. In the axial gauge the numerator of each free quantum propagator is
\begin{eqnarray}
&&d_{\mu\nu}(\hat{p}) = g_{\mu\nu} - \frac{\bar{n}_\mu \hat{p}_\nu + \hat{p}_\mu \bar{n}_\nu}{\bar{n}\cdot \hat{p}} \,,
\end{eqnarray}
where we choose a light-cone vector $\bar{n}^+ = 1$. 

In Eq. (\ref{prax}) we use the Schwinger’s notation, see Ref. \cite{Schwinger:1951nm}. We consider a coherent state $|x)$ which is an  eigenvector of the position operators. For an arbitrary function of the momentum operator, we have
\begin{eqnarray}
&&(x|f(\hat{p})|y) = \int \frac{d^4p}{(2\pi)^4}e^{-ip(x-y)}f(p)\,.
\end{eqnarray}
For brevity, in the following discussion we neglect the ``hat" notation for the operators.

Taking into account that the background field satisfies $B^{\rm bg}_+ = 0$, we can rewrite the contraction in the following form,
\begin{eqnarray}
&&i\langle B^{{\rm q}a}_\mu(x) B^{{\rm q}b}_\nu(y) \rangle_{B^{\rm bg}} = (x| (g_{\mu l}-\frac{\bar{n}_\mu}{\bar{n}\cdot p}p_l)\frac{1}{p^2}(\delta^l_\nu-p^l\frac{\bar{n}_\nu}{\bar{n}\cdot p}) - \frac{\bar{n}_\mu \bar{n}_\nu}{(\bar{n}\cdot p)^2} |y)^{ab} + g (x| (-\frac{\bar{n}_\mu }{\bar{n}\cdot p} B^{\rm bg}_l) \frac{1}{p^2} (\delta^l_\nu-p^l\frac{\bar{n}_\nu}{\bar{n}\cdot p}) |y)^{ab}
\nonumber\\
&&+ g (x| (g_{\mu l}-\frac{\bar{n}_\mu}{\bar{n}\cdot p}p_l) \frac{1}{p^2} ( - B^{{\rm bg}l} \frac{ \bar{n}_\nu}{\bar{n}\cdot p} ) |y)^{ab} - g (x| (g_{\mu l}-\frac{\bar{n}_\mu}{\bar{n}\cdot p}p_l) \frac{1}{p^2} \{ p_\alpha, B^{{\rm bg}\alpha} \} \frac{1}{p^2} (\delta^l_\nu-p^l\frac{\bar{n}_\nu}{\bar{n}\cdot p}) |y)^{ab} 
\nonumber\\
&&- 2 i g (x| \frac{1}{p^2} (g_{\mu l}-\frac{\bar{n}_\mu}{\bar{n}\cdot p}p_l)   (\partial^l B^{{\rm bg}m} - \partial^m B^{{\rm bg}l}) ( g_{m\nu}-p_m\frac{\bar{n}_\nu}{\bar{n}\cdot p}) \frac{1}{p^2} |y)^{ab} + O(g^2) \,,
\label{prop:oneglue}
\end{eqnarray}
which we are going to use in our calculation.

It is instructive to note that this form of the contraction suggests the following general expression,
\begin{eqnarray}
&&i\langle B^{{\rm q}a}_\mu(x) B^{{\rm q}b}_\nu(y) \rangle_{B^{\rm bg}} = (x| (g_{\mu l}-\frac{\bar{n}_\mu}{\bar{n}\cdot p}P_l)\frac{1}{P^2}(\delta^l_\nu-P^l\frac{\bar{n}_\nu}{\bar{n}\cdot p}) - \frac{\bar{n}_\mu \bar{n}_\nu}{(\bar{n}\cdot p)^2} |y)^{ab}
\nonumber\\
&&- 2 i g (x| (g_{\mu l}-\frac{\bar{n}_\mu}{\bar{n}\cdot p}P_l) \frac{1}{P^2} F^{lm} \frac{1}{P^2} ( g_{m\nu}-P_m\frac{\bar{n}_\nu}{\bar{n}\cdot p}) |y)^{ab} + \dots \, ,
\end{eqnarray}
where $P_\mu = p_\mu + g B^{\rm bg}_\mu$ and an ellipsis stands for terms non-linear in a fully transverse strength tensor $F^{lm}$.

To calculate the emission vertex we need a contraction in the mixed representation
\begin{eqnarray}
&&i\lim_{k^2\to 0}k^2\langle B^{{\rm q}a}_\mu(k) B^{{\rm q} b}_\nu(y) \rangle_{B^{\rm bg}} = (g_{\mu l}-\frac{\bar{n}_\mu}{k^-}k_l)(\delta^l_\nu-k^l\frac{\bar{n}_\nu}{k^-}) e^{ik\cdot y}\delta^{ab}
\nonumber\\
&&- g (g_{\mu l}-\frac{\bar{n}_\mu}{k^-}k_l) (k| B^{{\rm bg}l} \frac{ \bar{n}_\nu}{p^-} |y)^{ab} - g (g_{\mu l}-\frac{\bar{n}_\mu}{k^-}k_l) (k| \{ p_\alpha, B^{{\rm bg}\alpha} \} \frac{1}{p^2} (\delta^l_\nu-p^l\frac{\bar{n}_\nu}{p^-}) |y)^{ab} 
\nonumber\\
&&- 2 i g (g_{\mu l}-\frac{\bar{n}_\mu}{k^-}k_l) (k| (\partial^l B^{{\rm bg}m} - \partial^m B^{{\rm bg}l}) ( g_{m\nu}-p_m\frac{\bar{n}_\nu}{p^-}) \frac{1}{p^2} |y)^{ab}\Big|_{k^+ = \frac{k^2_\perp}{2k^-}} + O(g^2) \,.
\end{eqnarray}

Taking poles in $1/p^2$ we can further rewrite this equation as
\begin{eqnarray}
&&i\lim_{k^2\to 0}k^2\langle B^{{\rm q}a}_\mu(k) B^{{\rm q}b}_\nu(y) \rangle_{B^{\rm bg}} = (\delta^l_\mu-\frac{\bar{n}_\mu}{k^-}k^l) \Big((g_{l\nu}-k_l\frac{\bar{n}_\nu}{k^-}) \delta^{ab} - g B^{{\rm bg}ab}_l(y^-, y_\perp) \frac{ \bar{n}_\nu}{k^-} \Big) e^{ik\cdot y} \Big|_{k^+ = \frac{k^2_\perp}{2k^-}}
\\
&& + \frac{ig}{2k^-} (\delta^l_\mu-\frac{\bar{n}_\mu}{k^-}k^l) (k_\perp| \Big(\int^\infty_{y^-} dz^- e^{i\frac{k^2_\perp}{2k^-}z^-} B^{{\rm bg}ab}_{\alpha}(z^-)  e^{-i\frac{p^2_\perp}{2k^-}z^-}\Big)e^{i\frac{p^2_\perp}{2k^-}y^-} (k+p)^\alpha (g_{l\nu}-p_l\frac{\bar{n}_\nu}{k^-}) |y_\perp) e^{ik^-  y^+} - \frac{g}{k^-}
\nonumber\\
&&\times (\delta^l_\mu-\frac{\bar{n}_\mu}{k^-}k^l) (k_\perp|  \Big( \int^\infty_{y^-} dz^- e^{i\frac{k^2_\perp}{2k^-}z^-}(\partial_l B^{ab}_m - \partial_m B^{ab}_l)^{\rm bg}(z^-) e^{-i\frac{p^2_\perp}{2k^-}z^-} \Big) e^{i\frac{p^2_\perp}{2k^-}y^-} ( \delta^m_\nu - p^m\frac{\bar{n}_\nu}{k^-}) |y_\perp) e^{ik^-  y^+} + O(g^2) \,.
\nonumber
\end{eqnarray}

Substituting this contraction into the emission vertex (\ref{ev:initial}) and performing integration over $y^-$ we obtain
\begin{eqnarray}
&&L^{ab}_{\mu j}(k,y_\perp,x_B)=  (\delta^l_\mu-\frac{\bar{n}_\mu}{k^-}k^l) (k_\perp| - i x_BP^+ \Big\{ g_{lj} 2\pi\delta(x_BP^+ + \frac{k^2_\perp}{2k^-}) \delta^{ab}
\nonumber\\
&&- g \Big( \int^\infty_{-\infty} dz^- e^{i (x_BP^+ + \frac{k^2_\perp}{2k^-})z^-} B^{{\rm bg}ab}_-(z^-) \Big) \Big( \frac{2 k^- g_{lj} }{ 2x_BP^+k^- + k^2_\perp} - \frac{2 k^- g_{lj} }{ 2x_BP^+k^- + p^2_\perp}\Big)
\nonumber\\
&& + g \Big(\int^\infty_{-\infty} dz^- e^{i (x_BP^+ + \frac{k^2_\perp}{2k^-})z^-} B^{{\rm bg}ab}_m(z^-)  \Big) \frac{(k+p)^m}{2k^-} \frac{2 k^- g_{lj} }{ 2x_BP^+k^- + p^2_\perp}
\nonumber\\
&&+ i g\Big( \int^\infty_{-\infty} dz^- e^{i (x_BP^+ + \frac{k^2_\perp}{2k^-})z^-}(\partial_l B^{ab}_m - \partial_m B^{ab}_l)^{\rm bg}(z^-) \Big) \frac{1}{k^-} \frac{2 k^- \delta^m_j }{ 2x_BP^+k^- + p^2_\perp} \Big\}
\nonumber\\
&&- \partial_j \Big\{ - \frac{ k_l }{k^-} 2\pi\delta(x_BP^+ + \frac{k^2_\perp}{2k^-}) \delta^{ab}
+ g \Big( \int^\infty_{-\infty} dz^- e^{i (x_BP^+ + \frac{k^2_\perp}{2k^-})z^-} B^{{\rm bg}ab}_-(z^-) \Big) \Big( \frac{2k_l }{ 2x_BP^+k^- + k^2_\perp} - \frac{2p_l }{ 2x_BP^+k^- + p^2_\perp } \Big)
\nonumber\\
&& + g \Big(\int^\infty_{-\infty} dz^- e^{i (x_BP^+ + \frac{k^2_\perp}{2k^-})z^-} B^{{\rm bg}ab}_m(z^-)  \Big) \frac{(k+p)^m}{2k^-} \frac{ - 2p_l }{ 2x_BP^+k^- + p^2_\perp}
 - g \Big( \int^{\infty}_{-\infty} dz^- e^{i(x_BP^+ + \frac{k^2_\perp}{2k^-})z^-} B^{{\rm bg}ab}_l(z^-) \Big) \frac{1}{k^-} 
\nonumber\\
&&+ i g \Big( \int^\infty_{-\infty} dz^- e^{i (x_BP^+ + \frac{k^2_\perp}{2k^-})z^-}(\partial_l B^{ab}_m - \partial_m B^{ab}_l)^{\rm bg}(z^-) \Big) \frac{1}{k^-} \frac{ - 2p^m  }{ 2x_BP^+k^- + p^2_\perp}\Big\}
 \nonumber\\
 &&- g \Big( \int^\infty_{-\infty} dz^- e^{i (x_BP^+ + \frac{k^2_\perp}{2k^-})z^-} (\partial_- B^{ab}_j - \partial_j B^{ab}_-)^{\rm bg}(z^-) \Big) \frac{2k_l}{ 2x_BP^+k^- + k^2_\perp }
\nonumber\\
 && + g \Big( \int^{\infty}_{-\infty} dz^- e^{i (x_BP^+ + \frac{k^2_\perp}{2k^-} )z^-} (\partial_- B^{ab}_j - \partial_j B^{ab}_- )^{\rm bg}(z^-) \Big) \frac{2k_l}{k^2_\perp} |y_\perp)  + O(g^2) \,.
\end{eqnarray}
For an arbitrary operator $\mathcal{O}$ we can commute the transverse momentum operator as $p_i \mathcal{O} =  \mathcal{O}p_i + i\partial_i \mathcal{O}$. Using this, we can further rewrite the vertex as
\begin{eqnarray}
&&L^{ab}_{\mu j}(k,y_\perp,x_B)=  (\delta^l_\mu-\frac{\bar{n}_\mu}{k^-}k^l) (k_\perp| - i x_BP^+ \Big\{ g_{lj} 2\pi\delta(x_BP^+ + \frac{k^2_\perp}{2k^-}) \delta^{ab}
\nonumber\\
&& -i g \Big(\int^\infty_{-\infty} dz^- e^{i (x_BP^+ + \frac{k^2_\perp}{2k^-})z^-} (\partial_- B^{ab}_m - \partial_m B^{ab}_-)^{\rm bg}(z^-)  \Big) \frac{(p + k)_m}{2x_BP^+k^- + k^2_\perp} \frac{2 k^- g_{lj} }{ 2x_BP^+k^- + p^2_\perp}
\nonumber\\
&&+ i g\Big( \int^\infty_{-\infty} dz^- e^{i (x_BP^+ + \frac{k^2_\perp}{2k^-})z^-}(\partial_l B^{ab}_m - \partial_m B^{ab}_l)^{\rm bg}(z^-) \Big) \frac{1}{k^-} \frac{2 k^- \delta^m_j }{ 2x_BP^+k^- + p^2_\perp} \Big\}
\nonumber\\
&&- \partial_j \Big\{ - \frac{ k_l }{k^-} 2\pi\delta(x_BP^+ + \frac{k^2_\perp}{2k^-}) \delta^{ab}
- i g \Big( \int^{\infty}_{-\infty} dz^- e^{i(x_BP^+ + \frac{k^2_\perp}{2k^-})z^-} (\partial_- B^{ab}_l - \partial_l B^{ab}_-)^{\rm bg}(z^-) \Big) \frac{2}{2x_BP^+k^- + p^2_\perp} 
\nonumber\\
&& + i g \Big(\int^\infty_{-\infty} dz^- e^{i (x_BP^+ + \frac{k^2_\perp}{2k^-})z^-} (\partial_-B^{ab}_m - \partial_mB^{ab}_-)^{\rm bg}(z^-)  \Big) \frac{(p+k)_m}{2x_BP^+k^- + k^2_\perp} \frac{ 2k_l }{ 2x_BP^+k^- + p^2_\perp}
\nonumber\\
&& - i g \Big(\int^\infty_{-\infty} dz^- e^{i (x_BP^+ + \frac{k^2_\perp}{2k^-})z^-} (\partial_l B^{ab}_m - \partial_mB^{ab}_l)^{\rm bg}(z^-)  \Big) \frac{(p+k)_m - 2p_m }{2k^-} \frac{ 2 }{ 2x_BP^+k^- + p^2_\perp} \Big\}
 \nonumber\\
 &&- g \Big( \int^\infty_{-\infty} dz^- e^{i (x_BP^+ + \frac{k^2_\perp}{2k^-})z^-} (\partial_- B^{ab}_j - \partial_j B^{ab}_-)^{\rm bg}(z^-) \Big) \frac{2k_l}{ 2x_BP^+k^- + k^2_\perp }
\nonumber\\
 && + g \Big( \int^{\infty}_{-\infty} dz^- e^{i (x_BP^+ + \frac{k^2_\perp}{2k^-} )z^-} (\partial_- B^{ab}_j - \partial_j B^{ab}_- )^{\rm bg}(z^-) \Big) \frac{2k_l}{k^2_\perp} |y_\perp)  + O(g^2) \,.
\end{eqnarray}

Following our strategy we replace $\partial_\mu B^{\rm bg}_\nu - \partial_\nu b^{\rm bg}_\mu \to F^{\rm bg}_{\mu\nu}\equiv F_{\mu\nu}$ and restore gauge covariance by inserting appropriate gauge factors.\footnote{This can be done explicitly by calculation in the next to leading order in the coupling constant.} As a result, after some reorganization of terms, we obtain the following expression for the emission vertex in the dilute limit,
\begin{eqnarray}
&&L^{ab}_{\mu j}(k,y_\perp,x_B)=  (\delta^l_\mu-\frac{\bar{n}_\mu}{k^-}k^l) (k_\perp| \Big(- i x_BP^+ g_{lj} + \frac{ i k_l k_j}{k^-} \Big) 2\pi\delta(x_BP^+ + \frac{k^2_\perp}{2k^-}) \delta^{ab} - g\Big( \int^\infty_{-\infty} dz^- 
\label{ev:semi-final}\\
&&\times e^{i (x_BP^+ + \frac{k^2_\perp}{2k^-})z^-} [\infty, z^-]^{ae}F^{eb}_{-m}(z^-) \Big) \Big( \frac{(p + k)_m}{2x_BP^+k^- + k^2_\perp} \frac{2 x_BP^+ k^- g_{lj} - 2k_l k_j}{ 2x_BP^+k^- + p^2_\perp} 
- \frac{2 g_{ml} p_j + 2g_{mj}k_l}{2x_BP^+k^- + p^2_\perp} + \frac{2g_{mj} k_l}{k^2_\perp} \Big)
\nonumber\\
&&- g\Big( \int^\infty_{-\infty} dz^- e^{i (x_BP^+ + \frac{k^2_\perp}{2k^-})z^-} [\infty, z^-]^{ae}F^{eb}_{sm}(z^-) \Big) \Big( \frac{2\delta^s_l}{2k^-} \frac{2 x_BP^+ k^- g_{mj} - 2 p_m p_j }{ 2x_BP^+k^- + p^2_\perp} + \frac{(p+k)_m }{2k^-} \frac{ 2 \delta^s_l p_j + 2\delta^s_j k_l}{ 2x_BP^+k^- + p^2_\perp}\Big) |y_\perp)\,.
\nonumber
\end{eqnarray}

Note that
\begin{eqnarray}
k^\mu L^{ab}_{\mu j}(k,y_\perp,x_B) = 0
\end{eqnarray}
as required by gauge invariance. It's also easy to see that in a product of two vertexes (\ref{ev:semi-final}), see Eq. (\ref{re:twoLprod}), only transverse indexes $\mu$ contribute. For this reason it's sufficient to consider only transverse components of the vertex $L^{an}_{k j}(k,y_\perp,x_B)$.

We find that the emission vertex (\ref{ev:semi-final}) is in agreement with a linearized version of the emission vertex constructed in Refs. \cite{Balitsky:2015qba,Balitsky:2016dgz}. This is up to the fully transverse strength tensor $F_{sm}$ which was not included in \cite{Balitsky:2015qba,Balitsky:2016dgz}. In Refs. \cite{Balitsky:2015qba,Balitsky:2016dgz} the emission vertex was constructed as an interpolating solution between the light-cone expansion technique and the shock-wave approximation. From our calculation it is evident that the vertex can be derived by an explicit calculation in a general background field.

 The emission vertex (\ref{ev:semi-final}) introduces a mixing between the initial operator, c.f. Eq. (\ref{def:evL}),
 \begin{eqnarray}
&&\int^\infty_{-\infty} dy^- e^{i x_BP^+y^-} [\infty, y^-]F_{-j}(y^-) \,,
 \end{eqnarray}
 and an operator constructed from the fully transverse strength tensor
 \begin{eqnarray}
&&\int^\infty_{-\infty} dz^- e^{i (x_BP^+ + \frac{k^2_\perp}{2k^-})z^-} [\infty, z^-]F_{sm}(z^-) \,.
 \end{eqnarray}
 Such an operator doesn't contribute to the DGLAP evolution since it's associated with higher-twist effects at large-x. It's also neglected in the small-x formalism, e.g. doesn't contribute to the BFKL evolution, since it represents highly suppressed corrections. However, this operator is of leading order for spin effects at small-x which are of sub-eikonal nature, see Refs. \cite{Kovchegov:2015pbl,Kovchegov:2016zex,Cougoulic:2022gbk}. Since in this paper we aim to build a bridge between leading order large-x and small-x effects, we leave the study of mixing with this operator for future publications.

 Finally, assuming that $x_B>0$ we can also neglect a delta-function in the first term of (\ref{ev:semi-final}). As a result, we obtain the following form of the vertex
\begin{eqnarray}
&&L^{ab}_{k j}(k,y_\perp,x_B) = - 2g (k_\perp| \Big( \int^\infty_{-\infty} dz^- e^{i (x_BP^+ + \frac{k^2_\perp}{2k^-})z^-} [\infty, z^-]^{ae}F^{eb}_{-m}(z^-) \Big)
\label{evL:final}\\
&&\times \Big( \frac{(p + k)_m}{2x_BP^+k^- + k^2_\perp} \frac{ x_BP^+ k^- g_{kj} - k_k k_j}{ 2x_BP^+k^- + p^2_\perp} - \frac{ g_{mk} p_j + g_{mj}k_k}{2x_BP^+k^- + p^2_\perp} + \frac{g_{mj} k_k}{k^2_\perp} \Big) |y_\perp) \,.
\nonumber
\end{eqnarray}

A similar calculation can be done for the conjugated vertex with a result
\begin{eqnarray}
&&\tilde{L}^{\ k ba}_i(k,x_\perp,x_B) = - 2 g (x_\perp| \Big( \frac{(p + k)_l}{2 x_B P^+ k^-+k^2_\perp} \frac{ x_B P^+ k^-\delta^k_i - k^k k_i}{2 x_B P^+ k^-+p^2_\perp} - \frac{\delta^k_l p_i + g_{li} k^k }{2 x_B P^+ k^- + p^2_\perp} + \frac{g_{il} k^k }{k^2_\perp} \Big)
\label{evL:final-conj}\\
&&\times \Big( \int^\infty_{-\infty} dz^- e^{-i (x_BP^+ + \frac{k^2_\perp}{2k^-})z^-} [\infty, z^-]^{be}\tilde{F}^{ea}_{-l}(z^-) \Big) |k_\perp) \,.
\nonumber
\end{eqnarray}

\section{Calculation of the virtual diagrams\label{Ap:vd}}
The calculation of the virtual emission diagrams is similar to the calculation of the real emission diagrams presented in the previous section. Let's first calculate the emission diagrams presented in Figs. \ref{fig:ved}a-d We start with the corresponding amplitude,
\begin{eqnarray}
&&\langle \int^\infty_{-\infty} dy^- e^{i x_BP^+y^-} [\infty, y^-]^{an}_yF^{{\rm q+bg};n}_{-j}(y^-, y_\perp)\rangle_{B^{\rm bg}} = \int^{\infty}_{-\infty} dy^- e^{ix_BP^+y^-} \langle [\infty,y^-]^{an}_y (\delta^{nd}\partial_-
\nonumber\\
&& - i gB^{{\rm bg}nd}_-(y^-,y_\perp) ) B^{{\rm q}d}_j(y^-,y_\perp)\rangle_{B^{\rm bg}} - \int^{\infty}_{-\infty} dy^- e^{ix_BP^+y^-} \langle [\infty,y^-]^{an}_y (\delta^{nd }\partial_j
 - igB^{{\rm bg}nd}_j(y^-,y_\perp)) B^{{\rm q}d}_-(y^-,y_\perp)\rangle_{B^{\rm bg}}
\nonumber\\
&&+\int^{\infty}_{-\infty} dy^- e^{ix_BP^+y^-} \langle [\infty,y^-]^{an}_{y} \rangle_{B^{\rm bg}} F^n_{- j}(y^-, y_\perp)
+ \int^{\infty}_{-\infty} dy^- e^{ix_BP^+y^-} [\infty,y^-]^{an}_{y} \langle F^n_{- j}(y^-, y_\perp)\rangle_{B^{\rm bg}} \,,
\end{eqnarray}
where we need to substitute propagator (\ref{prop:oneglue}), which, assuming $x^- > y^-$, we rewrite in a form
\begin{eqnarray}
&&i\langle B^{{\rm q}a}_\mu(x) B^{{\rm q}b}_\nu(y) \rangle_{B^{\rm bg}}\Big|_{x^- > y^-} = - \frac{i}{2\pi}\int^\infty_0 \frac{dp^-}{2p^-} e^{-ip^-(x-y)^+}(x_\perp| (g_{\mu l}-\frac{\bar{n}_\mu}{p^-}p_l)e^{-i\frac{p^2_\perp}{2p^-}(x-y)^-} (\delta^l_\nu-p^l\frac{\bar{n}_\nu}{p^-}) |y_\perp)\delta^{ab} 
\nonumber\\
&&- (x| \frac{\bar{n}_\mu \bar{n}_\nu}{(p^-)^2} |y)\delta^{ab} + \frac{i g}{2\pi}\int^\infty_0 \frac{dp^-}{2p^-} e^{-ip^-(x-y)^+}(x_\perp| \frac{\bar{n}_\mu }{p^-} B^{{\rm bg}ab}_l(x^-, x_\perp) e^{-i\frac{p^2_\perp}{2p^-}(x-y)^-} (\delta^l_\nu-p^l\frac{\bar{n}_\nu}{p^-}) |y_\perp)+ \frac{i g}{2\pi}
\nonumber\\
&&\times \int^\infty_0 \frac{dp^-}{2p^-}  e^{-ip^-(x-y)^+} (x_\perp| e^{-i\frac{p^2_\perp}{2p^-}(x-y)^-} (\delta^l_\mu - \frac{\bar{n}_\mu}{p^-}p^l) B^{{\rm bg}ab}_l(y^-, y_\perp) \frac{ \bar{n}_\nu}{p^-} |y_\perp) + \frac{g}{2\pi} \int^\infty_0 \frac{dp^-}{(2p^-)^2} e^{-ip^- (x-y)^+ } \int^{x^-}_{y^-} dz^-
\nonumber\\
&&\times (x_\perp| (g_{\mu l}-\frac{\bar{n}_\mu}{p^-}p_l) e^{-i\frac{p^2_\perp}{2p^-}(x^- - z^-)} \{p^\alpha, B^{{\rm bg}ab}_\alpha (z^-)\} e^{-i\frac{p^2_\perp}{2p^-} (z^- - y^-)} (\delta^l_\nu-p^l\frac{\bar{n}_\nu}{ p^-}) |y_\perp) + \frac{ i g}{\pi} \int \frac{dp^-}{(2p^-)^2} e^{-ip^- (x-y)^+}
\nonumber\\
&&\times \int^{x^-}_{y^-} dz^- (x_\perp| (\delta^l_\mu-\frac{\bar{n}_\mu}{p^-}p^l) e^{-i\frac{p^2_\perp}{2p^-}(x-z)^-} (\partial_l B^{ab}_m - \partial_m B^{ab}_l)^{\rm bg}(z^-) e^{-i \frac{p^2_\perp}{2p^-} (z-y)^-} ( \delta^m_{\nu}-p^m\frac{\bar{n}_\nu}{p^-}) |y_\perp) + O(g^2) \,.
\end{eqnarray}

Performing calculation in the single gluon approximation and neglecting the contribution of a fully transverse strength tensor we arrive at the following result,
\begin{eqnarray}
\label{Virtual}
&&\langle \int^\infty_{-\infty} dy^- e^{i x_BP^+y^-} [\infty, y^-]^{an}_yF^{{\rm q+bg};n}_{-j}(y^-, y_\perp)\rangle_{B^{\rm bg}} 
\\
&&= \frac{g^2N_c}{2\pi} \Big[ \int^\infty_0 \frac{dk^-}{k^-} (y_\perp| \frac{  2x_BP^+k^- }{ p^2_\perp(2x_BP^+k^- + p^2_\perp)} |y_\perp) \Big( \int^{\infty}_{-\infty} dz^- e^{i x_BP^+ z^-} (\partial_jB^a_- - \partial_-B^a_j)^{\rm bg} (z^-, y_\perp)\Big) - i\int^\infty_0 \frac{dk^-}{k^-}
\nonumber\\ 
 &&\times (y_\perp| \frac{p^s}{p^2_\perp} (2\delta^k_j \delta^m_s - g_{sj} g^{mk})\partial_k \Big( \int^{\infty}_{-\infty} dz^- e^{i x_BP^+ z^-} (\partial_m B^a_- - \partial_-B^a_m)^{\rm bg} (z^-)\Big)  \frac{ 1 }{ 2x_BP^+k^- + p^2_\perp} |y_\perp)\Big] + O(g^3) \,.
 \nonumber
\end{eqnarray}

In the dilute limit, this result can be generalized to
\begin{eqnarray}
&&\langle \int^\infty_{-\infty} dy^- e^{i x_BP^+y^-} [\infty, y^-]^{an}_yF^{{\rm q+bg};n}_{-j}(y^-, y_\perp)\rangle_{B^{\rm bg}} 
\label{virt-v1}\\
&&= - \frac{g^2N_c}{2\pi} \int^\infty_0 \frac{dk^-}{k^-} (y_\perp| \frac{  2x_BP^+k^- }{ p^2_\perp(2x_BP^+k^- + p^2_\perp)} |y_\perp) \Big( \int^{\infty}_{-\infty} dz^- e^{i x_BP^+ z^-} [\infty, z^-]^{an}_yF^n_{-j}(z^-, y_\perp)\Big)
\nonumber\\ 
 &&+ \frac{ig^2N_c}{2\pi}\int^\infty_0 \frac{dk^-}{k^-} (y_\perp| \frac{p^s}{p^2_\perp} (2\delta^k_j \delta^m_s - g_{sj} g^{mk})\partial_k \Big( \int^{\infty}_{-\infty} dz^- e^{i x_BP^+ z^-} [\infty, z^-]^{an}F^n_{-m}(z^-)\Big)  \frac{ 1 }{ 2x_BP^+k^- + p^2_\perp} |y_\perp) \,,
 \nonumber
\end{eqnarray}
which we can rewrite as,
\begin{eqnarray}
&&\langle \int^\infty_{-\infty} dy^- e^{i x_BP^+y^-} [\infty, y^-]^{an}_yF^{{\rm q+bg};n}_{-j}(y^-, y_\perp)\rangle_{B^{\rm bg}} = - \frac{ g^2N_c}{2\pi}  \int^\infty_0\frac{d k^-}{k^-} \int \dhd^2k_\perp \frac{ 2x_BP^+k^- }{k^2_\perp(2x_BP^+k^- + k^2_\perp)}
\nonumber\\
&&\times \int^{\infty}_{-\infty} dz^- e^{i x_BP^+ z^-} [\infty, z^-]^{an}_y F^n_{-j}(z^-, y_\perp) + \frac{g^2 N_c}{2\pi}  \int^\infty_0\frac{d k^-}{k^-} \int \dhd^2 p_\perp \int \dhd^2 k_\perp e^{i (p_\perp-k_\perp ) y_\perp } \frac{k^s}{k^2_\perp} (2 \delta^k_j \delta^m_s  - g_{js}g^{mk})
\nonumber\\
&&\times \frac{ (k - p)_k }{ 2x_BP^+k^- + p^2_\perp} 
 \int d^2z_\perp e^{i ( k_\perp - p_\perp ) z_\perp } \int^{\infty}_{-\infty} dz^- e^{i x_BP^+ z^-} [\infty, z^-]^{an}_z F^n_{-m}(z^-, z_\perp) \,.
\end{eqnarray}

A similar result can be obtained for the diagrams conjugated to the diagrams presented in Figs. \ref{fig:ved}a-d. Taking a sum of these virtual diagrams we obtain
\begin{eqnarray}
&&\int^\infty_{-\infty} dz^- e^{-i x_BP^+ z^-}   \langle p | \tilde{F}^m_{-i}(z^-, b_\perp) [z^-, \infty]^{ma}_b [\infty, 0^-]^{an}_0 F^n_{-j}(0^-, 0_\perp) |p\rangle^{\rm virt}
\nonumber\\
&&= -\frac{g^2N_c}{2\pi} \int^\infty_0 \frac{d k^-}{k^-}\Big\{ - \int \dhd^2p_\perp e^{ ip_\perp b_\perp} \int \dhd^2k_\perp e^{- ik_\perp b_\perp} \frac{k^s}{k^2_\perp} (2\delta^k_j\delta^m_s-g_{js}g^{mk}) \frac{(k-p)_k}{2x_B P^+ k^- + p^2_\perp} 
\nonumber\\
&&\times \int d^2z_\perp e^{i (k_\perp - p_\perp ) z_\perp } \int^\infty_{-\infty} dz^- e^{-i x_BP^+ z^-}   \langle p | \tilde{F}^m_{-i}(z^-, z_\perp) [z^-, \infty]^{ma}_z [\infty, 0^-]^{an}_0 F^n_{-m}(0^-, 0_\perp) |p\rangle
\nonumber\\
&&+ \int \dhd^2k_\perp \frac{2x_B P^+ k^-}{k^2_\perp(2x_B P^+ k^- + k^2_\perp)} \int^\infty_{-\infty} dz^- e^{-i x_BP^+ z^-}   \langle p | \tilde{F}^m_{-i}(z^-, b_\perp) [z^-, \infty]^{ma}_b [\infty, 0^-]^{an}_0 F^n_{-j}(0^-, 0_\perp) |p\rangle \Big\}
\nonumber\\
&&-\frac{g^2N_c}{2\pi} \int^\infty_0 \frac{dk^-}{k^-} \Big\{ \int \dhd^2p_\perp e^{ip_\perp b_\perp} \int \dhd^2k_\perp e^{-ik_\perp b_\perp} \frac{(p-k)_k}{2x_B P^+ k^- + p^2_\perp} (2\delta^k_i\delta^l_s - g_{is}g^{kl}) \frac{k^s}{k^2_\perp} 
\nonumber\\
&&\times \int d^2z_\perp e^{i (k_\perp-p_\perp ) z_\perp} \int^\infty_{-\infty} dz^- e^{-i x_BP^+ z^-}   \langle p | \tilde{F}^m_{-l}(z^-, z_\perp) [z^-, \infty]^{ma}_z [\infty, 0^-]^{an}_0 F^n_{-j}(0^-, 0_\perp) |p\rangle
\nonumber\\
&&+ \int \dhd^2k_\perp \frac{2x_B P^+ k^-}{k^2_\perp(2x_B P^+ k^- + k^2_\perp)}  \int^\infty_{-\infty} dz^- e^{-i x_BP^+ z^-}   \langle p | \tilde{F}^m_{-i}(z^-, b_\perp) [z^-, \infty]^{ma}_b [\infty, 0^-]^{an}_0 F^n_{-j}(0^-, 0_\perp) |p\rangle \Big\} \,.
\label{virt:kmin}
\end{eqnarray}

Introducing the variable $z$, see Eq.  (\ref{var:z}), we rewrite this equation in the following form
\begin{eqnarray}
&&\int^\infty_{-\infty} dz^- e^{-i x_BP^+ z^-}   \langle p | \tilde{F}^m_{-i}(z^-, b_\perp) [z^-, \infty]^{ma}_b [\infty, 0^-]^{an}_0 F^n_{-j}(0^-, 0_\perp) |p\rangle^{\rm virt} = \frac{g^2N_c}{2\pi} \int^1_0 \frac{dz}{z}\int \dhd^2p_\perp e^{ ip_\perp b_\perp}
\nonumber\\
&&\times \int \dhd^2k_\perp e^{- ik_\perp b_\perp} \Big[ \delta^l_i k^s (2\delta^k_j\delta^m_s-g_{js}g^{mk}) (k-p)_k + (k - p)_k (2\delta^k_i\delta^l_s - g_{is}g^{kl}) \delta^m_j k^s \Big] \frac{1}{k^2_\perp(z k^2_\perp + (1-z)p^2_\perp)}
\nonumber\\
&&\times   \int d^2z_\perp e^{i ( k_\perp - p_\perp) z_\perp} \int^\infty_{-\infty} dz^- e^{-i x_BP^+ z^-}   \langle p | \tilde{F}^m_{-l}(z^-, z_\perp) [z^-, \infty]^{ma}_z [\infty, 0^-]^{an}_0 F^n_{-m}(0^-, 0_\perp) |p\rangle
\nonumber\\
&& - \frac{g^2N_c}{\pi} \int^1_0 \frac{dz}{1 - z} \int \frac{ \dhd^2k_\perp }{k^2_\perp }  \int^\infty_{-\infty} dz^- e^{-i x_BP^+ z^-}   \langle p | \tilde{F}^m_{-i}(z^-, b_\perp) [z^-, \infty]^{ma}_b [\infty, 0^-]^{an}_0 F^n_{-j}(0^-, 0_\perp) |p\rangle \,.
\end{eqnarray}
The last line of this equation contains a scaleless integral over transverse momentum which is set to be zero in the dimensional regularization. However, as we discuss in the main text, the role of this contribution is non-trivial and we keep it for now. As a result, the equation reads
\begin{eqnarray}
&&\int^\infty_{-\infty} dz^- e^{-i x_BP^+ z^-}   \langle p | \tilde{F}^m_{-i}(z^-, b_\perp) [z^-, \infty]^{ma}_b [\infty, 0^-]^{an}_0 F^n_{-j}(0^-, 0_\perp) |p\rangle^{\rm virt} = \frac{g^2N_c}{2\pi} \int^1_0 \frac{dz}{z}\int \dhd^2p_\perp e^{ ip_\perp b_\perp}
\nonumber\\
&&\times \int \dhd^2k_\perp e^{- ik_\perp b_\perp} \Big[ \delta^l_i k^s (2\delta^k_j\delta^m_s-g_{js}g^{mk}) (k-p)_k + (k - p)_k (2\delta^k_i\delta^l_s - g_{is}g^{kl}) \delta^m_j k^s \Big] \frac{1}{k^2_\perp(z k^2_\perp + (1-z)p^2_\perp)}
\nonumber\\
&&\times   \int d^2z_\perp e^{i ( k_\perp - p_\perp) z_\perp} \int^\infty_{-\infty} dz^- e^{-i x_BP^+ z^-}   \langle p | \tilde{F}^m_{-l}(z^-, z_\perp) [z^-, \infty]^{ma}_z [\infty, 0^-]^{an}_0 F^n_{-m}(0^-, 0_\perp) |p\rangle
\nonumber\\
&&- \frac{g^2N_c}{\pi} \int^1_0 \frac{dz}{1 - z} \int \frac{ \dhd^2k_\perp }{k^2_\perp }  \int^\infty_{-\infty} dz^- e^{-i x_BP^+ z^-}   \langle p | \tilde{F}^m_{-i}(z^-, b_\perp) [z^-, \infty]^{ma}_b [\infty, 0^-]^{an}_0 F^n_{-j}(0^-, 0_\perp) |p\rangle \,.
\label{virt:z}
\end{eqnarray}

\section{Calculation using background-Feynman gauge\label{App:FG}}
In this section, we shall repeat the above calculation with the choice of ``quantum" field gauge to be,
\begin{eqnarray}
&&D^{\mu}B^{\rm q}_{\mu}(x^-,x_{\perp})=0\,, 
\end{eqnarray}
where the covariant derivative is constructed from the background field $B^{\rm bg}_\mu$. This gauge is called the background-Feynman gauge. For this gauge choice, the propagator takes the form,
\begin{eqnarray}
&&i\langle B^{{\rm q} a}_{\mu}(x) B^{{\rm q} b}_{\nu}(y) \rangle_{B^{\rm bg}}=(x|\frac{1}{ P^2 g^{\mu\nu}+ 2i F^{\mu\nu}+i\epsilon}|y)^{ab}\,,
\end{eqnarray}
where $P_\mu\:=p_\mu + g B^{\rm bg}_\mu$. To simplify the calculation, in this section we'll also assume that the background field has only $B^{\rm bg}_-(x^-,x_{\perp})$ non-zero component.

Up to two gluon accuracy, the propagator can be written in the following form,
\begin{eqnarray}
&&i\langle B_{\mu}^{{\rm q}a}(x)B_{\nu}^{{\rm q} b}(y)\rangle_{B^{\rm bg}}\Big|_{x^- > y^-}
\label{feyn:propagator}\\
&&=-\frac{1}{2\pi}\int_0^{\infty} \frac{dp^-}{2p^-}e^{-ip^- (x-y)^+}(x_{\perp}|e^{-i \frac{p_{\perp}^2 }{ 2p^-}x^-}\Big(i g_{\mu\nu}-gg_{\mu\nu}\int_{y^-}^{x^-} dz^- e^{i\frac{p_{\perp}^2 }{ 2p^-} z^-} B^{\rm bg}_-(z^-)e^{-i\frac{p_{\perp}^2 }{ 2p^-} z^-}
\nonumber\\
&&- \frac{ig}{p^-}\int_{y^-}^{x^-} dz^- e^{i\frac{p_{\perp}^2 }{ 2p^-} z^-} F_{\mu\nu}(z^-) e^{-i\frac{p_{\perp}^2 }{ 2p^-} z^-} + \frac{gg_{\mu+}g_{\nu+} }{ 2 (p^-)^2}\int_{y^-}^{x^-} dz^- e^{i \frac{p_{\perp}^2 }{ 2 p^-}z^-} \partial^k F_{-k}e^{-i \frac{p_{\perp}^2 }{ 2 p^-}z^-}\Big) e^{i\frac{p_{\perp}^2 }{ 2p^-}y^-}|y_{\perp})^{ab}\,,
\nonumber
\end{eqnarray}
where the third term with $g_{\mu +}\:g_{\nu +}$ is the contribution from the quark background fields $\partial^k F^a_{-k} = g\bar{\psi}\gamma^+ t^a \psi$. The propagator which shall be used for real emissions takes the form,
\begin{eqnarray}
\label{feyn:propagator momentum}
&&i\lim_{k^2\rightarrow 0} k^2\langle B_{\mu}^{{\rm q}a}(k) B_{\nu}^{{\rm q} b}(y)\rangle = -ie^{ik^- y^+}(k_{\perp}|\Big(i g_{\mu\nu}-gg_{\mu\nu}\int_{y^-}^{\infty} dz^- e^{i\frac{p_{\perp}^2}{2k^-} z^-}B^{\rm bg}_-(z^-)\:e^{-i\frac{p_{\perp}^2}{2k^-} z^-}
\\
&&- \frac{ig}{k^-}\int_{y^-}^{\infty} dz^- e^{i\frac{p_{\perp}^2}{2k^-} z^-} F_{\mu\nu}(z^-) e^{-i\frac{p_{\perp}^2}{ 2k^-} z^-} + \frac{gg_{\mu+}g_{\nu+}}{ 2 (k^-)^2} \int_{y^-}^{\infty} dz^- e^{i \frac{p_{\perp}^2 }{ 2 k^-}z^-} \partial^k F_{-k}e^{-i \frac{p_{\perp}^2 }{ 2 k^-}z^-}\Big) e^{i\frac{p_{\perp}^2 }{ 2k^-}y^-}|y_{\perp})^{ab} \,.
\nonumber
\end{eqnarray}

First, we will compute the real emission vertex, see Fig. \ref{fig:evL}, using the above gauge and match it with (\ref{evL:final}). Apart for the gauge condition (\ref{gauge1}), we will also fix $B^{\rm bg}_i=0$, which means that in this calculation we only track the $\partial_i B^{\rm bg}_-$ component of the $F_{-i}$ strength tensor, which is sufficient to obtain the final result (\ref{evL:final}).

Writing (\ref{ev:initial}) slightly differently, we have
\begin{eqnarray}
&& L_{\mu j}^{ab}(k,y_{\perp},x_B)=\int_{-\infty}^{\infty}dy^- e^{ix_B P^+ y^-}\Big(-i x_B P^+ [\infty,y^-]^{bd}i\lim_{k^2\rightarrow 0}k^2\langle B_{\mu}^{{\rm q}a}(k)B_j^{{\rm q}d}(y^-,y_{\perp})\rangle_{B^{\rm bg}}
\label{feyn:initial}
\\
&&- [\infty,y^-]^{bd} i\lim_{k^2\rightarrow 0}k^2\langle B^{{\rm q} a}_{\mu}(k) \partial_j B^{{\rm q} d}_-(y^-,y_{\perp})\rangle_{B^{\rm bg}} 
\nonumber\\
&&+ ig \int_{y^-}^{\infty} dz^- \left([\infty,z^-] T^d [z^-,y^-]\right)^{bm} F^m_{-j}(y^-,y_{\perp}) i\lim_{k^2\rightarrow 0}k^2\langle B^{{\rm q}a}_{\mu}(k)B_-^{{\rm q}d}(z^-,y_{\perp})\rangle \Big)\,,
\nonumber
\end{eqnarray}
where the last propagator is free, without any background fields.

Substituting (\ref{feyn:propagator momentum}) into the expression (\ref{feyn:initial}), we have
\begin{eqnarray}
&& L_{\mu j}^{ab}(k,y_{\perp},x_B)= -i x_B P^+ \int_{-\infty}^{\infty}dy^- e^{i x_B P^+ y^-}(k_{\perp}|\Big( g_{\mu j} + i g g_{\mu j} \int_{y^-}^{\infty} dz^- e^{i \frac{p_{\perp}^2}{ 2 k^-} z^-} B^{\rm bg}_{-}(z^-) e^{-i\frac{p_{\perp}^2 }{ 2k^-}z^-}
\\
&&- \frac{ g }{ k^-} \int_{y^-}^{\infty} dz^- e^{i \frac{p_{\perp}^2 }{ 2 k^-}z^-} F_{\mu j}(z^-) e^{-i \frac{p_{\perp}^2 }{ 2k^-}z^-}\Big) e^{i \frac{p_{\perp}^2 }{ 2k^-}y^-} |y_{\perp})^{ab} 
+ g g_{\mu j} x_B P^+ \int_{-\infty}^{\infty}dy^- e^{i x_B P^+ y^-}
\nonumber\\
&&\times\int_{y^-}^{\infty} dz^- B_{-}^{{\rm bg}ba}(z^-) e^{i \frac{k_{\perp}^2 }{ 2k^-}y^-} e^{-i k_{\perp}y_{\perp}} 
- \int_{-\infty}^{\infty}dy^- e^{i x_B P^+ y^-}(k_{\perp}|\Big(g_{\mu-}+i g g_{\mu-}\int_{y^-}^{\infty} dz^- e^{i \frac{p_{\perp}^2 }{ 2k^-}z^-} B^{\rm bg}_-(z^-)e^{-i \frac{p_{\perp}^2 }{ 2k^-}z^-} 
\nonumber\\
&&- \frac{g }{ k^-}\int_{y^-}^{\infty} dz^- e^{i \frac{p_{\perp}^2 }{ 2 k^-}z^-} F_{\mu-}(z^-)e^{-i \frac{p_{\perp}^2 }{ 2 k^-}z^-}
- \frac{i g g_{\mu+}}{2 (k^-)^2} \int_{-\infty}^{\infty}dy^- e^{i x_B P^+ y^-}\int_{y^-}^{\infty} dz^- e^{i \frac{p_{\perp}^2 }{ 2 k^-}z^-} \partial^k F_{-k}(z^-) e^{-i \frac{p_{\perp}^2 }{ 2k^-}z^-}\Big)
\nonumber\\
&&\times e^{i \frac{p_{\perp}^2 }{ 2k^-}y^-}(ip_j) |y_{\perp})^{ab} - ig g_{\mu-} \int_{-\infty}^{\infty}dy^- e^{i x_B P^+ y^-} \int_{y^-}^{\infty} dz^- B^{{\rm bg}ab}_-(z^-) e^{i \frac{k_{\perp}^2 }{ 2 k^-}y^-} e^{-i k_{\perp}y_{\perp}}  (ik_j)
\nonumber\\
&&+ ig g_{\mu -}\int_{-\infty}^{\infty}dy^- e^{i x_B P^+ y^-} \int_{y^-}^{\infty} dz^- F_{-j}^{ab}(y^-,y_{\perp}) e^{i \frac{k_{\perp}^2 }{ 2 k^-} z^-} e^{-i k_{\perp}y_{\perp}}\,.
\nonumber
\end{eqnarray}
Carrying out the integral over $y^-$ and simplifying, we have
\begin{eqnarray}
&& L_{\mu j}^{ab}(k,y_{\perp},x_B) = -ig g_{\mu j} \int_{-\infty}^{\infty} dz^- e^{i(x_BP^+ + \frac{k_{\perp}^2 }{ 2k^-})z^-}(k_{\perp}|B^{\rm bg}_-(z^-)\frac{2 x_B P^+ k^-}{ 2x_B P^+k^- + p_{\perp}^2}|y_{\perp})^{ab} + \frac{g }{ k^-}
\\
&&\times \int_{-\infty}^{\infty} dz^- e^{i(x_BP^+ + \frac{k_{\perp}^2 }{ 2k^-})z^-} (k_{\perp}|F_{\mu j}(z^-) \frac{2 x_B P^+ k^-}{ 2 x_B P^+ k^- + p_{\perp}^2}|y_{\perp})^{ab}
-ig g_{\mu j} \int_{-\infty}^{\infty} dz^- e^{i(x_BP^+ + \frac{k_{\perp}^2 }{ 2k^-})z^-} B_-^{{\rm bg}ab}(z^-)
\nonumber\\
&&\times \frac{2x_B P^+ k^-}{ 2 x_B P^+ k^-+ k_{\perp}^2} e^{-ik_{\perp}y_{\perp}} -ig g_{\mu -} \int_{-\infty}^{\infty} dz^- e^{i(x_B P^+ + \frac{k_{\perp}^2 }{ 2k^-})z^-}(k_{\perp}|B^{\rm bg}_-(z^-) \frac{2k^- p_j }{ 2  x_B P^+ k^- + p_{\perp}^2}|y_{\perp})^{ab}
\nonumber\\
&&+\frac{g }{ k^-} \int_{-\infty}^{\infty} dz^- e^{i(x_B P^+ + \frac{k_{\perp}^2 }{ 2k^-})z^-}(k_{\perp}|F_{\mu-}(z^-) \frac{2p_j k^- }{ 2x_B P^+ k^-+ p_{\perp}^2}|y_{\perp})^{ab}
+ \frac{ig g_{\mu+} }{ 2 (k^-)^2} \int_{-\infty}^{\infty} dz^- e^{i(x_B P^+ + \frac{k_{\perp}^2 }{ 2k^-})z^-}
\nonumber\\
&&\times(k_{\perp}|\partial_k F_{-k}(z^-)\frac{2p_jk^- }{ 2 x_B P^+ k^-+ p_{\perp}^2}|y_{\perp})^{ab} +ig g_{\mu-} \int_{-\infty}^{\infty} dz^- e^{i(x_B P^+ + \frac{k_{\perp}^2 }{ 2k^-})z^-} B_{-}^{{\rm bg}ab}(z^-)\frac{2 k^- k_j }{ 2 x_B P^+ k^- + k_{\perp}^2} e^{-i k_{\perp}y_{\perp}}
\nonumber\\
&& - \frac{2g g_{\mu-}k^- }{ k_{\perp}^2}\int_{-\infty}^{\infty} dz^- e^{i(x_B P^+ + \frac{k_{\perp}^2 }{ 2k^-})z^-} F_{-j}^{ab}(z^-)e^{-i k_{\perp}y_{\perp}}\,.
\nonumber
\end{eqnarray}

Using $p_i \mathcal{O}\:=\:i\partial_i \mathcal{O}+\mathcal{O}p_i$ to simplify further,
\begin{eqnarray}
 &&L_{\mu j}^{ab}(k,y_{\perp},x_B) = g \int_{-\infty}^{\infty} dz^- e^{i(x_B P^+ + \frac{k_{\perp}^2 }{ 2k^-})z^-}(k_{\perp}|\frac{ 2 x_B P^+k^- g_{\mu j} + 2g_{\mu -}k^- k_j }{ 2 x_B P^+ k^- + k_{\perp}^2}\partial_k B^{\rm bg}_-(z^-)\frac{ p_k + k_k }{ 2x_B P^+k^- + p_{\perp}^2}|y_{\perp})^{ab} 
\nonumber\\
 &&+ \frac{g }{ k^-} \int_{-\infty}^{\infty} dz^- e^{i(x_BP^+ + \frac{k_{\perp}^2 }{ 2k^-})z^-}(k_{\perp}|F_{\mu j}(z^-)\frac{2 x_B P^+ k^- }{ 2 x_B P^+ k^-+ p_{\perp}^2}|y_{\perp})^{ab} -g g_{\mu-} (k_{\perp}|\int_{-\infty}^{\infty} dz^- e^{i(x_BP^+ + \frac{k_{\perp}^2 }{ 2k^-})z^-}
\nonumber\\
 &&\times \partial_j B^{\rm bg}_{-}(z^-)\frac{2k^- }{ 2x_B P^+k^- + p_{\perp}^2}|y_{\perp})^{ab} + \frac{gg_{\mu+} }{ 2 (k^-)^2} \int_{-\infty}^{\infty} dz^- e^{i(x_B P^+ + \frac{k_{\perp}^2 }{ 2k^-})z^-} (k_{\perp}|(k^k F_{-k}(z^-) - F_{-k}(z^-)p^k)
 \nonumber\\
 &&\times \frac{2p_j k^- }{ 2 x_B P^+ k^- + p_{\perp}^2}|y_{\perp})^{ab}
+ \frac{g }{ k^-} \int_{-\infty}^{\infty} dz^- e^{i(x_B P^+ + \frac{k_{\perp}^2 }{ 2k^-})z^-}(k_{\perp}|F_{\mu-}(z^-) \frac{2p_jk^- }{ 2x_B P^+ k^- + p_{\perp}^2}|y_{\perp})^{ab}
 \nonumber\\
 &&- \frac{2g g_{\mu-}k^- }{ k_{\perp}^2}(k_{\perp}|\int_{-\infty}^{\infty} dz^- e^{i(x_B P^+ + \frac{k_{\perp}^2 }{ 2k^-})z^-} F_{-j}(z^-)|y_{\perp})^{ab}\,.
\end{eqnarray}

Now, replacing $\partial_j B^{\rm bg}_{-} \to -F_{-j}$ and restoring the Wilson lines, we obtain
\begin{eqnarray}
&&L_{\mu j}^{ab}(k,y_{\perp},x_B)
\\
&&=-g (k_{\perp}|\frac{2 x_B P^+k^- g_{\mu j} + 2g_{\mu -}k^- k_j }{ 2x_B P^+k^- + k_{\perp}^2}\Big( \int^\infty_{-\infty} dz^- e^{i (x_BP^+ + \frac{k^2_\perp}{2k^-})z^-} [\infty, z^-]^{ae}F^{eb}_{-k}(z^-) \Big)\frac{ p_k + k_k }{ 2x_B P^+k^- + p_{\perp}^2}|y_{\perp})
\nonumber\\
&&+ \frac{gg_{\mu +}}{ k^-}(k_{\perp}|\Big( \int^\infty_{-\infty} dz^- e^{i (x_BP^+ + \frac{k^2_\perp}{2k^-})z^-} [\infty, z^-]^{ae}F^{eb}_{-j}(z^-) \Big)\frac{2k^- x_B P^+ }{ 2x_B P^+k^-  + p_{\perp}^2}|y_{\perp})
\nonumber\\
&&+g g_{\mu-}(k_{\perp}|\Big( \int^\infty_{-\infty} dz^- e^{i (x_BP^+ + \frac{k^2_\perp}{2k^-})z^-} [\infty, z^-]^{ae}F^{eb}_{-j}(z^-) \Big)\frac{2k^- }{ 2x_B P^+k^- + p_{\perp}^2}|y_{\perp})^{ab}
\nonumber\\
&&+ \frac{gg_{\mu+} }{ 2 (k^-)^2}(k_{\perp}|\Big( \int^\infty_{-\infty} dz^- e^{i (x_BP^+ + \frac{k^2_\perp}{2k^-})z^-} [\infty, z^-]^{ae}F^{eb}_{-k}(z^-) \Big)\frac{2p_jk^- (k^k - p^k) }{ 2k^- x_B P^+ + p_{\perp}^2}|y_{\perp})
\nonumber\\
&&- \frac{g }{ k^-}(k_{\perp}|\Big( \int^\infty_{-\infty} dz^- e^{i (x_BP^+ + \frac{k^2_\perp}{2k^-})z^-} [\infty, z^-]^{ae}F^{eb}_{-\mu}(z^-) \Big)\frac{2p_jk^- }{ 2k^-x_B P^+ + p_{\perp}^2}|y_{\perp})
\nonumber\\
&&-\frac{2gg_{\mu-}k^- }{ k_{\perp}^2}(k_{\perp}|\Big( \int^\infty_{-\infty} dz^- e^{i (x_BP^+ + \frac{k^2_\perp}{2k^-})z^-} [\infty, z^-]^{ae}F^{eb}_{-j}(z^-) \Big)|y_{\perp}) \,.
\nonumber
\end{eqnarray}
The above expression can be checked to be gauge covariant. As a result, since we aim to calculate a product (\ref{re:twoLprod}), we can make a replacement
\begin{eqnarray}
&&g_{\mu -}k^- \to g_{\mu -}k^- - k_{\mu}\:=\:-k^+g_{\mu+}-k^\perp_ \mu\,,
\end{eqnarray}
since one can add a term proportional to $k^{\mu}$ without any cost. We also notice that the terms proportional to $g_{\mu+}$ do not contribute to the square of two emission vertex (\ref{re:twoLprod}). Removing such terms, we get
\begin{eqnarray}
&&L_{k j}^{ab}(k,y_{\perp},x_B) = -2g(k_{\perp}|\Big( \int^\infty_{-\infty} dz^- e^{i (x_BP^+ + \frac{k^2_\perp}{2k^-})z^-} [\infty, z^-]^{ae}F^{eb}_{-m}(z^-) \Big)
\\
&&\times\Big(\frac{  x_B P^+ k^- g_{k j} - k_k k_j}{ 2 x_B P^+k^- + k_{\perp}^2}\frac{ p_m + k_m }{ 2x_B P^+k^- + p_{\perp}^2} - \frac{k_k g_{mj} + p_j g_{mk} }{ 2x_B P^+k^- + p_{\perp}^2} + \frac{k_k g_{mj} }{ k_{\perp}^2} \Big)|y_{\perp})
\nonumber
\end{eqnarray}
which is same as (\ref{evL:final}).\par
Now we calculate the virtual correction. To do that, we start with
\begin{eqnarray}
&&\langle \int^\infty_{-\infty} dy^- e^{i x_BP^+y^-} [\infty, y^-]^{an}_yF^{{\rm q+bg};n}_{-j}(y^-, y_\perp)\rangle_{B^{\rm bg}}
\nonumber\\
&&= gx_B P^+\int_{-\infty}^{\infty} dy^-e^{ix_B P^+ y^-} \int_{y^-}^{\infty} dz^- ([\infty,z^-]T^e[z^-,y^-])^{an} \langle B_{-}^{{\rm q}e}(z^-) B_j^{{\rm q}n}(y^-,y_{\perp})\rangle_{B^{\rm bg}}
\\
&&-ig \int_{-\infty}^{\infty} dy^- e^{ix_B P^+ y^-} \int_{y^-}^{\infty} dz^- ([\infty,z^-]T^e[z^-,y^-])^{an} \langle B_-^{{\rm q}e}(z^-)\partial_j B_-^{{\rm q}n}(y^-,y_{\perp}) \rangle_{B^{\rm bg}}\,.
\nonumber
\end{eqnarray}
Note, that a diagram in Fig. \ref{fig:ved}c doesn't contribute in the background-Feynman gauge. 

The explicit propagators (\ref{feyn:propagator}) which contribute are
\begin{eqnarray}
 &&i\langle B_-^{{\rm q}e}(z^-,y_{\perp}) B_j^{{\rm q}n}(y^-,y_{\perp}) \rangle_{B^{\rm bg}} 
 \\
 &&= \frac{ig}{ 2\pi} \int_0^{\infty} \frac{dp^- }{ 2 (p^-)^2} (y_{\perp}|e^{-i\frac{p_{\perp}^2 }{ 2p^-}z^-}
\Big(\int_{y^-}^{z^-} dx^-e^{i\frac{p_{\perp}^2 }{ 2p^-}x^-} F_{-j}(x^-) e^{-i\frac{p_{\perp}^2 }{ 2p^-}x^-}\Big) e^{i\frac{p_{\perp}^2 }{ 2p^-}y^-} |y_{\perp})^{en} \,,
\nonumber
\end{eqnarray}
 \begin{eqnarray}
 &&i \langle B_-^{{\rm q}e}(z^-,y_{\perp}) \partial_j B_-^{{\rm q}n}(y^-,y_{\perp}) \rangle_{B^{\rm bg}} 
 \\
 &&= -\frac{g }{ 2\pi} \int_0^{\infty} \frac{dp^- }{ 4 (p^-)^3} (y_{\perp}|e^{-i \frac{p_{\perp}^2 }{ 2p^-}z^-}
\Big(\int_{y^-}^{z^-} dx^-e^{i\frac{p_{\perp}^2 }{ 2p^-}x^-} \partial^kF_{-k}(x^-) e^{-i \frac{p_{\perp}^2 }{ 2p^-}x^-} \Big)e^{i \frac{p_{\perp}^2 }{ 2p^-}y^-} ip_j|y_{\perp})^{en}\,.
\nonumber
\end{eqnarray}

Integrating over $y^-$ and $z^-$, we get
\begin{eqnarray}
&&\langle \int^\infty_{-\infty} dy^- e^{i x_BP^+y^-} [\infty, y^-]^{an}_yF^{{\rm q+bg};n}_{-j}(y^-, y_\perp)\rangle_{B^{\rm bg}}
\\
&&= - \frac{g^2 N_c}{ 2\pi} \int_0^{\infty} \frac{dp^-}{ p^-} (y_{\perp}|\frac{1}{ p_{\perp}^2}\int_{-\infty}^{\infty} dx^- e^{i x_B P^+ x^-} F_{-j}^n(x^-) \frac{2x_B P^+p^- }{ 2x_B P^+ p^- + p_{\perp}^2}|y_{\perp})
\nonumber\\
&&- \frac{i g^2 N_c }{ 2\pi} \int_0^{\infty} \frac{dp^- }{ p^-} (y_{\perp}|\frac{1}{ p_{\perp}^2}\int_{-\infty}^{\infty} dx^- e^{ix_B P^+ x^-} \partial^k F^n_{-k}(x^-) \frac{p_j }{ 2x_B P^+p^- + p_{\perp}^2}|y_{\perp}) \,.
\end{eqnarray}
Again using the operator relation $p_i \mathcal{O}\:=\:i\partial_i \mathcal{O}+\mathcal{O}p_i$, and restoring the Wilson lines, we rewrite the equation in the following form
\begin{eqnarray}
&&\langle \int^\infty_{-\infty} dy^- e^{i x_BP^+y^-} [\infty, y^-]^{an}_yF^{{\rm q+bg};n}_{-j}(y^-, y_\perp)\rangle_{B^{\rm bg}}
\\
&&= - \frac{g^2 N_c }{ 2\pi} \int_0^{\infty} \frac{dp^-}{ p^-}  \Big( \int^\infty_{-\infty} dy^- e^{i x_BP^+y^-} [\infty, y^-]^{an}_yF^n_{-j}(y^-, y_\perp) \Big)(y_{\perp}|\frac{2x_B P^+ p^- }{ p_{\perp}^2 (2x_B P^+ p^- + p_{\perp}^2)}|y_{\perp})
\nonumber\\
&&+ \frac{ig^2 N_c }{ 2\pi} \int_0^{\infty} \frac{dp^-}{ p^-}(y_{\perp}|\frac{p^s(2 \delta^m_s \delta^k_j - g_{js} g^{km}) }{ p_{\perp}^2} \partial_k \Big( \int^\infty_{-\infty} dy^- e^{i x_BP^+y^-} [\infty, y^-]^{an}_yF^n_{-m}(y^-, y_\perp) \Big)\frac{1}{ 2x_B P^+ p^- + p_{\perp}^2}|y_{\perp})
\nonumber
\end{eqnarray}
which is same as (\ref{virt-v1}). Note that in this derivation we ignore the fully transverse strength tensor. 
\bibliographystyle{unsrt}
\bibliography{main}

\begin{thebibliography}{100}

\bibitem{Collins:2011zzd}
John Collins.
\newblock {\em {Foundations of perturbative QCD}}, volume~32.
\newblock Cambridge University Press, 11 2013.

\bibitem{Collins:1981uk}
John~C. Collins and Davison~E. Soper.
\newblock {Back-To-Back Jets in QCD}.
\newblock {\em Nucl. Phys. B}, 193:381, 1981.
\newblock [Erratum: Nucl.Phys.B 213, 545 (1983)].

\bibitem{Collins:1984kg}
John~C. Collins, Davison~E. Soper, and George~F. Sterman.
\newblock {Transverse Momentum Distribution in Drell-Yan Pair and W and Z Boson
  Production}.
\newblock {\em Nucl. Phys. B}, 250:199--224, 1985.

\bibitem{Collins:1987pm}
John~C. Collins and Davison~E. Soper.
\newblock {The Theorems of Perturbative QCD}.
\newblock {\em Ann. Rev. Nucl. Part. Sci.}, 37:383--409, 1987.

\bibitem{Collins:1989gx}
John~C. Collins, Davison~E. Soper, and George~F. Sterman.
\newblock {Factorization of Hard Processes in QCD}.
\newblock {\em Adv. Ser. Direct. High Energy Phys.}, 5:1--91, 1989.

\bibitem{Meng:1995yn}
Ruibin Meng, Fredrick~I. Olness, and Davison~E. Soper.
\newblock {Semiinclusive deeply inelastic scattering at small q(T)}.
\newblock {\em Phys. Rev. D}, 54:1919--1935, 1996.

\bibitem{Ji:2004wu}
Xiang-dong Ji, Jian-ping Ma, and Feng Yuan.
\newblock {QCD factorization for semi-inclusive deep-inelastic scattering at
  low transverse momentum}.
\newblock {\em Phys. Rev. D}, 71:034005, 2005.

\bibitem{Ji:2004xq}
Xiang-dong Ji, Jian-Ping Ma, and Feng Yuan.
\newblock {QCD factorization for spin-dependent cross sections in DIS and
  Drell-Yan processes at low transverse momentum}.
\newblock {\em Phys. Lett. B}, 597:299--308, 2004.

\bibitem{Boussarie:2023izj}
Renaud Boussarie et~al.
\newblock {TMD Handbook}.
\newblock 4 2023.

\bibitem{Echevarria:2014xaa}
Miguel~G. Echevarria, Ahmad Idilbi, Zhong-Bo Kang, and Ivan Vitev.
\newblock {QCD Evolution of the Sivers Asymmetry}.
\newblock {\em Phys. Rev. D}, 89:074013, 2014.

\bibitem{Kang:2015msa}
Zhong-Bo Kang, Alexei Prokudin, Peng Sun, and Feng Yuan.
\newblock {Extraction of Quark Transversity Distribution and Collins
  Fragmentation Functions with QCD Evolution}.
\newblock {\em Phys. Rev. D}, 93(1):014009, 2016.

\bibitem{Bacchetta:2017gcc}
Alessandro Bacchetta, Filippo Delcarro, Cristian Pisano, Marco Radici, and
  Andrea Signori.
\newblock {Extraction of partonic transverse momentum distributions from
  semi-inclusive deep-inelastic scattering, Drell-Yan and Z-boson production}.
\newblock {\em JHEP}, 06:081, 2017.
\newblock [Erratum: JHEP 06, 051 (2019)].

\bibitem{Bacchetta:2019sam}
Alessandro Bacchetta, Valerio Bertone, Chiara Bissolotti, Giuseppe Bozzi,
  Filippo Delcarro, Fulvio Piacenza, and Marco Radici.
\newblock {Transverse-momentum-dependent parton distributions up to N$^{3}$LL
  from Drell-Yan data}.
\newblock {\em JHEP}, 07:117, 2020.

\bibitem{Scimemi:2019cmh}
Ignazio Scimemi and Alexey Vladimirov.
\newblock {Non-perturbative structure of semi-inclusive deep-inelastic and
  Drell-Yan scattering at small transverse momentum}.
\newblock {\em JHEP}, 06:137, 2020.

\bibitem{Bertone:2019nxa}
Valerio Bertone, Ignazio Scimemi, and Alexey Vladimirov.
\newblock {Extraction of unpolarized quark transverse momentum dependent parton
  distributions from Drell-Yan/Z-boson production}.
\newblock {\em JHEP}, 06:028, 2019.

\bibitem{Bacchetta:2020gko}
Alessandro Bacchetta, Filippo Delcarro, Cristian Pisano, and Marco Radici.
\newblock {The 3-dimensional distribution of quarks in momentum space}.
\newblock {\em Phys. Lett. B}, 827:136961, 2022.

\bibitem{Echevarria:2020hpy}
Miguel~G. Echevarria, Zhong-Bo Kang, and John Terry.
\newblock {Global analysis of the Sivers functions at NLO+NNLL in QCD}.
\newblock {\em JHEP}, 01:126, 2021.

\bibitem{Kang:2020xgk}
Zhong-Bo Kang, Jared Reiten, Ding~Yu Shao, and John Terry.
\newblock {QCD evolution of the gluon Sivers function in heavy flavor dijet
  production at the Electron-Ion Collider}.
\newblock {\em JHEP}, 05:286, 2021.

\bibitem{Cammarota:2020qcw}
Justin Cammarota, Leonard Gamberg, Zhong-Bo Kang, Joshua~A. Miller, Daniel
  Pitonyak, Alexei Prokudin, Ted~C. Rogers, and Nobuo Sato.
\newblock {Origin of single transverse-spin asymmetries in high-energy
  collisions}.
\newblock {\em Phys. Rev. D}, 102(5):054002, 2020.

\bibitem{Bury:2021sue}
Marcin Bury, Alexei Prokudin, and Alexey Vladimirov.
\newblock {Extraction of the Sivers function from SIDIS, Drell-Yan, and
  $W^\pm/Z$ boson production data with TMD evolution}.
\newblock {\em JHEP}, 05:151, 2021.

\bibitem{Balitsky:2017flc}
I.~Balitsky and A.~Tarasov.
\newblock {Higher-twist corrections to gluon TMD factorization}.
\newblock {\em JHEP}, 07:095, 2017.

\bibitem{Balitsky:2017gis}
I.~Balitsky and A.~Tarasov.
\newblock {Power corrections to TMD factorization for Z-boson production}.
\newblock {\em JHEP}, 05:150, 2018.

\bibitem{Ebert:2018gsn}
Markus~A. Ebert, Ian Moult, Iain~W. Stewart, Frank~J. Tackmann, Gherardo Vita,
  and Hua~Xing Zhu.
\newblock {Subleading power rapidity divergences and power corrections for
  q$_{T}$}.
\newblock {\em JHEP}, 04:123, 2019.

\bibitem{Balitsky:2020jzt}
Ian Balitsky.
\newblock {Gauge-invariant TMD factorization for Drell-Yan hadronic tensor at
  small x}.
\newblock {\em JHEP}, 05:046, 2021.

\bibitem{Vladimirov:2023aot}
Alexey Vladimirov.
\newblock {Kinematic power corrections in TMD factorization theorem}.
\newblock 7 2023.

\bibitem{Collins:1981va}
John~C. Collins and Davison~E. Soper.
\newblock {Back-To-Back Jets: Fourier Transform from B to K-Transverse}.
\newblock {\em Nucl. Phys. B}, 197:446--476, 1982.

\bibitem{Polyakov:1980ca}
Alexander~M. Polyakov.
\newblock {Gauge Fields as Rings of Glue}.
\newblock {\em Nucl. Phys. B}, 164:171--188, 1980.

\bibitem{Korchemsky:1985xj}
G.~P. Korchemsky and A.~V. Radyushkin.
\newblock {Loop Space Formalism and Renormalization Group for the Infrared
  Asymptotics of {QCD}}.
\newblock {\em Phys. Lett. B}, 171:459--467, 1986.

\bibitem{Moch:2004pa}
S.~Moch, J.~A.~M. Vermaseren, and A.~Vogt.
\newblock {The Three loop splitting functions in QCD: The Nonsinglet case}.
\newblock {\em Nucl. Phys. B}, 688:101--134, 2004.

\bibitem{Baikov:2009bg}
P.~A. Baikov, K.~G. Chetyrkin, A.~V. Smirnov, V.~A. Smirnov, and
  M.~Steinhauser.
\newblock {Quark and gluon form factors to three loops}.
\newblock {\em Phys. Rev. Lett.}, 102:212002, 2009.

\bibitem{Li:2014afw}
Ye~Li, Andreas von Manteuffel, Robert~M. Schabinger, and Hua~Xing Zhu.
\newblock {Soft-virtual corrections to Higgs production at N$^3$LO}.
\newblock {\em Phys. Rev. D}, 91:036008, 2015.

\bibitem{Gehrmann:2010ue}
T.~Gehrmann, E.~W.~N. Glover, T.~Huber, N.~Ikizlerli, and C.~Studerus.
\newblock {Calculation of the quark and gluon form factors to three loops in
  QCD}.
\newblock {\em JHEP}, 06:094, 2010.

\bibitem{Lee:2010cga}
R.~N. Lee, A.~V. Smirnov, and V.~A. Smirnov.
\newblock {Analytic Results for Massless Three-Loop Form Factors}.
\newblock {\em JHEP}, 04:020, 2010.

\bibitem{Blumlein:2021enk}
J.~Bl\"umlein, P.~Marquard, C.~Schneider, and K.~Sch\"onwald.
\newblock {The three-loop unpolarized and polarized non-singlet anomalous
  dimensions from off shell operator matrix elements}.
\newblock {\em Nucl. Phys. B}, 971:115542, 2021.

\bibitem{Li:2016axz}
Ye~Li, Duff Neill, and Hua~Xing Zhu.
\newblock {An exponential regulator for rapidity divergences}.
\newblock {\em Nucl. Phys. B}, 960:115193, 2020.

\bibitem{Li:2016ctv}
Ye~Li and Hua~Xing Zhu.
\newblock {Bootstrapping Rapidity Anomalous Dimensions for Transverse-Momentum
  Resummation}.
\newblock {\em Phys. Rev. Lett.}, 118(2):022004, 2017.

\bibitem{Vladimirov:2016dll}
Alexey~A. Vladimirov.
\newblock {Correspondence between Soft and Rapidity Anomalous Dimensions}.
\newblock {\em Phys. Rev. Lett.}, 118(6):062001, 2017.

\bibitem{Scimemi:2018xaf}
Ignazio Scimemi and Alexey Vladimirov.
\newblock {Systematic analysis of double-scale evolution}.
\newblock {\em JHEP}, 08:003, 2018.

\bibitem{Collins:1981uw}
John~C. Collins and Davison~E. Soper.
\newblock {Parton Distribution and Decay Functions}.
\newblock {\em Nucl. Phys. B}, 194:445--492, 1982.

\bibitem{Kang:2012em}
Zhong-Bo Kang and Jian-Wei Qiu.
\newblock {QCD evolution of naive-time-reversal-odd parton distribution
  functions}.
\newblock {\em Phys. Lett. B}, 713:273--276, 2012.

\bibitem{Sun:2013hua}
Peng Sun and Feng Yuan.
\newblock {Transverse momentum dependent evolution: Matching semi-inclusive
  deep inelastic scattering processes to Drell-Yan and W/Z boson production}.
\newblock {\em Phys. Rev. D}, 88(11):114012, 2013.

\bibitem{Dai:2014ala}
Ling-Yun Dai, Zhong-Bo Kang, Alexei Prokudin, and Ivan Vitev.
\newblock {Next-to-leading order transverse momentum-weighted Sivers asymmetry
  in semi-inclusive deep inelastic scattering: the role of the three-gluon
  correlator}.
\newblock {\em Phys. Rev. D}, 92(11):114024, 2015.

\bibitem{Braun:2009mi}
V.~M. Braun, A.~N. Manashov, and B.~Pirnay.
\newblock {Scale dependence of twist-three contributions to single spin
  asymmetries}.
\newblock {\em Phys. Rev. D}, 80:114002, 2009.
\newblock [Erratum: Phys.Rev.D 86, 119902 (2012)].

\bibitem{Echevarria:2015byo}
Miguel~G. Echevarria, Ignazio Scimemi, and Alexey Vladimirov.
\newblock {Universal transverse momentum dependent soft function at NNLO}.
\newblock {\em Phys. Rev. D}, 93(5):054004, 2016.

\bibitem{Scimemi:2019gge}
Ignazio Scimemi, Andrey Tarasov, and Alexey Vladimirov.
\newblock {Collinear matching for Sivers function at next-to-leading order}.
\newblock {\em JHEP}, 05:125, 2019.

\bibitem{Catani:2011kr}
S.~Catani and M.~Grazzini.
\newblock {Higgs Boson Production at Hadron Colliders: Hard-Collinear
  Coefficients at the NNLO}.
\newblock {\em Eur. Phys. J. C}, 72:2013, 2012.
\newblock [Erratum: Eur.Phys.J.C 72, 2132 (2012)].

\bibitem{Catani:2012qa}
Stefano Catani, Leandro Cieri, Daniel de~Florian, Giancarlo Ferrera, and
  Massimiliano Grazzini.
\newblock {Vector boson production at hadron colliders: hard-collinear
  coefficients at the NNLO}.
\newblock {\em Eur. Phys. J. C}, 72:2195, 2012.

\bibitem{Gehrmann:2014yya}
Thomas Gehrmann, Thomas Luebbert, and Li~Lin Yang.
\newblock {Calculation of the transverse parton distribution functions at
  next-to-next-to-leading order}.
\newblock {\em JHEP}, 06:155, 2014.

\bibitem{Lubbert:2016rku}
Thomas L\"ubbert, Joel Oredsson, and Maximilian Stahlhofen.
\newblock {Rapidity renormalized TMD soft and beam functions at two loops}.
\newblock {\em JHEP}, 03:168, 2016.

\bibitem{Echevarria:2016scs}
Miguel~G. Echevarria, Ignazio Scimemi, and Alexey Vladimirov.
\newblock {Unpolarized Transverse Momentum Dependent Parton Distribution and
  Fragmentation Functions at next-to-next-to-leading order}.
\newblock {\em JHEP}, 09:004, 2016.

\bibitem{Dokshitzer:1977sg}
Yuri~L. Dokshitzer.
\newblock {Calculation of the Structure Functions for Deep Inelastic Scattering
  and e+ e- Annihilation by Perturbation Theory in Quantum Chromodynamics.}
\newblock {\em Sov. Phys. JETP}, 46:641--653, 1977.

\bibitem{Gribov:1972ri}
V.~N. Gribov and L.~N. Lipatov.
\newblock {Deep inelastic e p scattering in perturbation theory}.
\newblock {\em Sov. J. Nucl. Phys.}, 15:438--450, 1972.

\bibitem{Altarelli:1977zs}
Guido Altarelli and G.~Parisi.
\newblock {Asymptotic Freedom in Parton Language}.
\newblock {\em Nucl. Phys. B}, 126:298--318, 1977.

\bibitem{Scimemi:2017etj}
Ignazio Scimemi and Alexey Vladimirov.
\newblock {Analysis of vector boson production within TMD factorization}.
\newblock {\em Eur. Phys. J. C}, 78(2):89, 2018.

\bibitem{Landry:2002ix}
F.~Landry, R.~Brock, Pavel~M. Nadolsky, and C.~P. Yuan.
\newblock {Tevatron Run-1 $Z$ boson data and Collins-Soper-Sterman resummation
  formalism}.
\newblock {\em Phys. Rev. D}, 67:073016, 2003.

\bibitem{Konychev:2005iy}
Anton~V. Konychev and Pavel~M. Nadolsky.
\newblock {Universality of the Collins-Soper-Sterman nonperturbative function
  in gauge boson production}.
\newblock {\em Phys. Lett. B}, 633:710--714, 2006.

\bibitem{Becher:2011xn}
Thomas Becher, Matthias Neubert, and Daniel Wilhelm.
\newblock {Electroweak Gauge-Boson Production at Small $q_T$: Infrared Safety
  from the Collinear Anomaly}.
\newblock {\em JHEP}, 02:124, 2012.

\bibitem{DAlesio:2014mrz}
Umberto D'Alesio, Miguel~G. Echevarria, Stefano Melis, and Ignazio Scimemi.
\newblock {Non-perturbative QCD effects in $q_{T}$ spectra of Drell-Yan and
  Z-boson production}.
\newblock {\em JHEP}, 11:098, 2014.

\bibitem{Dominguez:2010xd}
Fabio Dominguez, Bo-Wen Xiao, and Feng Yuan.
\newblock {$k_t$-factorization for Hard Processes in Nuclei}.
\newblock {\em Phys. Rev. Lett.}, 106:022301, 2011.

\bibitem{Dominguez:2011wm}
Fabio Dominguez, Cyrille Marquet, Bo-Wen Xiao, and Feng Yuan.
\newblock {Universality of Unintegrated Gluon Distributions at small x}.
\newblock {\em Phys. Rev. D}, 83:105005, 2011.

\bibitem{Xiao:2017yya}
Bo-Wen Xiao, Feng Yuan, and Jian Zhou.
\newblock {Transverse Momentum Dependent Parton Distributions at Small-x}.
\newblock {\em Nucl. Phys. B}, 921:104--126, 2017.

\bibitem{Altinoluk:2014oxa}
Tolga Altinoluk, N\'estor Armesto, Guillaume Beuf, Mauricio Mart\'\i{}nez, and
  Carlos~A. Salgado.
\newblock {Next-to-eikonal corrections in the CGC: gluon production and spin
  asymmetries in pA collisions}.
\newblock {\em JHEP}, 07:068, 2014.

\bibitem{Altinoluk:2015gia}
Tolga Altinoluk, N\'estor Armesto, Guillaume Beuf, and Alexis Moscoso.
\newblock {Next-to-next-to-eikonal corrections in the CGC}.
\newblock {\em JHEP}, 01:114, 2016.

\bibitem{Kovchegov:2015pbl}
Yuri~V. Kovchegov, Daniel Pitonyak, and Matthew~D. Sievert.
\newblock {Helicity Evolution at Small-x}.
\newblock {\em JHEP}, 01:072, 2016.
\newblock [Erratum: JHEP 10, 148 (2016)].

\bibitem{Altinoluk:2015xuy}
Tolga Altinoluk and Adrian Dumitru.
\newblock {Particle production in high-energy collisions beyond the shockwave
  limit}.
\newblock {\em Phys. Rev. D}, 94(7):074032, 2016.

\bibitem{Balitsky:2015qba}
I.~Balitsky and A.~Tarasov.
\newblock {Rapidity evolution of gluon TMD from low to moderate x}.
\newblock {\em JHEP}, 10:017, 2015.

\bibitem{Kovchegov:2016zex}
Yuri~V. Kovchegov, Daniel Pitonyak, and Matthew~D. Sievert.
\newblock {Helicity Evolution at Small $x$: Flavor Singlet and Non-Singlet
  Observables}.
\newblock {\em Phys. Rev. D}, 95(1):014033, 2017.

\bibitem{Balitsky:2016dgz}
I.~Balitsky and A.~Tarasov.
\newblock {Gluon TMD in particle production from low to moderate x}.
\newblock {\em JHEP}, 06:164, 2016.

\bibitem{Agostini:2019avp}
Pedro Agostini, Tolga Altinoluk, and N\'estor Armesto.
\newblock {Non-eikonal corrections to multi-particle production in the Color
  Glass Condensate}.
\newblock {\em Eur. Phys. J. C}, 79(7):600, 2019.

\bibitem{Agostini:2019hkj}
Pedro Agostini, Tolga Altinoluk, and N\'estor Armesto.
\newblock {Effect of non-eikonal corrections on azimuthal asymmetries in the
  Color Glass Condensate}.
\newblock {\em Eur. Phys. J. C}, 79(9):790, 2019.

\bibitem{Cougoulic:2019aja}
Florian Cougoulic and Yuri~V. Kovchegov.
\newblock {Helicity-dependent generalization of the JIMWLK evolution}.
\newblock {\em Phys. Rev. D}, 100(11):114020, 2019.

\bibitem{Altinoluk:2020oyd}
Tolga Altinoluk, Guillaume Beuf, Alina Czajka, and Arantxa Tymowska.
\newblock {Quarks at next-to-eikonal accuracy in the CGC: Forward quark-nucleus
  scattering}.
\newblock {\em Phys. Rev. D}, 104(1):014019, 2021.

\bibitem{Altinoluk:2021lvu}
Tolga Altinoluk and Guillaume Beuf.
\newblock {Quark and scalar propagators at next-to-eikonal accuracy in the CGC
  through a dynamical background gluon field}.
\newblock {\em Phys. Rev. D}, 105(7):074026, 2022.

\bibitem{Kovchegov:2021iyc}
Yuri~V. Kovchegov and M.~Gabriel Santiago.
\newblock {Quark sivers function at small $x$: spin-dependent odderon and the
  sub-eikonal evolution}.
\newblock {\em JHEP}, 11:200, 2021.
\newblock [Erratum: JHEP 09, 186 (2022)].

\bibitem{Chirilli:2021lif}
Giovanni~Antonio Chirilli.
\newblock {High-energy operator product expansion at sub-eikonal level}.
\newblock {\em JHEP}, 06:096, 2021.

\bibitem{Cougoulic:2022gbk}
Florian Cougoulic, Yuri~V. Kovchegov, Andrey Tarasov, and Yossathorn Tawabutr.
\newblock {Quark and gluon helicity evolution at small x: revised and updated}.
\newblock {\em JHEP}, 07:095, 2022.

\bibitem{Agostini:2022oge}
Pedro Agostini, Tolga Altinoluk, and N\'estor Armesto.
\newblock {Finite width effects on the azimuthal asymmetry in proton-nucleus
  collisions in the Color Glass Condensate}.
\newblock {\em Phys. Lett. B}, 840:137892, 2023.

\bibitem{Fadin:1975cb}
Victor~S. Fadin, E.~A. Kuraev, and L.~N. Lipatov.
\newblock {On the Pomeranchuk Singularity in Asymptotically Free Theories}.
\newblock {\em Phys. Lett. B}, 60:50--52, 1975.

\bibitem{Kuraev:1976ge}
E.~A. Kuraev, L.~N. Lipatov, and Victor~S. Fadin.
\newblock {Multi - Reggeon Processes in the Yang-Mills Theory}.
\newblock {\em Sov. Phys. JETP}, 44:443--450, 1976.

\bibitem{Kuraev:1977fs}
E.~A. Kuraev, L.~N. Lipatov, and Victor~S. Fadin.
\newblock {The Pomeranchuk Singularity in Nonabelian Gauge Theories}.
\newblock {\em Sov. Phys. JETP}, 45:199--204, 1977.

\bibitem{Balitsky:1978ic}
I.~I. Balitsky and L.~N. Lipatov.
\newblock {The Pomeranchuk Singularity in Quantum Chromodynamics}.
\newblock {\em Sov. J. Nucl. Phys.}, 28:822--829, 1978.

\bibitem{Balitsky:1995ub}
I.~Balitsky.
\newblock {Operator expansion for high-energy scattering}.
\newblock {\em Nucl. Phys. B}, 463:99--160, 1996.

\bibitem{Balitsky:1997mk}
Ian Balitsky.
\newblock {Operator expansion for diffractive high-energy scattering}.
\newblock {\em AIP Conf. Proc.}, 407(1):953, 1997.

\bibitem{Kovchegov:1999yj}
Yuri~V. Kovchegov.
\newblock {Small x F(2) structure function of a nucleus including multiple
  pomeron exchanges}.
\newblock {\em Phys. Rev. D}, 60:034008, 1999.

\bibitem{Jalilian-Marian:1997qno}
Jamal Jalilian-Marian, Alex Kovner, Andrei Leonidov, and Heribert Weigert.
\newblock {The BFKL equation from the Wilson renormalization group}.
\newblock {\em Nucl. Phys. B}, 504:415--431, 1997.

\bibitem{Jalilian-Marian:1997ubg}
Jamal Jalilian-Marian, Alex Kovner, and Heribert Weigert.
\newblock {The Wilson renormalization group for low x physics: Gluon evolution
  at finite parton density}.
\newblock {\em Phys. Rev. D}, 59:014015, 1998.

\bibitem{Kovner:2000pt}
Alex Kovner, J.~Guilherme Milhano, and Heribert Weigert.
\newblock {Relating different approaches to nonlinear QCD evolution at finite
  gluon density}.
\newblock {\em Phys. Rev. D}, 62:114005, 2000.

\bibitem{Iancu:2000hn}
Edmond Iancu, Andrei Leonidov, and Larry~D. McLerran.
\newblock {Nonlinear gluon evolution in the color glass condensate. 1.}
\newblock {\em Nucl. Phys. A}, 692:583--645, 2001.

\bibitem{Ferreiro:2001qy}
Elena Ferreiro, Edmond Iancu, Andrei Leonidov, and Larry McLerran.
\newblock {Nonlinear gluon evolution in the color glass condensate. 2.}
\newblock {\em Nucl. Phys. A}, 703:489--538, 2002.

\bibitem{Kovner:2013ona}
Alex Kovner, Michael Lublinsky, and Yair Mulian.
\newblock {Jalilian-Marian, Iancu, McLerran, Weigert, Leonidov, Kovner
  evolution at next to leading order}.
\newblock {\em Phys. Rev. D}, 89(6):061704, 2014.

\bibitem{Kovner:2014lca}
Alex Kovner, Michael Lublinsky, and Yair Mulian.
\newblock {NLO JIMWLK evolution unabridged}.
\newblock {\em JHEP}, 08:114, 2014.

\bibitem{Lublinsky:2016meo}
Michael Lublinsky and Yair Mulian.
\newblock {High Energy QCD at NLO: from light-cone wave function to JIMWLK
  evolution}.
\newblock {\em JHEP}, 05:097, 2017.

\bibitem{Abbott:1980hw}
L.~F. Abbott.
\newblock {The Background Field Method Beyond One Loop}.
\newblock {\em Nucl. Phys. B}, 185:189--203, 1981.

\bibitem{Abbott:1981ke}
L.~F. Abbott.
\newblock {Introduction to the Background Field Method}.
\newblock {\em Acta Phys. Polon. B}, 13:33, 1982.

\bibitem{Chiu:2011qc}
Jui-yu Chiu, Ambar Jain, Duff Neill, and Ira~Z. Rothstein.
\newblock {The Rapidity Renormalization Group}.
\newblock {\em Phys. Rev. Lett.}, 108:151601, 2012.

\bibitem{Chiu:2012ir}
Jui-Yu Chiu, Ambar Jain, Duff Neill, and Ira~Z. Rothstein.
\newblock {A Formalism for the Systematic Treatment of Rapidity Logarithms in
  Quantum Field Theory}.
\newblock {\em JHEP}, 05:084, 2012.

\bibitem{Fleming:2014rea}
Sean Fleming.
\newblock {The role of Glauber exchange in soft collinear effective theory and
  the Balitsky\textendash{}Fadin\textendash{}Kuraev\textendash{}Lipatov
  Equation}.
\newblock {\em Phys. Lett. B}, 735:266--271, 2014.

\bibitem{Rothstein:2016bsq}
Ira~Z. Rothstein and Iain~W. Stewart.
\newblock {An Effective Field Theory for Forward Scattering and Factorization
  Violation}.
\newblock {\em JHEP}, 08:025, 2016.

\bibitem{Kang:2019ysm}
Zhong-Bo Kang and Xiaohui Liu.
\newblock {Power Counting the Small-$x$ Observables}.
\newblock 10 2019.

\bibitem{Liu:2022ijp}
Hao-yu Liu, Kexin Xie, Zhongbo Kang, and Xiaohui Liu.
\newblock {Single inclusive jet production in pA collisions at NLO in the
  small-x regime}.
\newblock {\em JHEP}, 07:041, 2022.

\bibitem{Echevarria:2011epo}
Miguel~G. Echevarria, Ahmad Idilbi, and Ignazio Scimemi.
\newblock {Factorization Theorem For Drell-Yan At Low $q_T$ And Transverse
  Momentum Distributions On-The-Light-Cone}.
\newblock {\em JHEP}, 07:002, 2012.

\bibitem{Chiu:2009yx}
Jui-yu Chiu, Andreas Fuhrer, Andre~H. Hoang, Randall Kelley, and Aneesh~V.
  Manohar.
\newblock {Soft-Collinear Factorization and Zero-Bin Subtractions}.
\newblock {\em Phys. Rev. D}, 79:053007, 2009.

\bibitem{Beneke:2003pa}
M.~Beneke and T.~Feldmann.
\newblock {Factorization of heavy to light form-factors in soft collinear
  effective theory}.
\newblock {\em Nucl. Phys. B}, 685:249--296, 2004.

\bibitem{Chiu:2007yn}
Jui-yu Chiu, Frank Golf, Randall Kelley, and Aneesh~V. Manohar.
\newblock {Electroweak Sudakov corrections using effective field theory}.
\newblock {\em Phys. Rev. Lett.}, 100:021802, 2008.

\bibitem{Becher:2011dz}
Thomas Becher and Guido Bell.
\newblock {Analytic Regularization in Soft-Collinear Effective Theory}.
\newblock {\em Phys. Lett. B}, 713:41--46, 2012.

\bibitem{Becher:2010tm}
Thomas Becher and Matthias Neubert.
\newblock {Drell-Yan Production at Small $q_T$, Transverse Parton Distributions
  and the Collinear Anomaly}.
\newblock {\em Eur. Phys. J. C}, 71:1665, 2011.

\bibitem{Becher:2012yn}
Thomas Becher, Matthias Neubert, and Daniel Wilhelm.
\newblock {Higgs-Boson Production at Small Transverse Momentum}.
\newblock {\em JHEP}, 05:110, 2013.

\bibitem{Leibbrandt:1983pj}
George Leibbrandt.
\newblock {The Light Cone Gauge in Yang-Mills Theory}.
\newblock {\em Phys. Rev. D}, 29:1699, 1984.

\bibitem{Altinoluk:2023dww}
Tolga Altinoluk, Guillaume Beuf, and Jamal Jalilian-Marian.
\newblock {Renormalization of the gluon distribution function in the background
  field formalism}.
\newblock 5 2023.

\bibitem{Mueller:2012uf}
A.~H. Mueller, Bo-Wen Xiao, and Feng Yuan.
\newblock {Sudakov Resummation in Small-$x$ Saturation Formalism}.
\newblock {\em Phys. Rev. Lett.}, 110(8):082301, 2013.

\bibitem{Mueller:2013wwa}
A.~H. Mueller, Bo-Wen Xiao, and Feng Yuan.
\newblock {Sudakov double logarithms resummation in hard processes in the
  small-x saturation formalism}.
\newblock {\em Phys. Rev. D}, 88(11):114010, 2013.

\bibitem{Caucal:2021ent}
Paul Caucal, Farid Salazar, and Raju Venugopalan.
\newblock {Dijet impact factor in DIS at next-to-leading order in the Color
  Glass Condensate}.
\newblock {\em JHEP}, 11:222, 2021.

\bibitem{Caucal:2022ulg}
Paul Caucal, Farid Salazar, Bj\"orn Schenke, and Raju Venugopalan.
\newblock {Back-to-back inclusive dijets in DIS at small x: Sudakov suppression
  and gluon saturation at NLO}.
\newblock {\em JHEP}, 11:169, 2022.

\bibitem{Caucal:2023nci}
Paul Caucal, Farid Salazar, Bj\"orn Schenke, Tomasz Stebel, and Raju
  Venugopalan.
\newblock {Back-to-back inclusive dijets in DIS at small x: gluon
  Weizs\"acker-Williams distribution at NLO}.
\newblock {\em JHEP}, 08:062, 2023.

\bibitem{Caucal:2023fsf}
Paul Caucal, Farid Salazar, Bj\"orn Schenke, Tomasz Stebel, and Raju
  Venugopalan.
\newblock {Back-to-back inclusive dijets in DIS at small $x$: Complete NLO
  results and predictions}.
\newblock 7 2023.

\bibitem{Stasto:2018rci}
Anna Stasto, Shu-Yi Wei, Bo-Wen Xiao, and Feng Yuan.
\newblock {On the Dihadron Angular Correlations in Forward $pA$ collisions}.
\newblock {\em Phys. Lett. B}, 784:301--306, 2018.

\bibitem{Balitsky:2022vnb}
Ian Balitsky and Giovanni~A. Chirilli.
\newblock {Rapidity evolution of TMDs with running coupling}.
\newblock {\em Phys. Rev. D}, 106(3):034007, 2022.

\bibitem{Balitsky:2023hmh}
Ian Balitsky.
\newblock {Rapidity-only TMD factorization at one loop}.
\newblock {\em JHEP}, 03:029, 2023.

\bibitem{Zhou:2016tfe}
Jian Zhou.
\newblock {The evolution of the small x gluon TMD}.
\newblock {\em JHEP}, 06:151, 2016.

\bibitem{Zhou:2018lfq}
Jian Zhou.
\newblock {Scale dependence of the small x transverse momentum dependent gluon
  distribution}.
\newblock {\em Phys. Rev. D}, 99(5):054026, 2019.

\bibitem{Hentschinski:2020tbi}
Martin Hentschinski, Krzysztof Kutak, and Andreas van Hameren.
\newblock {Forward Higgs production within high energy factorization in the
  heavy quark limit at next-to-leading order accuracy}.
\newblock {\em Eur. Phys. J. C}, 81(2):112, 2021.
\newblock [Erratum: Eur.Phys.J.C 81, 262 (2021)].

\bibitem{Hentschinski:2021lsh}
Martin Hentschinski.
\newblock {Transverse momentum dependent gluon distribution within high energy
  factorization at next-to-leading order}.
\newblock {\em Phys. Rev. D}, 104(5):054014, 2021.

\bibitem{Celiberto:2022fgx}
Francesco~Giovanni Celiberto, Michael Fucilla, Dmitry~Yu. Ivanov, Mohammed
  M.~A. Mohammed, and Alessandro Papa.
\newblock {The next-to-leading order Higgs impact factor in the infinite
  top-mass limit}.
\newblock {\em JHEP}, 08:092, 2022.

\bibitem{Neill:2023jcd}
Duff Neill, Aditya Pathak, and Iain~W. Stewart.
\newblock {Small-x factorization from effective field theory}.
\newblock {\em JHEP}, 09:089, 2023.

\bibitem{Stewart:2023lwz}
Iain Stewart and Varun Vaidya.
\newblock {Power Counting to Saturation}.
\newblock 5 2023.

\bibitem{Catani:1990eg}
S.~Catani, M.~Ciafaloni, and F.~Hautmann.
\newblock {High-energy factorization and small x heavy flavor production}.
\newblock {\em Nucl. Phys. B}, 366:135--188, 1991.

\bibitem{Catani:1990xk}
S.~Catani, M.~Ciafaloni, and F.~Hautmann.
\newblock {GLUON CONTRIBUTIONS TO SMALL x HEAVY FLAVOR PRODUCTION}.
\newblock {\em Phys. Lett. B}, 242:97--102, 1990.

\bibitem{Catani:1994sq}
S.~Catani and F.~Hautmann.
\newblock {High-energy factorization and small x deep inelastic scattering
  beyond leading order}.
\newblock {\em Nucl. Phys. B}, 427:475--524, 1994.

\bibitem{Salam:1998tj}
G.~P. Salam.
\newblock {A Resummation of large subleading corrections at small x}.
\newblock {\em JHEP}, 07:019, 1998.

\bibitem{Altarelli:1999vw}
Guido Altarelli, Richard~D. Ball, and Stefano Forte.
\newblock {Resummation of singlet parton evolution at small x}.
\newblock {\em Nucl. Phys. B}, 575:313--329, 2000.

\bibitem{Ciafaloni:1999au}
M.~Ciafaloni, D.~Colferai, and G.~P. Salam.
\newblock {A collinear model for small x physics}.
\newblock {\em JHEP}, 10:017, 1999.

\bibitem{Ciafaloni:1999yw}
M.~Ciafaloni, D.~Colferai, and G.~P. Salam.
\newblock {Renormalization group improved small x equation}.
\newblock {\em Phys. Rev. D}, 60:114036, 1999.

\bibitem{Altarelli:2001ji}
Guido Altarelli, Richard~D. Ball, and Stefano Forte.
\newblock {Factorization and resummation of small x scaling violations with
  running coupling}.
\newblock {\em Nucl. Phys. B}, 621:359--387, 2002.

\bibitem{Altarelli:2003hk}
Guido Altarelli, Richard~D. Ball, and Stefano Forte.
\newblock {An Anomalous dimension for small x evolution}.
\newblock {\em Nucl. Phys. B}, 674:459--483, 2003.

\bibitem{Ciafaloni:2003rd}
M.~Ciafaloni, D.~Colferai, G.~P. Salam, and A.~M. Stasto.
\newblock {Renormalization group improved small x Green's function}.
\newblock {\em Phys. Rev. D}, 68:114003, 2003.

\bibitem{Ciafaloni:2003ek}
M.~Ciafaloni, D.~Colferai, D.~Colferai, G.~P. Salam, and A.~M. Stasto.
\newblock {Extending QCD perturbation theory to higher energies}.
\newblock {\em Phys. Lett. B}, 576:143--151, 2003.

\bibitem{Ciafaloni:2003kd}
Marcello Ciafaloni, Dimitri Colferai, Gavin~P. Salam, and Anna~M. Stasto.
\newblock {The Gluon splitting function at moderately small x}.
\newblock {\em Phys. Lett. B}, 587:87--94, 2004.

\bibitem{Marzani:2015oyb}
Simone Marzani.
\newblock {Combining $Q_T$ and small-$x$ resummations}.
\newblock {\em Phys. Rev. D}, 93(5):054047, 2016.

\bibitem{Colferai:2023dcf}
Dimitri Colferai, Wanchen Li, and Anna~M. Stasto.
\newblock {Renormalization group improved photon impact factors and the high
  energy virtual photon scattering}.
\newblock 11 2023.

\bibitem{Ji:2014hxa}
Xiangdong Ji, Peng Sun, Xiaonu Xiong, and Feng Yuan.
\newblock {Soft factor subtraction and transverse momentum dependent parton
  distributions on the lattice}.
\newblock {\em Phys. Rev. D}, 91:074009, 2015.

\bibitem{Ji:2018hvs}
Xiangdong Ji, Lu-Chang Jin, Feng Yuan, Jian-Hui Zhang, and Yong Zhao.
\newblock {Transverse momentum dependent parton quasidistributions}.
\newblock {\em Phys. Rev. D}, 99(11):114006, 2019.

\bibitem{Ebert:2018gzl}
Markus~A. Ebert, Iain~W. Stewart, and Yong Zhao.
\newblock {Determining the Nonperturbative Collins-Soper Kernel From Lattice
  QCD}.
\newblock {\em Phys. Rev. D}, 99(3):034505, 2019.

\bibitem{Ebert:2019okf}
Markus~A. Ebert, Iain~W. Stewart, and Yong Zhao.
\newblock {Towards Quasi-Transverse Momentum Dependent PDFs Computable on the
  Lattice}.
\newblock {\em JHEP}, 09:037, 2019.

\bibitem{Ebert:2019tvc}
Markus~A. Ebert, Iain~W. Stewart, and Yong Zhao.
\newblock {Renormalization and Matching for the Collins-Soper Kernel from
  Lattice QCD}.
\newblock {\em JHEP}, 03:099, 2020.

\bibitem{Ji:2019sxk}
Xiangdong Ji, Yizhuang Liu, and Yu-Sheng Liu.
\newblock {TMD soft function from large-momentum effective theory}.
\newblock {\em Nucl. Phys. B}, 955:115054, 2020.

\bibitem{Ji:2019ewn}
Xiangdong Ji, Yizhuang Liu, and Yu-Sheng Liu.
\newblock {Transverse-momentum-dependent parton distribution functions from
  large-momentum effective theory}.
\newblock {\em Phys. Lett. B}, 811:135946, 2020.

\bibitem{Ebert:2020gxr}
Markus~A. Ebert, Stella~T. Schindler, Iain~W. Stewart, and Yong Zhao.
\newblock {One-loop Matching for Spin-Dependent Quasi-TMDs}.
\newblock {\em JHEP}, 09:099, 2020.

\bibitem{Ji:2020jeb}
Xiangdong Ji, Yizhuang Liu, Andreas Sch\"afer, and Feng Yuan.
\newblock {Single Transverse-Spin Asymmetry and Sivers Function in Large
  Momentum Effective Theory}.
\newblock {\em Phys. Rev. D}, 103(7):074005, 2021.

\bibitem{Ji:2021znw}
Xiangdong Ji and Yizhuang Liu.
\newblock {Computing light-front wave functions without light-front
  quantization: A large-momentum effective theory approach}.
\newblock {\em Phys. Rev. D}, 105(7):076014, 2022.

\bibitem{Ebert:2022fmh}
Markus~A. Ebert, Stella~T. Schindler, Iain~W. Stewart, and Yong Zhao.
\newblock {Factorization connecting continuum \& lattice TMDs}.
\newblock {\em JHEP}, 04:178, 2022.

\bibitem{Deng:2022gzi}
Zhi-Fu Deng, Wei Wang, and Jun Zeng.
\newblock {Transverse-momentum-dependent wave functions and soft functions at
  one-loop in large momentum effective theory}.
\newblock {\em JHEP}, 09:046, 2022.

\bibitem{Tarasov:2021yll}
Andrey Tarasov and Raju Venugopalan.
\newblock {Role of the chiral anomaly in polarized deeply inelastic scattering.
  II. Topological screening and transitions from emergent axionlike dynamics}.
\newblock {\em Phys. Rev. D}, 105(1):014020, 2022.

\bibitem{Tarasov:2020cwl}
Andrey Tarasov and Raju Venugopalan.
\newblock {Role of the chiral anomaly in polarized deeply inelastic scattering:
  Finding the triangle graph inside the box diagram in Bjorken and Regge
  asymptotics}.
\newblock {\em Phys. Rev. D}, 102(11):114022, 2020.

\bibitem{Adamiak:2023yhz}
Daniel Adamiak, Nicholas Baldonado, Yuri~V. Kovchegov, W.~Melnitchouk, Daniel
  Pitonyak, Nobuo Sato, Matthew~D. Sievert, Andrey Tarasov, and Yossathorn
  Tawabutr.
\newblock {Global analysis of polarized DIS \& SIDIS data with improved
  small-$x$ helicity evolution}.
\newblock 8 2023.

\bibitem{Schwinger:1951nm}
Julian~S. Schwinger.
\newblock {On gauge invariance and vacuum polarization}.
\newblock {\em Phys. Rev.}, 82:664--679, 1951.

\end{thebibliography}

\end{document}